\documentclass[traditabstract]{aa}  
\usepackage[varg]{txfonts}
\usepackage{graphicx}
\usepackage{subfigure}
\usepackage{rotating}
\usepackage{color}
\usepackage{ragged2e}
\usepackage{soul} 
\usepackage{float}
\usepackage{amsmath} 
\usepackage{wrapfig,lipsum}

\usepackage[flushleft]{threeparttable} 

\begin{document}
\defcitealias{2020A&A...641A.171M}{M20}

\title{Central star formation in double-peak gas rich radio galaxies}
\titlerunning{Central star formation in double-peak gas rich radio galaxies}
\authorrunning{Maschmann et al.}

\author{Daniel Maschmann\inst{1}, Anne-Laure Melchior\inst{1}, Francoise Combes\inst{1, 2}, Barbara Mazzilli Ciraulo\inst{1}, Jonathan Freundlich\inst{3}, Ana\"elle Halle\inst{1} and Alexander Drabent\inst{4}\\ 
}
\institute{Sorbonne {Universit\'e},
LERMA, Observatoire de Paris, PSL university, CNRS, F-75014, Paris, France\\
\email{Daniel.Maschmann@observatoiredeparis.psl.eu, A.L.Melchior@oobservatoiredeparis.psl.eu}
\and
Coll\`ege de France, 11, Place Marcelin Berthelot, F-75005, Paris, France
\and
Universit\'e de Strasbourg, CNRS, Observatoire astronomique de Strasbourg, UMR 7550, F-67000 Strasbourg, France
\and
Th\"uringer Landessternwarte, Sternwarte 5, 07778 Tautenburg, Germany
}
\date{Received 14 September 2021/ accepted 11 May 2022}
\abstract {
The respective contributions of gas accretion, galaxy interactions and mergers to the mass assembly of galaxies, as well as the evolution of their molecular gas and star formation activity are still not fully understood.    
In a recent work, a large sample of double-peak (DP) emission line galaxies have been identified from the SDSS. 
While the two peaks could represent two kinematic components, they may be linked to the large bulges which their host galaxies tend to have. Star-forming DP galaxies display a central star formation enhancement and have been discussed to be compatible with a sequence of recent minor mergers. 
In order to probe merger induced star formation mechanisms, we conducted observations of the molecular gas content of 35 star-forming DP galaxies in the upper part of the main sequence (MS) of star formation with the IRAM 30m telescope. Including similar galaxies 0.3 dex above MS and 
with existing molecular gas observations from the literature, we finally obtain a sample of 52 such galaxies. 
We succeed in fitting the same kinematic parameters to the optical ionised and molecular gas emission lines for 10 ($19\,\%$) galaxies.
We find a central star formation enhancement resulting most likely from a galaxy merger or galaxy interaction, which is indicated by an excess of gas extinction found in the centre. 
This star formation is traced by radio continuum emissions at 150\,MHz, 1.4\,GHz and 3\,GHz, which are all three linearly correlated in log with the CO luminosity with the same slope.
The 52 DP galaxies are found to have a significantly larger amount of molecular gas and larger depletion times, hence a lower star formation (SF) efficiency, than the expected values at their distance of the MS. The large bulges in these galaxies might be stabilising the gas, hence reducing the SF efficiency. This is consistent with a scenario of minor mergers increasing the mass of bulges and driving gas to the centre.
We also exclude the inwards directed gas migration and central star formation enhancement to be the origin of a bar morphology.
Hence, these 52 DP galaxies could be the results of recent minor mergers that funnelled molecular gas towards their centre, triggering star formation, but with a moderate efficiency.
}

\maketitle
\section{Introduction}\label{sect:introduction}
The evolutionary state of galaxies depends mostly on their growth rate and their efficiency to transform gas into stars. Galaxy interactions, smooth accretion of gas, and internal mechanisms such as AGN feedback all affect the gas content and the star formation. 
Galaxy interactions and mergers are well-known to enhance the star formation rate \citep{1986ApJ...301...57B, 2002MNRAS.331..333P}. However, while they tend to increase the molecular gas content \citep{1994A&A...281..725C, 2018MNRAS.476.2591V, 2019A&A...627A.107L}, their effect on the evolution of the neutral hydrogen gas fraction is still an open question. Some studies find little difference in close galaxy pairs \citep[e.g.][]{2018ApJS..237....2Z, 1993A&A...269....7B} or post-merger galaxies \citep[e.g.][]{2015MNRAS.448..221E} compared to the general population of similar galaxies. Other studies find an enhancement of the atomic gas fraction in recently merged galaxies \citep{2008A&A...492..367H, 2018MNRAS.478.3447E}, or a deficit in the final stages of merging \citep{1996AJ....111..655H}. The environment can be also responsible for the final quenching of a galaxy \citep{1998ApJ...504L..75B}. 
Interactions and mergers can also drive gas towards the centre and hence fuel a nuclear black hole, enhancing active galactic nuclei (AGN) activity and feedback \citep{2006MNRAS.365...11C, 2005MNRAS.361..776S}, which can then influence star formation in the host galaxy \citep{2017ApJ...850...27B, 2017A&A...606A..36C, 2017ApJ...839..120W}. 
In cases of powerful AGN, the radiation can shut down the star formation entirely \citep{2005Natur.433..604D, 2006MNRAS.365...11C, 2009Natur.460..213C}.
Relying on simulations, \citet{2021ApJ...911..116S} discussed that two successive minor merger events can quench Milky-Way like galaxies through AGN feedback. 
Based on the projected distances between galaxies and projected velocities, \citet{2008AJ....135.1877E} and \citet{2011MNRAS.412..591P} conducted studies on large galaxy pairs samples and the associated effects. 
They found an increase of central star formation with decreasing galaxy separation. 
By extending the pair search with quasi stellar objects and AGN, \citet{2011MNRAS.418.2043E} found that AGN activity can be triggered by galaxy interactions before the final coalescence. 

To explain the overall growth of galaxies over cosmic time, \citet{2016MNRAS.457.2790T} described a scenario of recurring episodes of gas in-fall and depletion phases. Gas is accreted into a galaxy in large amounts through streams from the surroundings \citep{2009Natur.457..451D} or through minor merger events, causing a contraction of the gas disc with efficient star formation sites and a central enhancement \citep{2014MNRAS.438.1870D}. This shifts the galaxy above the star formation main sequence (MS), before gas depletion lets the galaxy descend underneath the MS.
Smooth gas accretion from filaments 
\citep{2010ApJ...718.1001B, 2011MNRAS.415...11D, 2012MNRAS.421...98D, 2013MNRAS.433.1910F, 2013ApJ...772..119L, 2013MNRAS.435..999D, 2014MNRAS.443.3643P, 2014MNRAS.438.1870D} can explain that most galaxies on the MS exhibit a disc morphology \citep{2006ApJ...645.1062F, 2006Natur.442..786G, 2008ApJ...687...59G, 2008Natur.455..775S, 2010ApJ...713..686D, 2011ApJ...738..106W} and that star-forming galaxies at $z = 1-2$ experience long sustained star formation cycles \citep{2005ApJ...626..680D, 2007ApJ...670..173D, 2006ApJ...637..727C}. 

The occurrence of double-peak (DP) emission lines in spectra of galaxies can have different causes, amongst which galaxy mergers. As predicted by \citet{1980Natur.287..307B}, galaxy mergers lead at one point to the final coalescence of the super-massive black holes of their progenitors. Earlier stages of this scenario have been reported many times in the form of dual AGN \citep[e.g.][]{2001ApJ...563..527G,2016ApJ...824L...4K,2018Natur.563..214K,2019ApJ...879L..21G}. 
Such galaxy mergers can be identified through kinematic signatures with spectroscopic observations.
Post-coalescence mergers can create two separated gas populations, which can be observed as DP emission lines. This has been studied in works focusing on AGN \citep[e.g.][]{2009ApJ...698..956C, 2011ApJ...737..101L, 2012ApJ...746L..22K, 2013ApJ...762..110L, 2013ApJ...777...64C, 2015ApJ...799...72F}. DP signatures were found to be related to merging processes in \citet{2018ApJ...867...66C} and \citet{2019A&A...627L...3M}. In a recent study, \citet{2021A&A...653A..47M} succeeded in resolving two independent kinematic components using integrated field spectroscopy of a DP emission line galaxy and identified two galaxies in the process of merging, superimposed in projection along the line of sight.

In order to discuss the nature of DP emission line galaxies, \citet{2020A&A...641A.171M} (hereafter \citetalias{2020A&A...641A.171M}) developed an automated selection procedure and found 5663 DP galaxies, including non-AGN galaxies. A systematic search for DP emission-line galaxies had also been conducted by \citet{2012ApJS..201...31G} including also non-AGN galaxies such as star-forming or composite galaxies. \citetalias{2020A&A...641A.171M} relied on reduced spectra provided by the value-added Reference Catalogue of Spectral Energy Distributions (RCSED) \citep{2017ApJS..228...14C}, and compared them to single-peaked galaxies with the same emission line signal-to-noise (S/N) properties and with the same redshift and stellar mass distributions. 
They found that most of the DP galaxies are massive star-forming galaxies characterised by a central enhancement of their star formation rate. In addition, they exhibit a large bulge with a Sersic index larger than for the single-peaked galaxies comparison sample. While this configuration could result from repetitive minor mergers as discussed by \citet{2007A&A...476.1179B}, it could also correspond to a rotating inner structure. However, as discussed in details in \citet{2021A&A...653A..47M}, integrated field spectroscopy is needed to identify two individual gas components and conclusively identify galaxy mergers. Here, we will study statistical properties of a sub-sample of 52 galaxies with SDSS spectra. 

In this work, we explore the most extreme part of this DP sample, focusing on DP galaxies located more than 0.3 dex above the MS. We perform new molecular gas observations at IRAM-30m telescope and complete the sample with existing molecular gas observations from the literature.
We aim at studying the relation between the molecular gas content and the star formation activity. In order to test possible biases due to dust, we also use the radio continuum emission, extensively studied as a tracer of SF \citep{1992ARA&A..30..575C, 2003ApJ...586..794B, 2006ApJ...643..173S, 2011ApJ...737...67M}.
We also rely on the kinematics to explore the possible connection between the ionised and molecular gas. We use these combined analyses to probe the relation between galaxy merging and star formation mechanisms.

This work is organised as follows. In Sect.\,\ref{sect:data}, the sample is defined with a description of the CO observations performed at IRAM-30m and the data selection from the literature.
In Sect.\,\ref{sect:data_analysis}, we describe the emission line fitting and the characteristics of the galaxy sample. We analyse the sample in Sect.\,\ref{sect:results} with different star formation tracers, and calculate the molecular gas content. We present the Kennicutt-Schmidt relation and explore the connection between the CO luminosity and radio continuum emission. Lastly, we discuss our results in Sect.\,\ref{sect:discussion} and present the conclusion in Sect.\,\ref{sect:conclusion}. 

A cosmology of $\Omega_{m} = 0.3$, $\Omega_{\Lambda} = 0.7$ and $h = 0.7$ is assumed in this work.

\section{Data}\label{sect:data}
We focus on a sample of 52 DP galaxies lying more than 0.3 dex above the main sequence of star formation and gather a few comparison galaxy samples. In Sect. \ref{ssect:sample_selection},  the 52-galaxies sample is presented: the selection of 35 galaxies of \citetalias{2020A&A...641A.171M} in Sect. \ref{ssect:dp_selection},  their observation in CO at IRAM-30m in \ref{sssect:observation}, and the selection of 17 additional galaxies with CO observations available from the literature in Sect. \ref{ssect:sample_selection:lit}. In Sect. \ref{ssect:existing_co}, different comparison galaxy samples obtained from existing CO and SFR measurements are described. Lastly, in \ref{ssect:ms}, all the galaxies of the different samples are displayed with respect to distance to the main sequence of star formation as a function of their stellar mass. In Sect. \ref{ssect:radio_continuum}, the different samples are cross-identified with existing radio-continuum surveys.
\begin{figure}[h]
\centering 
\includegraphics[width=0.48\textwidth]{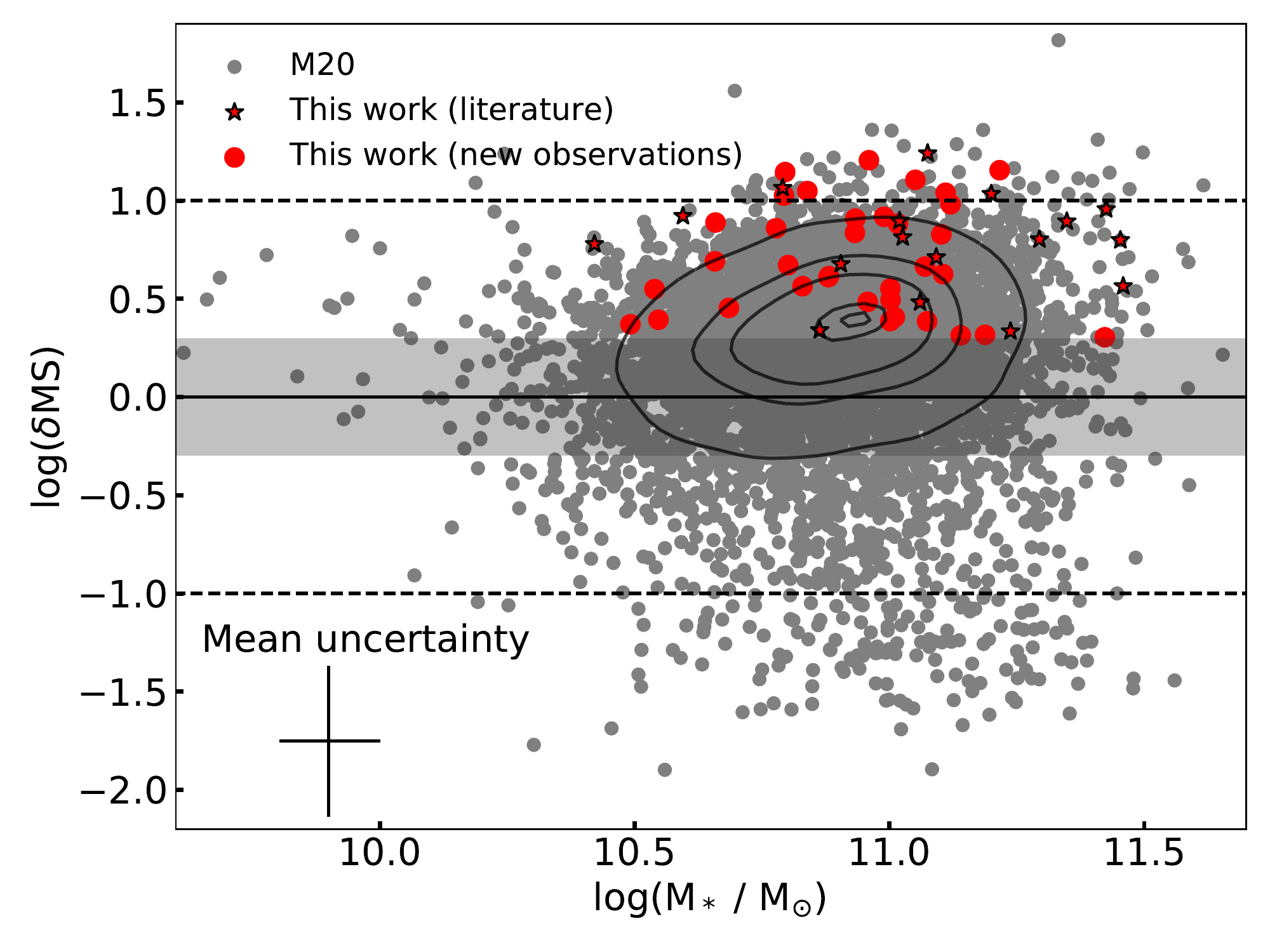}
\caption{Offset from the MS for all CO samples as a function of stellar mass (${\rm M_{*}}$) using the parametrisation of the MS found by \citet{2014ApJS..214...15S}. The shaded area marks the 0.3\,dex and the dashed lines the $\pm 1\,{\rm dex}$ scatter. We use the SFR computed by \citet{2004MNRAS.351.1151B} and the stellar mass from \citet{2003MNRAS.341...33K}.
We show the \citetalias{2020A&A...641A.171M} sample with grey dots and mark the DP sample (definition in Sect.\,\ref{ssect:sample_selection}) with red circles and stars. Circles represent galaxies with new CO observations and stars represent galaxies for which we obtained the CO measurements from the literature. The contour lines show the density of the \citetalias{2020A&A...641A.171M} sample. In the bottom left, we show the mean uncertainties.}
\label{fig:simple:ms}%
\end{figure}

\subsection{Sample of 52 DP galaxies 0.3 dex above the MS}\label{ssect:sample_selection}

The main sample of this work consists of \citetalias{2020A&A...641A.171M} DP galaxies lying more than 0.3\,dex above the MS, observed during two observing runs with the IRAM 30m telescope in April and December 2020.

\subsubsection{Selection of \citetalias{2020A&A...641A.171M} DP galaxies 0.3 dex above the MS}\label{ssect:dp_selection}
We compute the star formation rate of the MS ${\rm SFR_{MS} = SFR(MS; z, M_{*})}$ as parametrised by \citet{2014ApJS..214...15S}, at the redshifts $z$ and stellar masses M$_{*}$ of the DP galaxies of \citetalias{2020A&A...641A.171M}, and compute their offset from the MS as ${\rm \delta MS = SFR / SFR_{MS}}$, using the SFR computed by \citet{2004MNRAS.351.1151B} and the stellar masses from \citep{2003MNRAS.341...33K}.
The MS is estimated from observations with a typical scatter of ${\rm \delta MS \sim 0.3\,dex}$ \citep[e.g][]{2007ApJ...660L..43N, 2011ApJ...739L..40R, 2012ApJ...754L..29W, 2015A&A...575A..74S}.
To target galaxies with increased star formation activity in comparison to the MS, we thus select galaxies which are located at least ${\rm \delta MS = 0.3\,dex}$ above the MS. With this criterion, we aim to focus our work on galaxies with ongoing star formation which can either be recently induced by galaxy interaction or gas accretion \citep[e.g.][]{1986ApJ...301...57B, 2002MNRAS.331..333P} or be the remaining of a faded starburst event \citep{2009ApJ...690.1672S, 2015ApJ...801....1F}. We select 35 DP galaxies, which we observed with the IRAM 30m telescope. These galaxies correspond to the red dots of Fig.\,\ref{fig:simple:ms}, on which the whole parent \citetalias{2020A&A...641A.171M} sample is shown in grey dots.

\subsubsection{IRAM-30m observations of the selected \citetalias{2020A&A...641A.171M} galaxies}\label{sssect:observation}
We observed the 35 DP galaxies during two observing runs from the 21st to the 24th of April 2020 and from the 23rd to the 29th of December 2020 with the IRAM 30m telescope at Pico Veleta in Spain. Galaxies with a redshift $z < 0.144$ could be observed simultaneously in the CO(1-0) and CO(2-1) lines, and 5 galaxies at higher redshift were only observable in CO(1-0) at the time. The mean emission line wavelengths were about $\sim 3\,{\rm mm}$ for the CO(1-0) line and $\sim 1.5\,{\rm mm}$ for the CO(2-1) line. Thick clouds and snow prevented us from observing during 1.5 nights during the first run and 2 nights during the second run but we were able to observe all proposed galaxies during the remaining time under excellent conditions. 

The galaxies were observed using the broad-band EMIR receiver, tuned in single-band mode with a total band-width of 3.715\,GHz per polarisation. This allows us to observe an average velocity range of 11\,140\,km\,s$^{-1}$ for the CO(1-0) line, and 5\,570\,km\,s$^{-1}$ for the CO(2-1) line. The Wobbler switching mode was used to carry out the observations and the backends WILMA and FTS were used in parallel with a channel width of 2\,MHz and 0.195\,MHz, respectively.

We pointed on average one hour at each galaxy and reached noise levels between 0.1 and 1.8\,mK (main-beam temperature), smoothed over 60\,km/s. 
Focus measurements were performed at the beginning of the night and at dawn, as well as pointing measurements every 2 hours. 
The temperature scale we use here is main-beam temperature and the beam size is $\lambda / {\rm D} = 22^{\prime\prime}$ at 2.8\,mm and $12^{\prime\prime}$ at 1.4\,mm wavelength with an average beam efficiency of $\eta_{\rm mb} = {\rm T^{*}_{A} / T_{mb}} = 0.76$ and 0.56 respectively. The observation data were reduced using the CLASS/GILDAS software. We transformed the observed main-beam temperature into units of spectral flux density by using the IRAM 30m antenna factor of ${\rm 5 \, Jy / K}$, in order to compare our observations with other CO samples. 

\subsubsection{Inclusion of galaxies with CO data from the literature}\label{ssect:sample_selection:lit}
In order to enlarge our sample, we further select DP galaxies lying more than 0.3 dex above the MS from published CO observations. An emission line fit with the method described in \citetalias{2020A&A...641A.171M} is shown in Fig.\,\ref{fig:line:fit} for DP-8\footnote{ Note that the continuum of the [OI]$\lambda 6302$ line shows a small dip. This is most likely due to the fact that for the stellar continuum fit the emission lines are masked and structures close to the emission lines cannot be accurately modelled. Since we fit all emission lines simultaneously with the same kinematic parameters, this has no effect on the emission line fit.}, one of the 35 galaxies taken from \citetalias{2020A&A...641A.171M}. For galaxies from the literature, we perform a simplified DP selection procedure compared to the one of \citetalias{2020A&A...641A.171M}, especially with no emission line stacking nor multiple selection stages: Our present algorithm consists of a simultaneous fit of multiple emission lines and selection criteria but we finally rely on a visual inspection, to exclude some noisy spectra but also to enlarge the selection to galaxies with strongly perturbed gas kinematics making emission lines deviate from pure double Gaussian profiles.  
The identification of DP emission line galaxies in literature samples is hence as follows. The best fitting stellar continuum template provided by \citet{2017ApJS..228...14C} is first subtracted from the SDSS spectrum to get the pure emission line spectrum. Then we fit single and double Gaussian functions to the emission lines H$\beta$, [OIII]$\lambda 4960$, [OIII]$\lambda 5008$, [OI]$\lambda 6302$, [NII]$\lambda 6550$, H$\alpha$, [NII]$\lambda 6585$, [SII]$\lambda 6718$ and [SII]$\lambda 6733$ simultaneously and use a global velocity $\mu$ (resp. $\mu_1$ and $\mu_2$ for the double Gaussian fit) and a Gaussian standard deviation $\sigma$ (resp. $\sigma_1$ and $\sigma_2$) for all emission lines but keep the individual emission line amplitudes as free parameters. We also include the spectral instrumental broadening $\sigma_{\rm inst}$ into the fitted $\sigma$ for each observed emission line individually to get the observed Gaussian velocity dispersion $\sigma_{\rm obs} = \sqrt{\sigma_{\rm inst}^2 + \sigma^2}$. We pre-select galaxies which are selected by the F-test criterion, with an emission line separation $\Delta v = |\mu_1 - \mu_2|$ of at least $200\,{\rm km\,s}^{-1}$ and an amplitude ratio of the [OIII]$\lambda 5008$ or H$\alpha$ line to be between 1/3 and 3, as described in detail in \citetalias{2020A&A...641A.171M}. 

We select 17 DP galaxies from the literature:
\begin{itemize}
    \item 11 galaxies observed with the Combined Array for Research in Millimeter-wave Astronomy (CARMA) by \citet{2013ApJ...768..132B}
    \item 3 ultra luminous infrared galaxies (ULIRG) observed with the 14m telescope of the Five College Radio Astronomy Observatory (FCRAO) observed by \citet{2009AJ....138..858C}
    \item 2 galaxies observed with the IRAM 30m telescope as part of the COLD GASS survey \citep{2011MNRAS.415...32S, 2017ApJS..233...22S}
    \item 1 galaxy observed by \protect\citet{1997ApJ...478..144S}, which is known as Arp 220. We found ALMA CO(1-0) observation for this galaxy in the ESO-archives\footnote{\url{http://archive.eso.org/scienceportal}}. With the high spatial resolution of 37\,pc, \citet{2017ApJ...836...66S} succeeded to precisely locate the two nuclei and study their nuclear gas discs.
    We extract the molecular gas observation from the exact same location as the $3^{\prime\prime}$ SDSS fibre and also from the entire galaxy. We note that the majority of the molecular gas coincides with the two nuclei of Arp 220. However, these two nuclei are strongly obscured by dust and the SDSS $3^{\prime\prime}$ fibre observation is centred about $4^{\prime\prime}$ north of the two nuclei \citep{2017ApJ...836...66S}.
\end{itemize}

We thus gather a DP galaxy sample with CO observations of 52 galaxies lying more than 0.3 dex above the MS: the 35 galaxies from  \citetalias{2020A&A...641A.171M} for which we present new CO observations and the 17 galaxies selected from the literature. This sample is presented in Table\,\ref{table:sample} with characteristic measurements such as the redshift, stellar mass, SFR, radio continuum fluxes, galaxy size and inclination.
\begin{table*}
\caption{Characteristics of DP galaxy sample}\label{table:sample}
\centering
\begin{tabular}{ l c c c c c c c c c}        
\hline\hline
\multicolumn{1}{c}{ID} & \multicolumn{1}{c}{SDSS Designation} & \multicolumn{1}{c}{z} & \multicolumn{1}{c}{log(M$_{*}$/M$_{\odot}$)} & \multicolumn{1}{c}{SFR} & \multicolumn{1}{c}{F$_{\rm 150\, MHz}$} & \multicolumn{1}{c}{F$_{\rm 1.4\, GHz}$} & \multicolumn{1}{c}{F$_{\rm 3\, GHz}$} & \multicolumn{1}{c}{D$_{25}$} & \multicolumn{1}{c}{$i$} \\ 
\hline
\multicolumn{1}{c}{} & \multicolumn{1}{c}{} & \multicolumn{1}{c}{} & \multicolumn{1}{c}{} & \multicolumn{1}{c}{M$_{\odot}$ yr$^{-1}$} & \multicolumn{1}{c}{mJy} & \multicolumn{1}{c}{mJy} & \multicolumn{1}{c}{mJy} & \multicolumn{1}{c}{arcsec} & \multicolumn{1}{c}{$^{\rm o}$} \\ 
\hline
DP-1$^{a}$ & J153457.21+233013.3 & 0.0181 & 10.9 & 4.7 & - & 316.1 & 242.8 & 90.8 & - \\
DP-2 & J143117.98+075640.7 & 0.0269 & 11.0 & 7.7 & - & 5.8 & 3.1 & 65.8 & 66 \\
DP-3 & J142129.76+050423.6 & 0.0271 & 10.5 & 3.9 & - & 5.4 & 2.8 & 82.2 & 61 \\
DP-4 & J130702.99+130429.3 & 0.0274 & 10.5 & 3.5 & - & 5.0 & 2.3 & 61.6 & 58 \\
DP-5 & J141721.07+265126.8 & 0.0367 & 11.1 & 13.4 & - & 21.0 & 10.6 & 60.9 & - \\
DP-6 & J141916.59+261755.0 & 0.0368 & 11.1 & 7.4 & - & 9.7 & 8.1 & 59.7 & 50 \\
DP-7 & J160457.92+140815.3 & 0.0372 & 10.5 & 5.8 & - & 2.6 & 1.8 & 43.5 & 51 \\
DP-8 & J143713.73+143954.5 & 0.0379 & 10.9 & 10.1 & - & 9.1 & 5.4 & 25.2 & - \\
DP-9 & J141238.95+273740.7 & 0.0386 & 10.7 & 9.3 & - & 2.7 & 1.6 & 44.6 & 64 \\
DP-10 & J090007.20+600458.1 & 0.0390 & 11.0 & 43.5 & 8.5 & 4.8 & 2.9 & 38.5 & 52 \\
DP-11 & J132357.46+120233.3 & 0.0390 & 11.0 & 7.3 & - & 4.1 & - & 41.4 & 56 \\
DP-12 & J135309.67+143920.9 & 0.0405 & 11.2 & 7.4 & - & 18.0 & 3.9 & 47.0 & - \\
DP-13$^{b}$ & J020359.16+141837.3 & 0.0427 & 10.9 & 12.3 & - & - & 3.3 & 47.4 & 62 \\
DP-14 & J105716.67+283230.0 & 0.0457 & 10.9 & 10.4 & 8.4 & 1.6 & 1.8 & 43.7 & 56 \\
DP-15 & J113507.51+295327.7 & 0.0462 & 11.0 & 7.2 & 15.2 & 3.3 & 2.8 & 49.6 & - \\
DP-16 & J145415.59+254121.3 & 0.0479 & 10.8 & 24.8 & - & 3.3 & 1.9 & 25.3 & 48 \\
DP-17$^{b}$ & J091954.54+325559.8 & 0.0490 & 10.4 & 9.1 & 36.4 & 22.3 & 7.8 & 65.8 & - \\
DP-18 & J094142.69+283555.6 & 0.0538 & 11.1 & 7.5 & 10.7 & 2.8 & 2.0 & 45.2 & 58 \\
DP-19$^{c}$ & J233455.24+141731.1 & 0.0621 & 11.0 & 21.5 & - & - & 1.7 & 27.5 & - \\
DP-20 & J120854.47+472833.3 & 0.0677 & 10.7 & 17.0 & 7.2 & 2.4 & - & 19.6 & 36 \\
DP-21 & J134316.22+524742.5 & 0.0690 & 11.1 & 36.7 & 6.0 & 5.1 & 2.8 & 27.1 & 46 \\
DP-22 & J110428.21+560131.4 & 0.0702 & 11.0 & 10.5 & 15.8 & 4.7 & 2.6 & 33.1 & 42 \\
DP-23 & J110746.31+552633.3 & 0.0715 & 10.9 & 25.2 & 6.9 & 2.9 & 1.2 & 25.2 & - \\
DP-24$^{d}$ & J121346.08+024841.5 & 0.0731 & 10.6 & 17.5 & - & 24.6 & 17.0 & 4.5 & - \\
DP-25$^{d}$ & J102142.79+130656.1 & 0.0763 & 11.1 & 19.8 & - & 16.4 & 7.1 & 19.4 & - \\
DP-26 & J120437.97+525717.2 & 0.0815 & 10.9 & 22.4 & 10.4 & 5.2 & 1.4 & 31.0 & - \\
DP-27$^{c}$ & J221938.11+134213.9 & 0.0835 & 11.1 & 11.7 & - & - & - & 33.4 & - \\
DP-28 & J114325.16+505154.7 & 0.0844 & 11.0 & 28.0 & 10.1 & 4.0 & 2.1 & 22.6 & 39 \\
DP-29 & J131943.32+515255.8 & 0.0902 & 11.1 & 18.5 & 18.8 & 4.1 & 2.2 & 23.2 & 61 \\
DP-30 & J114050.50+561335.3 & 0.1065 & 10.8 & 43.7 & 6.9 & 1.1 & - & 19.3 & - \\
DP-31$^{d}$ & J135609.99+290535.1 & 0.1087 & 11.2 & 55.8 & 13.1 & 10.7 & 5.2 & 20.1 & - \\
DP-32 & J150439.86+501100.4 & 0.1124 & 10.8 & 15.2 & 7.8 & 1.5 & - & 15.0 & 45 \\
DP-33 & J150911.71+482041.2 & 0.1194 & 10.8 & 12.7 & 5.6 & 1.9 & - & 18.2 & 40 \\
DP-34 & J124406.54+503940.3 & 0.1211 & 10.8 & 23.7 & 8.4 & 2.0 & - & 13.2 & - \\
DP-35 & J130704.53+485845.5 & 0.1230 & 11.2 & 80.6 & 8.7 & 1.7 & 1.2 & 13.8 & 39 \\
DP-36 & J130847.69+504259.8 & 0.1244 & 11.0 & 35.6 & 8.7 & 3.5 & - & 14.6 & 29 \\
DP-37 & J143616.57+554822.0 & 0.1400 & 11.1 & 63.8 & 11.5 & 2.1 & 1.2 & 12.2 & - \\
DP-38 & J135705.89+523532.3 & 0.1437 & 11.1 & 36.5 & 5.4 & 2.8 & 1.0 & 19.0 & - \\
DP-39 & J113703.72+504420.7 & 0.1601 & 10.8 & 47.1 & 4.8 & 1.0 & - & 15.3 & 61 \\
DP-40 & J141803.61+534104.0 & 0.1638 & 11.1 & 65.9 & 10.9 & 2.2 & 1.0 & 14.1 & 39 \\
DP-41$^{c}$ & J100518.63+052544.2 & 0.1657 & 10.8 & 47.1 & - & - & - & 9.8 & - \\
DP-42$^{c}$ & J105527.19+064015.0 & 0.1731 & 11.0 & 44.1 & - & - & - & 11.9 & - \\
DP-43$^{c}$ & J091426.24+102409.6 & 0.1762 & 11.5 & 61.5 & - & 1.1 & - & 9.8 & - \\
DP-44$^{c}$ & J114649.18+243647.7 & 0.1767 & 11.1 & 106.1 & - & - & - & 6.7 & - \\
DP-45$^{c}$ & J134322.28+181114.1 & 0.1781 & 11.3 & 67.7 & - & - & - & 10.0 & - \\
DP-46 & J144017.35+563503.7 & 0.1801 & 11.4 & 19.3 & 5.3 & 1.6 & - & 19.3 & 21 \\
DP-47$^{c}$ & J223528.64+135812.7 & 0.1830 & 11.4 & 88.2 & - & - & 0.0 & 15.7 & - \\
DP-48 & J110333.10+475932.7 & 0.1906 & 11.0 & 21.1 & 6.5 & 1.4 & - & 9.9 & 42 \\
DP-49$^{c}$ & J002353.97+155947.9 & 0.1918 & 11.3 & 54.6 & - & - & - & 13.6 & - \\
DP-50 & J143211.77+495535.8 & 0.1938 & 10.7 & 11.4 & 3.1 & 1.5 & - & 9.8 & - \\
DP-51$^{c}$ & J092831.94+252313.9 & 0.2830 & 11.2 & 25.4 & - & - & - & 14.8 & - \\
DP-52$^{c}$ & J132047.14+160643.7 & 0.3124 & 11.5 & 64.9 & - & - & - & 10.2 & - \\
\hline
\end{tabular}
\tablefoot{The DP emission line galaxies sample consisting of 52 galaxies, observed in CO and lying more than 0.3 dex above the MS. We conducted CO observations for 35 galaxies and mark observations from the literature for the 17 remaining ones. We note galaxies taken from \citet{1997ApJ...478..144S} with the footnote $a$, from \citet{2017ApJS..233...22S} with $b$, from \citet{2013ApJ...768..132B} with $c$, and from \citet{2009AJ....138..858C} with $d$. We show the SDSS designation, redshift, stellar mass \citep{2003MNRAS.341...33K} and SFR \citep{2004MNRAS.351.1151B}. Radio fluxes at 150\,MHz are taken from \citet{2019A&A...622A...1S}, at 1.4\,GHz from \citep{1997ApJ...475..479W} and at 3\,GHz from \citep{2020PASP..132c5001L}. D$_{25}$ is the optical diameter at the 25\,mag isophote taken from the NASA/IPAC Extra galactic Database (NED)\footnote{https://ned.ipac.caltech.edu/}. We used the radii computed from the SDSS $r$-band observation or, if available, from photometric B-band observation. We further present the galaxy inclination calculated from a 2D S\'ersic profile fit, as described in detail in Sect.\,\ref{sssect:aperture}.}
\end{table*}

In Sect.\,\ref{ssect:existing_co}, we further discuss the total DP detection rate for each public CO galaxy sample included in this work. 

\subsection{Comparison samples}\label{ssect:existing_co}
In order to discuss the peculiarities of our DP sample of 52 galaxies, we assemble here complementary galaxy samples from existing CO observations in the literature at different redshifts, star-forming activities and evolutionary states. For each of these galaxy samples, we perform single and double Gaussian fits to the SDSS emission line spectra, when available, as described in Sect.\,\ref{ssect:sample_selection:lit} and present an overview of the DP fractions in Table\,\ref{table:dp_rate}.
The DP galaxies lying more than 0.3 dex above the MS have been included in the DP sample of 52 galaxies, as discussed in Sect. \ref{ssect:sample_selection:lit}.

\subsubsection{Sample from the COLD GASS survey}\label{ssect:gold_gass}
We use 213 CO(1-0) detected galaxies from the final COLD GASS sample \citep{2011MNRAS.415...32S, 2017ApJS..233...22S}, observed with the IRAM 30m telescope with M$_{*}$ greater than ${\rm M_{*} > 10^{10} \, M_{\odot}}$ and $0.025 < z < 0.050$. These constraints exclude the COLD GASS-low extension, which is composed of galaxies of ${\rm 10^{9} \, M_{\odot} < M_{*} < 10^{10} \, M_{\odot}}$. We discard these galaxies since they have M$_{*}$ of about $\sim 1-2\,{\rm dex}$ smaller than the discussed DP sample. Due to their smaller gravitational potential, these galaxies play a different role in terms of merger-induced star formation. The selected sample represents the local galaxy population, since it was selected randomly out of the complete parent sample of the SDSS within the ALFALFA footprint. We find 13 galaxies to be identified with a DP and include the 2 of them which are situated more than 0.3\,dex above the MS in our present DP sample ( Sect.\,\ref{ssect:sample_selection:lit}).

\subsubsection{M sample}\label{ssect:ms_sample}
To characterise galaxies which are scattered around the MS at higher redshift ($z = 0.5 - 3.2$), we compose a CO detected sample which is a part of the sample used in \citet{2018ApJ...853..179T}. This sample is associated with the MS at higher redshift and we name it the M sample. The purpose of this sample is to compare the molecular gas content and scaling relations of gas depletion time and molecular gas fractions of DP galaxies with galaxies associated with the MS. We gather 51 MS galaxies from the PHIBSS1 survey \citep{2013ApJ...768...74T} observed with IRAM Plateau de Bure Interferometer (PdBI) in CO(3-2) at two groups of redshift $z= 1-1.5$ and $z = 2-2.5$, 87 MS galaxies from the PHIBSS2 survey \citep{2018ApJ...853..179T, 2019A&A...622A.105F} observed with NOEMA in CO(2-1) or (3-2) at $z = 0.5 - 2.7$, 9 MS galaxies at $z = 0.5 - 3.2$ observed by with IRAM PdBI in CO(2-1) or (3-2) \citep{2010ApJ...713..686D, 2012ApJ...758L...9M}, 6 MS galaxies from the \textit{Herschel}-PACS Evolutionary Probe (PEP) survey \citep{2011A&A...532A..90L} observed with the IRAM PdBI in CO(2-1) at redshift $z = 1 - 1.2$ \citep{2012A&A...548A..22M} and 8 MS gravitationally lensed galaxies observed with the IRAM PdBI in CO(3-2) at $z = 1.4 - 3.2$ \citep[and references therein]{2013ApJ...778....2S}. 
As shown in Fig.\,\ref{fig:ms}, this sample is scattered around the MS with some outliers of up to ${\rm \delta MS = 1\,dex}$. Contrary to the sample used in \citet{2018ApJ...853..179T} we discuss the COLD GASS sample, the EGNOG and ULIRG samples separately and exclude all sub samples of galaxies situated above the MS. We compose the M sample with 161 galaxies. Even though this sample lies at higher redshift than our DP sample, it allows us to discuss underlying mechanisms accounting for deviation from the scaling relations found by \citet{2018ApJ...853..179T} and which contribute to various stages of cosmic evolution.
Due to their high redshifts, we do not have optical spectra of the M sample galaxies and are thus unable to estimate their DP fraction.

\subsubsection{Sample from the EGNOG survey}\label{ssect:egnog}
We use 31 CO(1–0) or (3–2) galaxies detected above the MS from the EGNOG survey \citep{2013ApJ...768..132B} at redshift $z = 0.06 - 0.5$. These galaxies are mainly characterised by star formation enhancement and show starbursts in some cases. We have SDSS spectra for 26 of these galaxies and find 11 galaxies exhibiting a DP, which we select in our present DP sample (discussed in Sect.\,\ref{ssect:sample_selection:lit}). To discuss the remaining single-peaked (SP) galaxies of this sample, we gather them in the SP-EGNOG sample. The DP galaxies of the EGNOG sample are similar to the present DP sample ones in terms of SFR \citep{2004MNRAS.351.1151B}, M$_*$ \citep{2003MNRAS.341...33K} and redshift. One main difference is the absence of radio continuum observations for the most part of this sample.

\subsubsection{ULIRG sample}\label{ssect:ulirg}
To compare our galaxies with the brightest infra-red (IR) galaxies, we assemble a sample of ultra luminous infrared galaxies (ULIRG) with existing CO detections performed with the IRAM 30m and the FCRAO 14m telescope. These galaxies exhibit a starburst or are identified as strong quasars. We select: 18 ULIRGs detected in CO(1-0), (2-1) or (3-2) at $z = 0.2 - 0.6$ with far IR luminosities of ${\rm log(L_{FIR} / L_{\odot}) > 12.45}$ \citep{2011A&A...528A.124C}, 15 ULIRGs detected in CO(2-1), (3-2) or (4-3) at $z = 0.6 - 1.0$ with ${\rm log(L_{FIR} / L_{\odot}) > 12}$ \citep{2013A&A...550A..41C}, 27 ULIRGs detected in CO(1-0) at $z = 0.04 - 0.11$  with ${\rm L_{FIR} = 0.24 - 1.60 \,10^{12} \, L_{\odot}}$ \citep{2009AJ....138..858C} and 37 CO(1-0) detected ULIRGs at $z < 0.3$ with ${\rm L_{FIR} = 0.29 - 3.80 \,10^{12} \, L_{\odot}}$ \citep{1997ApJ...478..144S}. We identify 3 DP galaxies out of 8 SDSS galaxies published by \citet{2009AJ....138..858C} which are also part of our present sample (defined in Sect.\,\ref{ssect:sample_selection:lit}). One DP galaxy out of the 8 SDSS galaxies in \citet{1997ApJ...478..144S} is Arp 220, part of our present DP sample. This provides us 93 ULIRGs, enabling us to compare our DP sample with strong IR and radio sources.

\subsubsection{Low SF sample}\label{ssect:low_sf_sample}
To study the difference between star-forming galaxies and galaxies at late stages of a starburst, or even with quenched SF, we gather a low SF sample. Therefore, we use 11 galaxies from \citet{2009ApJ...690.1672S} which are CO(1-0) detected with the IRAM 30m telescope. These galaxies are early-type galaxies at a redshift of $0.05 < z < 0.10$, currently undergoing the process of quenching or showing late-time star formation. We further select 17 CO(1-0) and (2-1) detected post-starburst galaxies with little ongoing star formation (${\rm \sim 1 M_{\odot}\,yr^{-1}}$) at $0.01 < z < 0.12$ \citep{2015ApJ...801....1F}, of which 4 are exhibiting DP emission lines in the SDSS spectra. We add 15 bulge-dominated, quenched galaxies with large dust lanes detected in CO(1-0) and (2-1) with the IRAM 30m telescope at $ 0.025 < z < 0.133$ \citep{2015MNRAS.449.3503D}, of which 3 have DP emission line in the SDSS spectra. Finally, we add 2 quenched massive spiral galaxies at $z \sim 0.1$ detected in CO(1-0) with the IRAM 30m telescope \citep{2020MNRAS.496L.116L}. The low SF sample therefore consists of 38 galaxies, creating a well suited counterpart to MS and above-MS galaxies. 

\subsubsection{MEGAFLOW sample}\label{ssect:megaflow}
Furthermore, we aim to discuss our observations with respect to recent NOEMA observations, conducted by \citet{2021MNRAS.501.1900F}. In a pilot program of the MusE GAs FLOw and Wind (MEGAFLOW) survey, they measured CO(3-2) and (4-3) detection limits for 6 galaxies at $z = 0.6-1.1$ with confirmed in- and outflows in the circum-galactic medium, 
to test the quasi-equilibrium model and the compaction scenario describing the evolution of galaxies along the MS, implying a tight relation between SF activity, the gas content and in-/outflows. This sample will help us discuss different mechanisms of compaction due to filaments or merger driven inflows, increasing both the molecular gas content and the star formation efficiency which is discussed in Sect.\,\ref{ssect:discussion:central:sf}. 

\subsubsection{Fraction of DP galaxies in the comparison samples}\label{ssect:dp_francrion}
\begin{table}
\caption{DP rates in samples from the literature}\label{table:dp_rate}
\centering
\begin{tabular}{ l c c c}        
\hline\hline
\multicolumn{1}{c}{Sample} & \multicolumn{1}{c}{Size} & \multicolumn{1}{c}{SDSS spectrum} & \multicolumn{1}{c}{confirmed DP} \\ 
\hline
EGNOG & 31 & 27 & 11(35\,\%)\\
COLD GASS & 213 & 213 & 13(6\,\%)\\
ULIRG & 97 & 18 & 4(4\,\%)\\
Low SF & 47 & 47 & 7(15\,\%)\\
\hline
\end{tabular}
\tablefoot{To determine the DP rate we show for each sample the size, the number of galaxies with a SDSS spectra and the number of galaxies with confirmed DP emission lines. We do not show the MEGAFLOW nor the M sample as they do not have SDSS spectra.}
\end{table}
43 DP galaxies have been idenfied in the EGNOG, COLD GASS, ULIRG and the Low SF samples. Table\,\ref{table:dp_rate} shows the fraction of DP galaxies in each sample. The M and the MEGAFLOW samples are not part of the SDSS and it is not possible to derive a DP fraction for them.
Furthermore, only 19\,\% of the ULIRG sample is covered by the SDSS, which makes it difficult to estimate a DP fraction.
As described in Sect.\,\ref{ssect:sample_selection:lit}, only galaxies situated more than 0.3\,dex above the MS  have been selected for the present DP sample, restricting to 17 galaxies. 
Hence, the remaining 26 DP galaxies are excluded from the subsequent analysis of the DP sample.

\subsection{Distributions of M$_{*}$ and distance to the MS for all samples}\label{ssect:ms}
\begin{figure*}
\centering 
\includegraphics[width=0.98\textwidth]{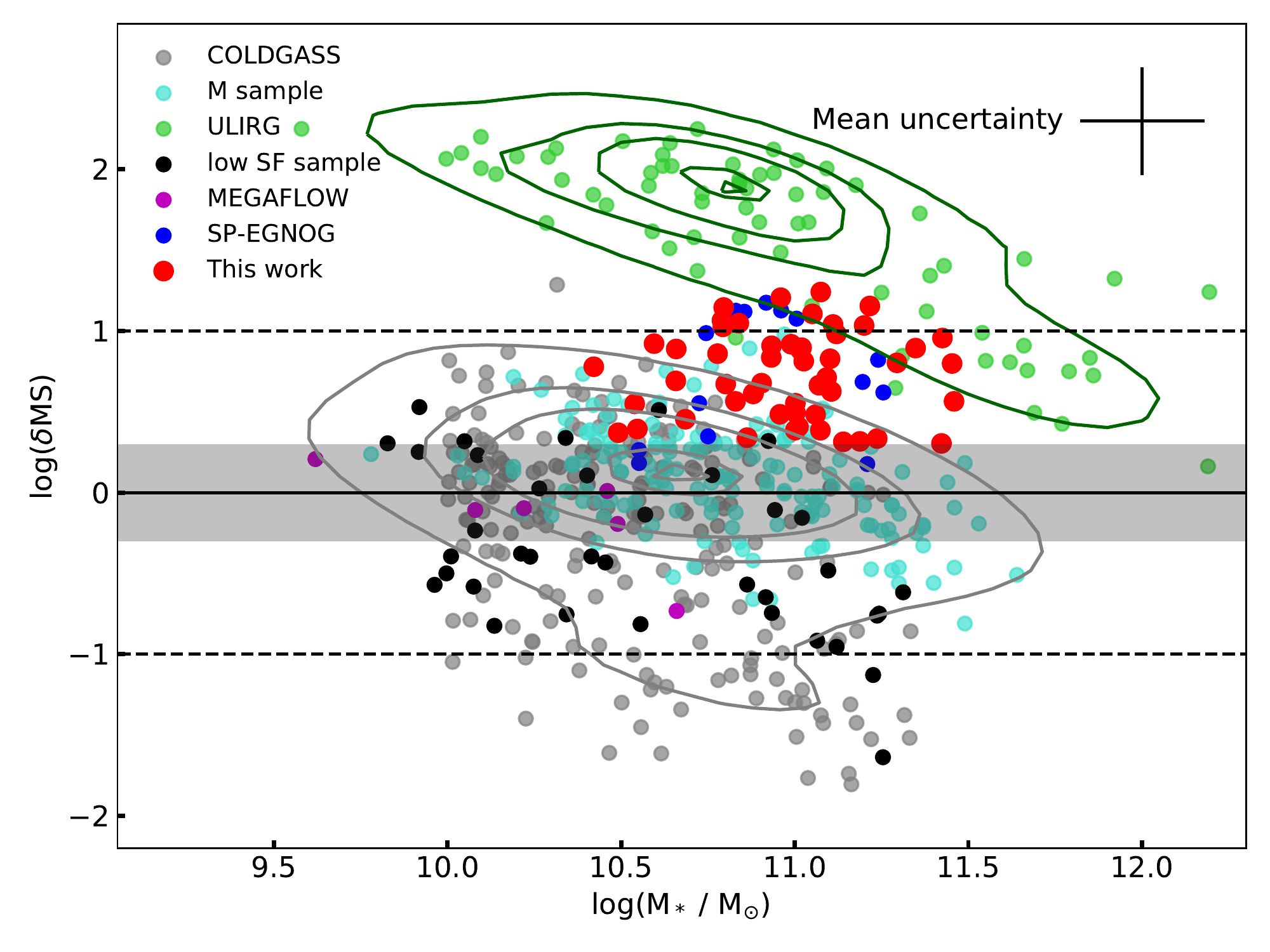}
\caption{
Offset from the MS as in Fig.\,\ref{fig:simple:ms}.
We show the DP sample with red dots, the SP-EGNOG sample with blue dots, the COLD GASS sample with grey dots, the low SF sample with black dots, the ULIRG sample with green dots, the M sample with turquoise dots and the MEGAFLOW sample with magenta dots. The literature samples are introduced in Sect.\,\ref{ssect:gold_gass}-\ref{ssect:megaflow} and a detailed description of the MS is done in Sect.\,\ref{ssect:ms}. 
We show contour lines for the ULIRG sample in green and for the COLD GASS sample combined with the M sample in grey. In the top right we show the mean uncertainties of all samples and discuss the individual uncertainties for each sample in the text.}
\label{fig:ms}%
\end{figure*}
Figure\,\ref{fig:ms} displays the location of the galaxies from all the samples with respect to the MS, as defined in Sect. \ref{ssect:sample_selection}. 
The estimated uncertainty of ${\rm SFR_{MS}}$ is 0.2\,dex \citep{2014ApJS..214...15S}. 
We use the SFR estimates provided by \citet{2004MNRAS.351.1151B} and M$_{*}$ by \citet{2003MNRAS.341...33K} for our DP sample, the SF-EGNOG sample, the COLD GASS sample and the low SF sample, if available. 
We estimate in all these samples a mean uncertainty of 0.1\,dex for M$_{*}$. For the SFR measurement, the average uncertainties are 0.3\,dex for the DP sample, 0.15\,dex for the SF-EGNOG sample, 0.45\,dex for the low SF sample and 0.4\,dex for the COLD GASS sample. The high mean uncertainties for the latter two samples are mainly influenced by quenched galaxies, as they show large SFR uncertainties \citep{2004MNRAS.351.1151B}.
For the M and the MEGAFLOW samples, we use SFR and ${\rm M_{*}}$ values provided in the literature. An estimate of the mean uncertainties is 0.25 and 0.2\,dex for the SFR and ${\rm M_{*}}$ respectively \citep{2018ApJ...853..179T,2019A&A...622A.105F, 2021MNRAS.501.1900F}.
For the ULIRG sample, we use literature ${\rm M_{*}}$ estimate if available and compute the SFR from the $\rm L_{FIR}$ following \citet{1998ApJ...498..541K}. 
The uncertainties for the SFR and ${\rm M_{*}}$ are 0.2 and 0.15\,dex, respectively, as discussed in \citet{2015ApJ...800...20G}. 
However, many of these galaxies are known to host powerful AGN which can contribute substantially to the IR flux. Furthermore, the aperture effects and possible contribution of companions can also lead to a systematic overestimation of both the SFR and the stellar mass \citep{1996ARA&A..34..749S}. 
Since we cannot quantify systematic uncertainties, we use these estimates with caution. 

We find that the M sample, the majority of the COLD GASS sample, the MEGAFLOW sample, and parts of the low SF sample are situated within the MS. We observe that parts of the COLD GASS and the low SF samples are shifted below the MS. As expected due to their high IR luminosities, we find the ULIRG sample to be located far above the MS and in some cases of even more than 2\,dex. Since their SFR is estimated using $\rm L_{FIR}$, it is possible, that in some cases, non-stellar gas heating from the AGN dominates the IR emission, biasing the SFR estimate as shown in \citet{2015A&A...576A..10C}. We find the DP and EGNOG samples situated in the same environment: in the upper MS or above with high stellar masses of $\sim {\rm 10^{11}\,M_{*}}$, and below the ULIRG sample.

\subsection{Radio continuum for all samples}\label{ssect:radio_continuum}
To discuss the star-forming activity based on synchrotron emission, we cross match the different samples with radio continuum observation catalogues at ${\rm 150\,MHz}$, ${\rm 1.4\,GHz}$ or ${\rm 3\,GHz}$. These measurements would also be sensitive to the contribution of a possible hidden AGN. 
We thus select galaxies observed by the LOFAR Two-metre Sky Survey (LoTSS) at ${\rm 150\,MHz}$ \citep[see ][for DR1]{2019A&A...622A...1S}, the Faint Images of the Radio Sky at Twenty-Centimeters (FIRST) at ${\rm 1.4\,GHz}$ \citep{1997ApJ...475..479W} or the Very Large Array Sky Survey (VLASS) at ${\rm 3\,GHz}$ \citep{2020PASP..132c5001L}.
We use the integrated radio flux measured for each source. 
We have used the LoTSS DR2 (early access) fluxes, as the DR2 offers a larger coverage of SDSS DR7 footprint than the DR1. One can note that the DP galaxies covered by LOFAR observations have been detected. 

We include radio measurements at 150\,MHz provided by the LoTSS DR2 \citep[see for DR1][]{2019A&A...622A...1S} and at 3\,GHz taken from the VLASS \citep{2020PASP..132c5001L}. We use the 1.4\,GHz observations from the FIRST survey \citep{1997ApJ...475..479W} or the NVSS \citep{1998AJ....115.1693C}. In Table\,\ref{table:radio_detection}, we present the fraction of available radio measurements for the different samples. 
We had early access to the LoTSS DR2, which is not covering the entire northern hemisphere. We thus can only take  into account galaxies which are within the observed footprint.
\begin{table}
\caption{Number of available radio measurements for the CO samples}.\label{table:radio_detection}
\centering
\begin{tabular}{ l c c c c}        
\hline\hline
\multicolumn{1}{c}{Sample} & \multicolumn{1}{c}{CO} & \multicolumn{1}{c}{150\,MHz} & \multicolumn{1}{c}{1.4\,GHz} & \multicolumn{1}{c}{3\,GHz} \\ 
\hline
DP sample & 52 & 26 (50\,\%) & 41 (79\,\%) & 32 (62\,\%) \\
SP-EGNOG & 13 & 1 (8\,\%) & 5 (38\,\%) & 3 (23\,\%) \\
M sample & 161 & 3(1.9\,\%) & 0(0\,\%) & 0(0\,\%) \\
COLD GASS & 204 & 16 (8\,\%) & 40 (20\,\%) & 28 (14\,\%) \\
ULIRG & 93 & 41 (44\,\%) & 52 (56\,\%) & 72 (77\,\%) \\
Low SF & 38 & 21 (55\,\%) & 11 (29\,\%) & 8 (21\,\%) \\
\hline
\end{tabular}
\tablefoot{We present the number of available radio continuum measurements in all three radio frequencies used in Fig.\,\ref{fig:radio}. The percentages do not represent the radio detection rates, since not all samples are located in the observed footprints of the used radio surveys.
}
\end{table}
We compute the radio luminosity as:

\begin{equation}\label{eq:radio_lum}
    {\rm \left( \frac{L_{\nu}}{W \, Hz^{-1}} \right) = \frac{36 \, \pi \times 10^{18}}{(1 + z)^{1 + \alpha}}  \left( \frac{F_{\nu}}{Jy} \right)  \left( \frac{D_{L}}{Mpc} \right)^{2}},
\end{equation}

  ${\rm F_{\nu}}$ is the integrated radio continuum flux at the observed frequency, ${\rm D_L}$ the luminosity distance, and $\alpha$ the spectral index \citep{2019ApJ...872..148C}. We calculate the spectral index using 2 radio measurements $\nu_1$ and $\nu_2$ at two different frequencies:
\begin{equation}\label{eq:spectral_index}
     {\rm \alpha = \frac{log(F_{\nu_1} / F_{\nu_2})}{log(\nu_1 / \nu_2)}}
\end{equation}
We preferably use the radio measurements at 150\,MHz and 1.4\,GHz, if available, otherwise a combination with the 3\,GHz measurement. For galaxies where we only have a single measurement, we use $\alpha = -0.7$ \citep{2019ApJ...872..148C} as an approximation.

\section{Data Analysis}\label{sect:data_analysis}
In Sect. \ref{ssect:fitting}, we describe the fit applied to the CO emission lines. A combined fit, performed on optical and molecular gas spectra, enables to identify 10 DP galaxies with identical kinematic parameters, suggesting a compact molecular gas configuration. For the remaining galaxies, a single, double and a triple Gaussian functions are fitted and the best fit is selected in order to accurately measure the emission line integral. 
In Sect.\,\ref{ssect:co_h2}, the CO luminosity and the aperture corrected gas mass are computed.
In Sect. \ref{ssect:classification}, the characteristics of the DP sample is compared with the literature samples, namely, the BPT diagram, the morphology and environment and the galaxy inclination. In Sect. \ref{ssect:sf}, the star formation rates measured for the entire galaxy and only the SDSS $3^{\prime\prime}$ fibre are compared.

\subsection{CO line fitting}\label{ssect:fitting}
%
\begin{figure*}
\centering 
\includegraphics[width=0.98\textwidth]{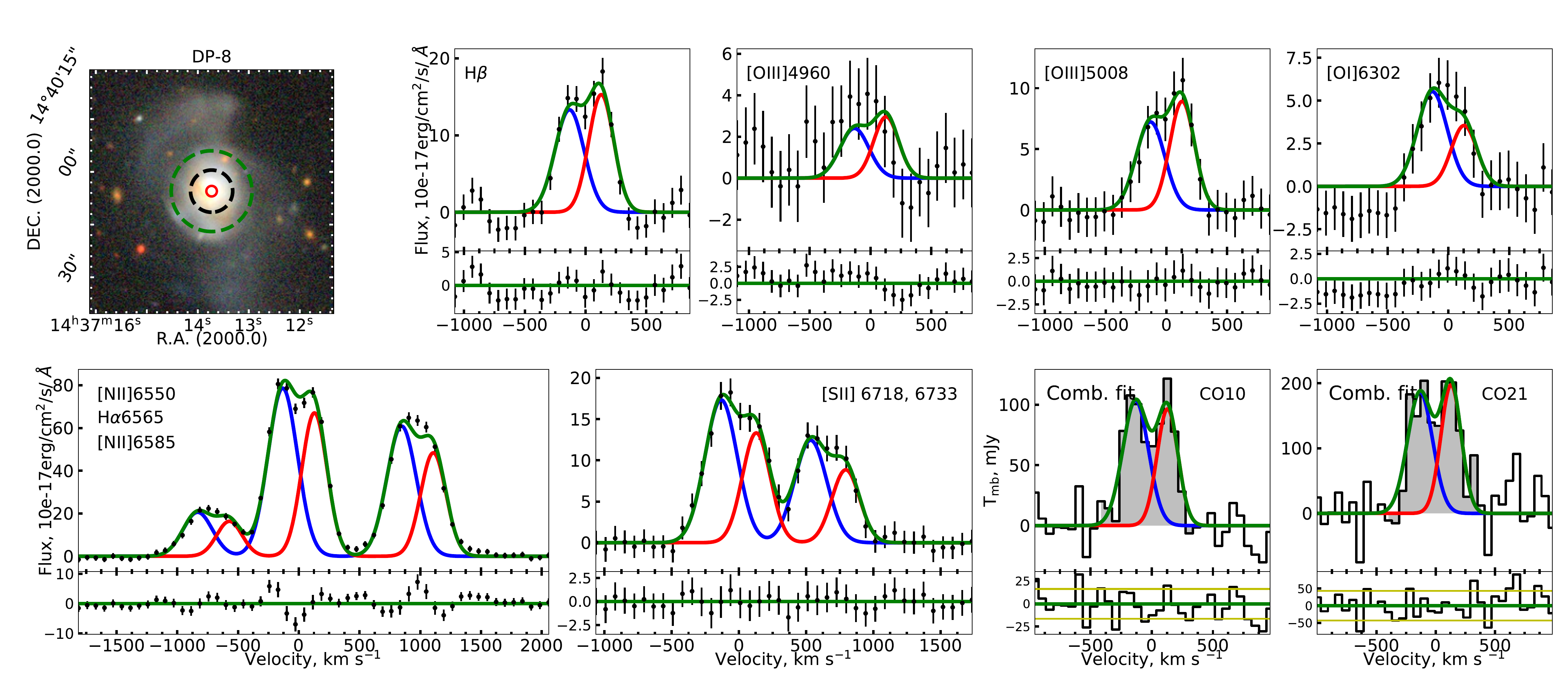}
\caption{Example of combined emission line fit for DP-8. We show the Legacy survey snapshot on the top left panel with a red circle for the SDSS 3$^{\prime \prime}$ fibre in red, and  dashed green (resp. black) circles for the FWHM of the CO(1-0) (resp. CO(2-1)) beam of the IRAM 30m telescope. The top row displays, next to the snapshot, the H$\beta$, [OIII]$\lambda 4960$, [OIII]$\lambda 5008$ and [OI]$\lambda 6302$ emission lines. The bottom row displays the [NII]$\lambda 6550$, H$\alpha$, [NII]$\lambda 6585$, [SII]$\lambda 6718$, [SII]$\lambda 6733$ emission lines, and two CO emission lines: CO(1-0) and CO(2-1). We show the double Gaussian fit with the blueshifted (resp. redshifted) component in blue (resp. red), and the total fitted function in green. 
For each line, we show the residuals below. We display the RMS level in yellow for the CO lines, estimated beside the lines.
The $x$-axis measure the deviations from the velocity calculated using the redshift.
For the H$\alpha$ line and the [NII]$\lambda 6550,6585$ doublet, we display the lines with respect to the expected H$\alpha$ line velocity, and, for the [SII]$\lambda 6718,6733$, with respect to the [SII]$\lambda 6718$ line velocity.}
\label{fig:line:fit}%
\end{figure*}
The SDSS 3$^{\prime \prime}$ fibre only probes the central few kpc of a galaxy in comparison to the IRAM 30m CO(1-0) beam of 23$^{\prime \prime}$ which covers roughly the entire galaxy at a redshift $z > 0.05$. So these two measurements probe not only different types of gas, they also probe different regions of the galaxy. 
However, in a scenario where a merger event, a galaxy interaction or the accretion of a large amount of gas have funnelled the gas into the central region fuelling the star formation, we would expect the molecular gas to follow similar kinematics as the ionised gas with the latter tracing these star-forming regions. 
Such a scenario would motivate a combined fit approach where we expect similar kinematics in the molecular and ionised gas measurements.

\subsubsection{Combined fit of ionised and molecular gas spectra}\label{ssect:combined_fit}
We test if the same kinematic parameters can be fitted for the ionised gas and for the molecular gas emission lines.
Therefore,  the same Gaussian kinematic parameters $\mu_{1, 2}$ and $\sigma_{1, 2}$ obtained from the optical ionised gas emission lines (as described in Sect.\,\ref{ssect:sample_selection:lit}) are fixed for the CO lines fit. Thus, only the emission line amplitudes are fitted.
Then, we check if the ratios, between the blueshifted and redshifted Gaussian fit components, ${\rm A_b / A_B}$, for the CO and the H$\alpha$ emissions are compatible within 3\,$\sigma$.
In order to test the significance of the fitted components, we furthermore compute the RMS outside the CO emission lines and check if the residuals of the performed fit exceed 3 times the RMS value. If this is the case, a significant deviation from the residuals would indicate a molecular gas component which cannot be represented by the velocity distribution found in the ionised gas. 
In addition, we demand that each of the two fit components have a signal-to-noise ratio of at least 3. Secondly, if these criteria are met, we adopt this fit and flag the CO line to indicate a successful combined fit.

When available, we first check the CO(2-1) spectra since this observation probes a smaller region than the CO(1-0) observation. Therefore, if we do not manage to perform a combined fit in the CO(2-1) line, we do not fit the CO(1-0) with this approach. 
In four galaxies, we succeed in fitting a combined fit only in the CO(2-1) line but not in the CO(1-0) line.

We finally find 10 (19\,\%) galaxies with a successful combined fit and show, on Fig.\,\ref{fig:line:fit}, an example of combined fit results with all included lines for DP-8. 

\subsubsection{Individual CO emission lines fit}\label{ssect:independent_co_fit}
In order to estimate the CO emission lines of those galaxies where the combined fit approach failed, we fit these spectra individually. To accurately model also clumpy line shapes, we fit each emission line with a single, a double and a triple Gaussian function and select the best fit through an F-test, as performed for the ionised gas emission line fit in Sect.\,\ref{ssect:sample_selection:lit}. This allows us to model also complex line shapes such as a double horn or asymmetric emission lines.
To further provide an uniform estimation for the entire DP sample, we perform a single Gaussian fit for each emission line. This allows us to compare for example the full width half maximum (FWHM) value of each galaxy.
In Fig.\,\ref{fig:spec_1}, we show all results with only the H$\alpha$ line and the [NII]$\lambda 6550, 6585$ doublet and the CO lines.
We mark a successful combined fit with a flag in each molecular emission line. The CO fit results are presented in Tab.\,\ref{table:CO(1-0)-parameters} and \ref{table:CO(2-1)-parameters}.

\subsection{CO luminosity and H$_{2}$ mass}\label{ssect:co_h2}
To derive the total H$_2$ mass, we first compute the intrinsic CO luminosity with the velocity integrated transition line flux ${\rm F_{CO(J \rightarrow J-1)}}$ and calculate
\begin{equation}\label{eq:l_co}
     {\rm \left( \frac{L'_{CO(J \rightarrow J-1)}}{K \, km \, s^{-1} \, pc^{2}} \right) = \frac{3.25 \times 10^{7}}{(1 + z)} \left( \frac{F_{CO(J \rightarrow J-1)}}{Jy \, km \, s^{-1}} \right)  \left( \frac{\nu_{rest}}{GHz} \right)^{-2}  \left( \frac{D_{L}}{Mpc} \right)^{2}},
\end{equation}
where $\nu_{\rm rest}$ is the rest CO line frequency and ${\rm D_L}$ the luminosity distance \citep{1997ApJ...478..144S}. We can thus derive the total molecular gas mass including a correction of $36\,\%$ for interstellar helium using:
\begin{equation}\label{eq:co2mh2}
    {\rm M_{H_2} = \alpha_{CO} \, L'_{CO(J \rightarrow J-1)} / r_{J1}},
\end{equation}
where the mass-to-light ratio $\alpha_{\rm CO}$ denotes the CO(1-0) luminosity-to-molecular-gas-mass conversion factor and ${\rm r_{J1} = F_{CO\,(1\rightarrow 0)} / F_{CO\,(J \rightarrow J-1)}}$ is the CO line ratio.

\subsubsection{Conversion factor}\label{sssect:conversion_factor}
The conversion factor estimated for the Milky Way and nearby star-forming galaxies with similar stellar metallicities to the Milky Way, including a correction for interstellar helium, is $\alpha_{\rm G} = 4.36 \pm 0.9 \, {\rm M_{\odot} / (K\,km\,s^{-1}\, pc^2)}$ \citep{1996A&A...308L..21S, 2010ApJ...710..133A}. As discussed in \citet{2010ApJ...716.1191W} and \citet{2013ARA&A..51..207B}, the CO conversion factor depends on the metallicity and we use a mean value for the correction established by \citet{2012ApJ...746...69G} and \citet{2013ARA&A..51..207B}, and adopted by \citet{2015ApJ...800...20G}, \citet{2018ApJ...853..179T} and \citet{2021MNRAS.501.1900F}:
\begin{equation}\label{eq:alpha_co}
    \alpha_{\rm CO} = \sqrt{0.67 \times {\rm exp}(0.36 \times 10^{\rm 8.67 - log\,Z}) \times 10^{\rm -1.27\times(8.67 - log\,Z)}},
\end{equation}
where ${\rm log\,Z = 12 + log(O/H)}$ is the gas-phase metallicity on the \citet{2004MNRAS.348L..59P} scale, which we can estimate from the stellar mass using
\begin{equation}\label{eq:mass_met}
    {\rm log\,Z} = 8.74 - 0.087 \times ({\rm log(M_{*}) - b})^2,
\end{equation}
with ${\rm b = 10.4 + 4.46 \times log(1 + z) - 1.78 \times (log(1 + z))^2}$ \citep[and references therein]{2015ApJ...800...20G}. 
The gas-phase metallicity can be estimated using optical emission line ratios, as discussed in \citet{2004MNRAS.348L..59P}. However, the SDSS central $3^{\prime\prime}$ spectral observation is only covering the central part of the galaxy, depending on the redshift. Here, we use equation\,\ref{eq:mass_met} in order to get an estimate for the entire galaxies and to be consistent with previous works \citep{2015ApJ...800...20G, 2018ApJ...853..179T, 2021MNRAS.501.1900F}.
Furthermore, this approach enables us us to compute the gas-phase metallicity for galaxies with no available spectral measurements.
We find a mean conversion factor for the DP sample of $\alpha_{\rm CO} = 3.85 \pm 0.08 {\rm \,M_{\odot} / (K\,km\,s^{-1}\, pc^2)}$, which is similar to the conversion factor we find for the EGNOG sample, of $3.86 \pm 0.09{\rm \,M_{\odot} / (K\,km\,s^{-1}\, pc^2)}$, the low SF sample ($3.86 \pm 0.12{\rm \,M_{\odot} / (K\,km\,s^{-1}\, pc^2)}$) or the COLD GASS sample ($3.84 \pm 0.10{\rm \,M_{\odot} / (K\,km\,s^{-1}\, pc^2)}$). For the ULIRG sample, we find a slightly higher conversion factor $\alpha_{\rm CO} = 4.00 \pm 0.39{\rm \,M_{\odot} / (K\,km\,s^{-1}\, pc^2)}$. In case we do not have the stellar mass of a galaxy, we use the mean stellar mass of the sample to compute the conversion factor and then the molecular gas mass. 
This estimation is adapted for MS galaxies \citep[e.g.][]{2018ApJ...853..179T} and might be overestimated in comparison to the conversion factor of $\alpha_{\rm CO} = 0.80 {\rm \,M_{\odot} / (K\,km\,s^{-1}\, pc^2)}$ for ULIRGs as discussed in \citet{1997ApJ...478..144S} and we therefore use this conversion for these galaxies.
The molecular gas mass of the three ULIRGs which we adapted for our DP sample from \citet{2009AJ....138..858C} are calculated using Eq.\,\ref{eq:alpha_co} in order to keep a consistent molecular gas mass estimate.

To compare the calculated molecular gas masses, we need to assume the line ratio r$_{J1}$. In \citet{2015ApJ...800...20G}, \citet{2018ApJ...853..179T} and \citet{2021MNRAS.501.1900F}, a line ratio of r$_{21} = 0.77$ and r$_{31} = 0.5$ was assumed, which is here used for the M sample. For the ULIRG sample, we choose ratios of r$_{21} = 0.83$, r$_{31} = 0.52$ and r$_{41} = 0.42$, which are empirically motivated by recent observations \citep[see][and references therein]{2015ApJ...800...20G}.                                                  

\subsubsection{Aperture correction}\label{sssect:aperture}
The closest galaxies that we observed are not entirely covered by the CO(1-0) $22^{\prime\prime}$ beam, resulting in an incomplete measurement of the molecular gas. To account for the gas content outside the telescope beam, we perform an aperture correction following \citet{2011A&A...534A.102L}. Relying on CO maps of local spiral galaxies \citep{2001PASJ...53..757N,2001ApJ...561..218R,2008AJ....136.2782L}, these authors assume an exponential distribution function of the CO gas.
Hence, we first define the scale and geometry of each galaxy. To approximate the apparent galaxy size, we extract the optical radius at the 25\,mag isophote $r_{25}$ (see Table\,\ref{table:sample}). As discussed in \citet{2011A&A...534A.102L}, we can assume $r_{e} / r_{25} = 0.2$ where $r_e$ is the CO scale length. We measure the galaxy inclination using the minor-to-major axial ratio $b/a$ estimated from a 2D S\'ersic profile fit using the photometric diagnostic software {\sc statmorph}\footnote{\url{https://statmorph.readthedocs.io}} \citep{2019MNRAS.483.4140R}. We compute the inclination $i$ as:
\begin{equation}\label{eq:inclination}
    \cos i  = \sqrt{\frac{(b/a)^2 - q_0^2}{1 - q_0^2}},
\end{equation}
where $q_0$ describes the intrinsic axial ratio of an edge-on observation and is set to $q_0 = 0.2$ \citep{2012MNRAS.420.1959C, 2018MNRAS.479.2133A}. For galaxies classified as mergers, we set the inclination to $0^{\circ}$, since we cannot identify their orientation with a S\'ersic profile. Last, following \citet{2011A&A...534A.102L}, the aperture correction factor is computed as:
\begin{equation}\label{eq:aperture}
\begin{split}
    {\rm f_{a}} = \pi r_{e}^{2} \Bigg\{
    \int^{\infty}_{0} dx 
    \int^{\infty}_{0} dy \,
    {\rm exp} \left(-{\rm ln}(2)
    \left[ 
    \left( \frac{2 \, x}{ \Theta_{\rm B}} \right)^2 + 
    \left( \frac{2 \, y \, {\rm cos}(i)}{ \Theta_{\rm B}} \right)^2 
    \right]\right) \\
    {\rm exp} \left(-\frac{\sqrt{x^2 + y^2}}{r_e}  \right)
    \Bigg\}^{-1} ,
    \end{split}
\end{equation}
where $\Theta_{\rm B}$ is the FWHM of the observation beam. We carry out the integration numerically. 

We present the aperture correction factors and the corrected molecular gas masses in Table\,\ref{table:molecular_mass}. We set the correction factor for galaxies which are observed using interferometry to 1 since we have accurate molecular gas mass measurements. We measure a mean correction factor for the DP sample of ${\rm f_a = 1.27}$. 
We present the CO luminosities, the molecular gas mass and the aperture correction in Table\,\ref{table:molecular_mass}.

\subsection{DP sample characteristics}
\label{ssect:classification}
The properties of the DP galaxies are discussed here, namely: their position on the BPT diagram (Sect. \ref{sssect:bpt}); their morphology and their environment (Sect. \ref{sssect:environment_morph}) and their inclination (Sect. \ref{sssect:inclination}).

\subsubsection{BPT diagram}\label{sssect:bpt}
\begin{figure}
\centering 
\includegraphics[width=0.48\textwidth]{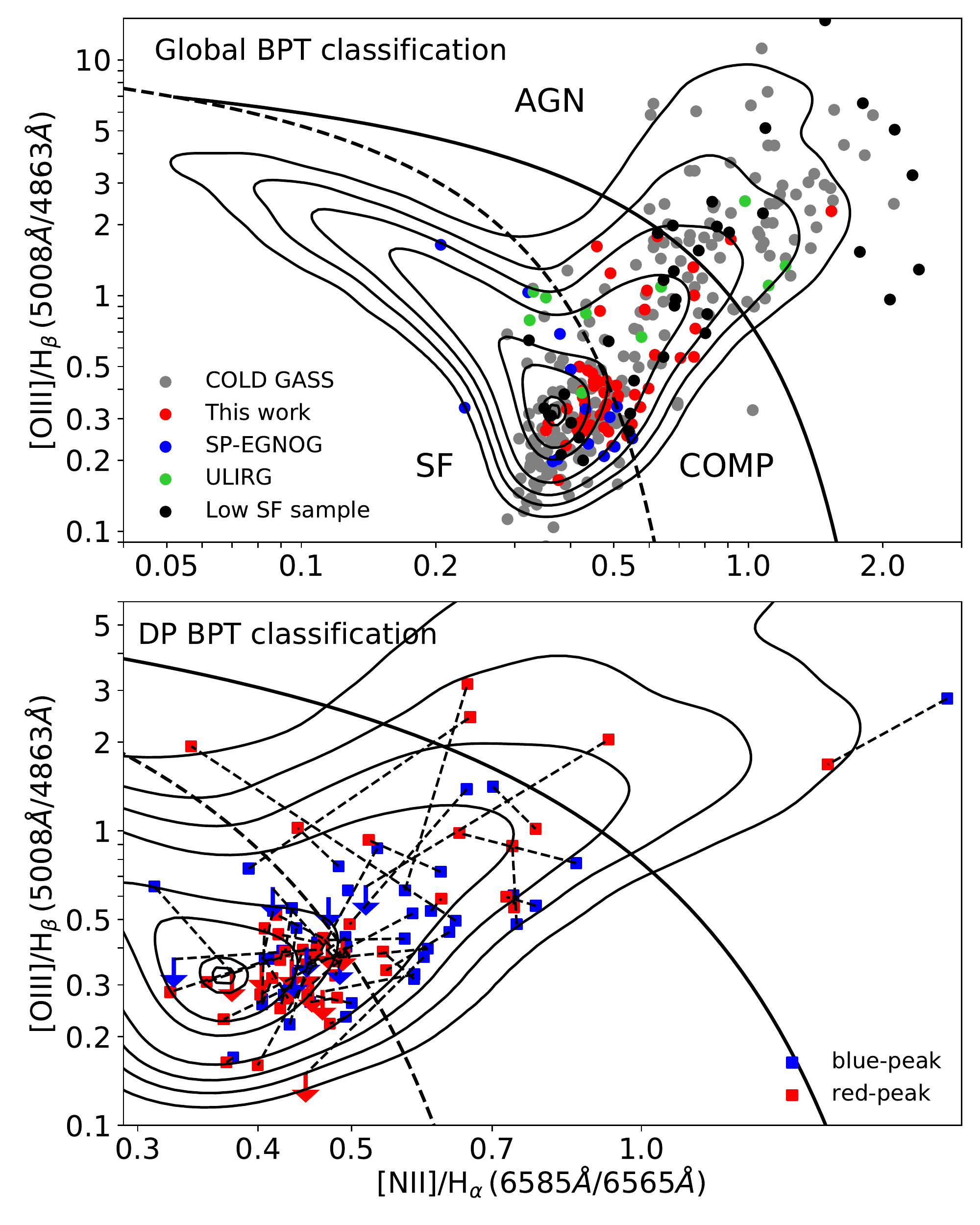}
\caption{BPT diagrams to classify our galaxy samples into different galaxy types based on ionised gas emission line ratios \citep{2006MNRAS.372..961K}. The black solid line separates active galactic nuclei (AGN) and composite (COMP) galaxies and the dashed black line star-forming (SF) galaxies from COMP galaxies. 
In the \textit{top} panel, we show galaxies of all samples with existing SDSS spectra. We use Gaussian or non-parametric emission line fits, in case of a non Gaussian emission line shape, provided by \citet{2017ApJS..228...14C}.
In the \textit{bottom} panel, we show, for each galaxy of the DP sample, the blueshifted and redshifted peaks represented by blue and red squares, respectively, and connect them by a black dashed line. In comparison to the \textit{top} panel, we zoomed into the region where we detect DP galaxies, to resolve both components. For galaxies with one of the needed emission lines below $3\,\sigma$, we indicate the emission line ratio limits by arrows.
In both panels, we show contour lines representing galaxies of the RCSED catalogue which have a ${\rm S/N\,>\,3}$ in all required emission lines.}
\label{fig:bpt}%
\end{figure}
We use the BPT diagnostic diagram \citep{1981PASP...93....5B} to classify our galaxy samples based on optical emission line ratios [OIII]$\lambda 5008$/H$\beta$ on the $y$-axis and [NII]$\lambda 6585$/H$\alpha$ on the $x$-axis. Relying on the criteria empirically found by \citet{2006MNRAS.372..961K}, we differentiate between star-forming (SF) galaxies, active galactic nuclei (AGN) and composite (COMP) galaxies, which are characterised by both mechanisms: SF and AGN.
In the top panel of Fig.\,\ref{fig:bpt}, we show the position on the BPT diagram of the DP, COLD GASS, the SP-EGNOG and the low SF samples. Depending on which function fits the data better, we use the Gaussian or the non-parametric emission line estimate provided by \citet{2017ApJS..228...14C}. For the DP sample e.g. we used the latter which gives us a global estimation of the entire emission line.
In order to characterise each emission line component individually, we classify each of them separately. 
We present both classifications in the lower panel of Fig.\,\ref{fig:bpt} and list their classification in Tab.\,\ref{table:obs_characteristics}. We also show the classification using the non-parametric fit. 
In order to unambiguously classify each emission line component, the H$\beta$ and [OIII]$\lambda 5008$ lines must be detected, with at least a ${\rm S/N > 3}$, which is not always the case, as for DP-45 and DP-51.
However, using the non-parametric fit we are able to classify DP-51.

Using the non-parametric fit, we classify 56\,\% of the DP sample as SF, 37\,\% as COMP, and 4\,\% as AGN. The DP sample is dominated by SF galaxies which is consistent with the fact that the DP sample was selected 0.3 dex above the MS. When each emission line component is considered individually,  9 galaxies (17\,\%) have their two components classified differently. In particular, we find 7 SF + COMP, 1 SF + AGN, and 1 COMP + AGN. However, we do not find any trend in molecular gas mass, morphological type or success of combined fit for these peculiar galaxies.

We classify galaxies of all samples, if possible, with the BPT diagram and build up SF-COMP subsamples of all galaxies classified as SF or COMP, and an active galaxies sample for those classified as AGN. For the ULIRG sample, we use classifications provided in the literature since we have SDSS spectra of 10 galaxies enabling a detailed BPT classification. We classify all Low Intensity Narrow Emission-line Regions (LINER), Seyfert galaxies and quasars as an active/AGN subsample. Galaxies of the ULIRG sample show large fractions of strong radio galaxies, we thus do not select any SF-COMP subsample for them since we are not able to correctly characterise their AGN contribution. This classification allows us to test the radio flux as a tracer of molecular gas in SF galaxies and to discuss the behaviour of active/AGN galaxies.

\subsubsection{Morphology and galaxy environment}\label{sssect:environment_morph}
To further characterise the evolutionary state of the galaxies, we visually inspect the legacy survey images \citep{2019AJ....157..168D} and categorise them as mergers if we see an optical perturbation, as late-type galaxies (LTG) if we can identify a spiral disc, or as S0 if we can identify a disc with the bulge dominating the shape.
The results found by \citet{2018MNRAS.476.3661D}, with a machine-learning based classification used in \citetalias{2020A&A...641A.171M}, inspired this classification.

We also flag LTG and S0 galaxies which have tidal features, since this can be the sign of a recent merger or interaction. We present in Table\,\ref{table:obs_characteristics} the morphological type of each galaxy of the DP sample.

We find $27\,\%$ to be classified as LTG, $38\,\%$ as S0 galaxies and $35\,\%$ as mergers. We also find $13\,\%$ to be either S0 galaxies or LTG with notable tidal features. 
A close examination of all DP LTGs reveals that they are all bulge-dominated and are of Sa type.
In order to compare the obtained merger fraction to the one of single-peaked emission line galaxies, we select 1000 single-peak galaxies from \citet{2020A&A...641A.171M} which are situated at 0.3\,dex above the MS and classify them in the exact same way. This sample of single-peaked emission line galaxies was selected with the same thresholds of signal-to-noise ratios of the H$\alpha$ and O[III]$\lambda 5008$ emission lines, and follow the same redshift and stellar mass distribution as the DP galaxy sample of \citet{2020A&A...641A.171M}. We find for this galaxy sample only $10\,\%$ mergers and $14\,\%$ galaxies with tidal features. It is not straightforward to compare these two galaxy samples. On the one hand, we have selected 4 ULIRGs for the DP sample from the literature, which are all mergers, and on the other hand, galaxies with unusually high SFR values were selected for our CO observations, introducing a bias. However, we find $48\,\%$ of the DP sample to show either a visual merger or tidal features, which is about twice as much as we find for single-peak emission line galaxies.

To discuss the fact that we see more bulge-dominated galaxies in the DP sample, we perform a morphological classification of other nearby galaxies samples. Using the Legacy Survey images, we can classify the SP-EGNOG, the COLD GASS  and the low SF samples in the exact same way as for the DP sample without adding a bias of resolution due to larger redshifts to this classification. 
The SP-EGNOG sample shows a very similar distribution of stellar masses and SFR as discussed in Sect.\,\ref{ssect:ms}. We also find a very similar morphological composition of 30\,\% LTG, 40\,\% S0 and 10\,\% mergers. The remaining 20\,\% are at redshift 0.5 and thus not classifiable with the Legacy survey images. 
Interestingly, we find also the LTG galaxies of the SP-EGNOG sample to be bulge-dominated and of type Sa. Furthermore, we detect fewer mergers but identify 25\,\% of the LTG and S0 galaxies to have tidal features. 
In order to compare the DP sample to the literature galaxies, we classify the galaxies of the COLD GASS sample which are situated more than 0.3\,dex above the MS. These galaxies have a mean stellar mass of ${\rm log(M_* / M_{\odot}) = 10.5}$, which is 0.5 dex smaller than the mean stellar mass of the DP sample. We find 63\,\% of LTGs, 18\,\% of S0 galaxies, and 18\,\% of galaxy mergers. The LTGs exhibit smaller bulges which are of type Sb or Sc. 
The low SF sample, in contrast, consists of only 13\,\% LTGs and 18\,\% mergers. However, we find 68\,\% to be classified as bulge-dominated galaxies i.e. S0 or elliptical galaxies. While the merger rates are not discriminant, the low SF sample is dominated by early-type galaxies partly quenching explaining their low SFR, while the COLD GASS sample hosts more discy galaxies with smaller bulges than the DP sample.

To discuss the impact of the environment, we identify the associated group galaxies using \citet{2016A&A...596A..14S} for galaxies at $z < 0.11$, and \citet{2007ApJ...671..153Y} for galaxies at $z > 0.11$. In \citet{2016A&A...596A..14S}, galaxy groups were identified using a group finding algorithm which was calibrated with cosmological simulations. The group finding algorithm in \citet{2007ApJ...671..153Y} is a halo based friends-of-friends finding algorithm. Both algorithms provide the number of galaxies in the group and we can measure the projected distance to the closest neighbour. In Tab.\,\ref{table:obs_characteristics}, we present the environment parameters for each DP galaxy.

\begin{table*}
\caption{Characteristics of observed galaxies}
\label{table:obs_characteristics}
\centering
\begin{tabular}{ l c c c c c c c}        
\hline\hline
\multicolumn{1}{c}{ID} & \multicolumn{1}{c}{Morphology} & \multicolumn{1}{c}{BPT$_{1}$} & \multicolumn{1}{c}{BPT$_{2}$} & \multicolumn{1}{c}{BPT$_{\rm t}$} & \multicolumn{1}{c}{Comb. fit} & \multicolumn{1}{c}{N$_{\rm G}$} & \multicolumn{1}{c}{D$_{\rm N}$} \\ 
\hline
\multicolumn{1}{c}{} & \multicolumn{1}{c}{} & \multicolumn{1}{c}{} & \multicolumn{1}{c}{} & \multicolumn{1}{c}{} & \multicolumn{1}{c}{} & \multicolumn{1}{c}{} & \multicolumn{1}{c}{kpc} \\ 
\hline
DP-1$^{a}$ & Merger & agn & agn & agn & 0 & 2$^{\dagger}$ & 334$^{\dagger}$ \\
DP-2 & LTG & comp & sf & comp & 0 & 3$^{\dagger}$ & 269$^{\dagger}$ \\
DP-3 & LTG & comp & comp & comp & 1 & 1$^{\dagger}$ & - \\
DP-4 & LTG & sf & sf & sf & 1 & 2$^{\dagger}$ & 117$^{\dagger}$ \\
DP-5 & Merger & comp & comp & comp & 1 & 4$^{\dagger}$ & 27$^{\dagger}$ \\
DP-6 & LTG & comp & sf & comp & 0 & 12$^{\dagger}$ & 61$^{\dagger}$ \\
DP-7 & LTG & sf & sf & sf & 0 & 13$^{\dagger}$ & 115$^{\dagger}$ \\
DP-8 & Merger & comp & comp & comp & 1 & 1$^{\dagger}$ & - \\
DP-9 & LTG & sf & sf & sf & 0 & 1$^{\dagger}$ & - \\
DP-10 & LTG & comp & comp & comp & 1 & 3$^{\dagger}$ & 64$^{\dagger}$ \\
DP-11 & S0 & sf & sf & sf & 0 & 2$^{\dagger}$ & 313$^{\dagger}$ \\
DP-12 & Merger & comp & comp & comp & 0 & 20$^{\dagger}$ & 15$^{\dagger}$ \\
DP-13$^{b}$ & LTG & sf & sf & sf & 0 & 1$^{\dagger}$ & - \\
DP-14 & LTG & comp & comp & comp & 0 & 2$^{\dagger}$ & 146$^{\dagger}$ \\
DP-15 & Merger & sf & sf & sf & 1 & 1$^{\dagger}$ & - \\
DP-16 & S0 + T & sf & sf & sf & 1 & 6$^{\dagger}$ & 174$^{\dagger}$ \\
DP-17$^{b}$ & Merger & comp & comp & comp & 0 & 6$^{\dagger}$ & 161$^{\dagger}$ \\
DP-18 & S0 & sf & sf & sf & 0 & 4$^{\dagger}$ & 138$^{\dagger}$ \\
DP-19$^{c}$ & LTG + T & comp & sf & sf & 1 & 3$^{\dagger}$ & 85$^{\dagger}$ \\
DP-20 & S0 + T & sf & sf & sf & 0 & 3$^{\dagger}$ & 399$^{\dagger}$ \\
DP-21 & S0 & comp & sf & sf & 0 & 2$^{\dagger}$ & 104$^{\dagger}$ \\
DP-22 & S0 &  & agn & agn & 0 & 3$^{\dagger}$ & 707$^{\dagger}$ \\
DP-23 & Merger & comp & sf & sf & 0 & 3$^{\dagger}$ & 651$^{\dagger}$ \\
DP-24$^{d}$ & Merger & comp & comp & comp & 0 & 8$^{\dagger}$ & 129$^{\dagger}$ \\
DP-25$^{d}$ & Merger & comp & comp & comp & 0 & 4$^{\dagger}$ & 7$^{\dagger}$ \\
DP-26 & Merger & comp & sf & comp & 0 & 4$^{\dagger}$ & 42$^{\dagger}$ \\
DP-27$^{c}$ & LTG & sf & sf & sf & 0 & 1$^{\dagger}$ & - \\
DP-28 & S0 + T & sf & sf & sf & 0 & 1$^{\dagger}$ & - \\
DP-29 & S0 & comp & agn & comp & 0 & 2$^{\dagger}$ & 370$^{\dagger}$ \\
DP-30 & Merger & comp & comp & comp & 0 & 1$^{\dagger}$ & - \\
DP-31$^{d}$ & Merger & comp & comp & comp & 0 & 1$^{\dagger}$ & - \\
DP-32 & S0 & sf & sf & sf & 0 & 1$^{+}$ & - \\
DP-33 & S0 + T & sf & sf & sf & 1 & 1$^{+}$ & - \\
DP-34 & Merger & sf & sf & sf & 0 & 1$^{+}$ & - \\
DP-35 & LTG & comp & sf & sf & 0 & 2$^{+}$ & 303$^{+}$ \\
DP-36 & S0 & sf & sf & sf & 0 & 1$^{+}$ & - \\
DP-37 & Merger & sf & sf & sf & 0 & 1$^{+}$ & - \\
DP-38 & Merger & sf & sf & sf & 0 & - & - \\
DP-39 & LTG + T & sf & sf & sf & 0 & 1$^{+}$ & - \\
DP-40 & S0 & sf & sf & sf & 1 & 1$^{+}$ & - \\
DP-41$^{c}$ & S0 & sf & sf & sf & 0 & - & - \\
DP-42$^{c}$ & S0 & sf & sf & sf & 0 & 1$^{+}$ & - \\
DP-43$^{c}$ & Merger & sf & sf & sf & 0 & 1$^{+}$ & - \\
DP-44$^{c}$ & S0 & comp &  &  & 0 & 1$^{+}$ & - \\
DP-45$^{c}$ & S0 &  &  &  & 0 & 1$^{+}$ & - \\
DP-46 & S0 + T & comp & comp & comp & 0 & 1$^{+}$ & - \\
DP-47$^{c}$ & S0 & sf & sf & comp & 0 & 1$^{+}$ & - \\
DP-48 & S0 & sf & sf & sf & 0 & 1$^{+}$ & - \\
DP-49$^{c}$ & Merger & comp & sf & sf & 0 & 1$^{+}$ & - \\
DP-50 & Merger & sf & agn & comp & 0 & - & - \\
DP-51$^{c}$ & S0 &  &  & sf & 0 & - & - \\
DP-52$^{c}$ & LTG & sf & sf & comp & 0 & - & - \\

\hline
\end{tabular}
\tablefoot{Column 2 shows the morphological classification based on visual inspection. Galaxies which show tidal features are indicated with a `+ T´ We also show the BPT classification of the blueshifted and redshifted components in columns 3 and 4 respectively. In Column 5 we show the total BPT classification using the non-parametric fit. In column 6 we indicate with a 1 if we succeeded to perform a combined fit. We display the number of galaxies N$_{\rm G}$ associated in the same group in column 7 and the distance to the closest neighbour in the column 8. We use preferably \citet{2016A&A...596A..14S} which is denoted with a $\dagger$ and, for galaxies at $z > 0.11$, we use \citet{2007ApJ...671..153Y}, denoted with an $+$. 
The denotations $a, b$ and $c$ are the same as in Table \ref{table:sample}.}
\end{table*}

\subsubsection{Relation between inclination and kinematics}\label{sssect:inclination}
\begin{figure}
\centering 
\includegraphics[width=0.48\textwidth]{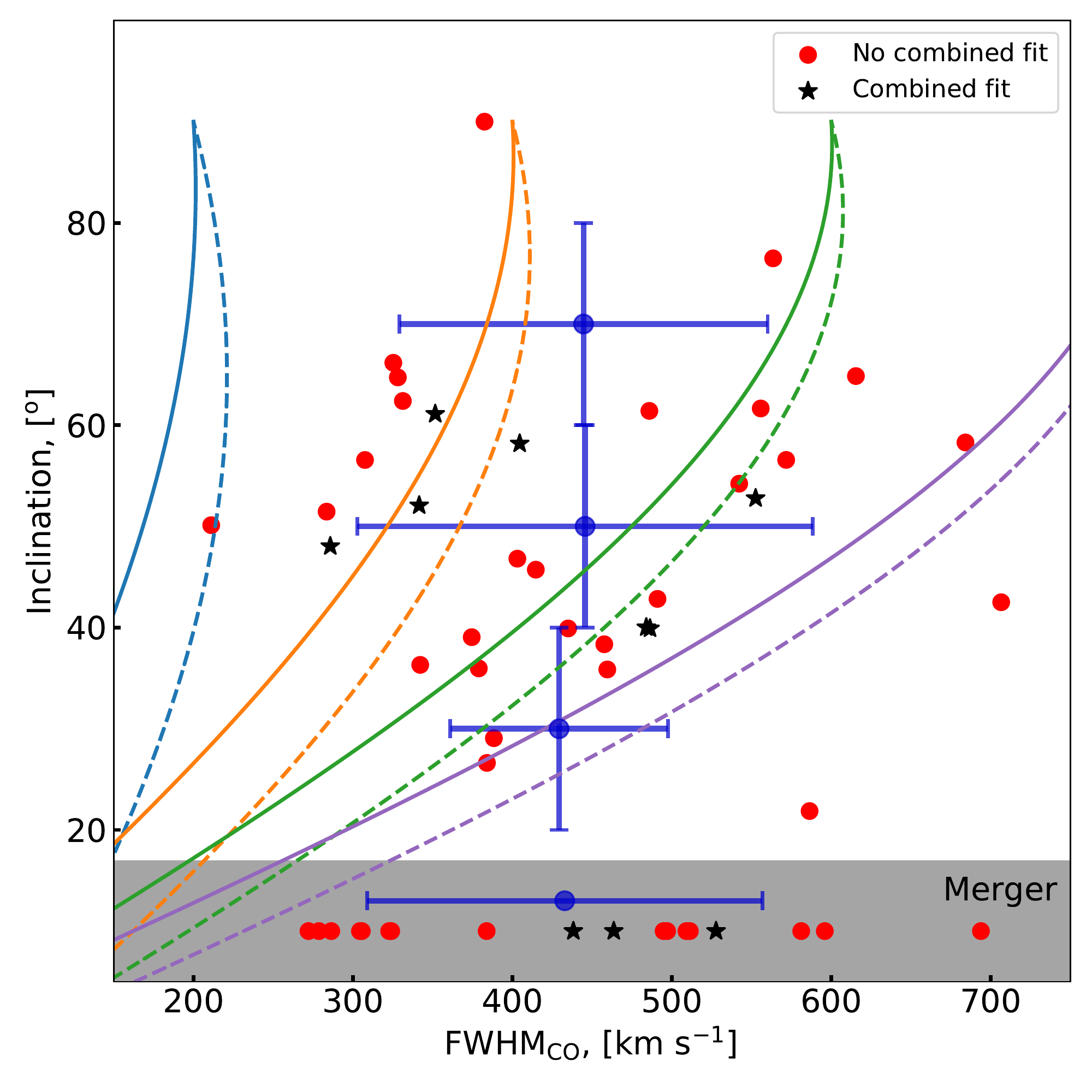}
\caption{Values of the CO FWHM ($x$-axis) for different galaxy inclinations \textit{i} ($y$-axis) of the DP sample. Black stars (resp. red dots) are galaxies for which we succeed (resp. fail) to apply a combined line fit (see Sect.\,\ref{ssect:fitting}). 
Since it is not possible to measure the inclination of galaxy mergers, we show their FHWM values separately in the grey area at the bottom of the plot.
Mean values of the FWHM and standard deviations are displayed with blue error bars for three groups of different inclinations and for the merger sub sample. The curves show the relation between FWHM and inclination obtained for the estimate of Eq~\ref{eq:fwhm} for a rotating disc, with $v_{rot}=$~100, 200, 300 and 400~km/s for the blue, orange, green and purple curves, respectively,  and
with a gas velocity dispersion of 10 and 40~km/s for solid and dashed lines.}
\label{fig:inclination}%
\end{figure}
A rotating disc creating different velocity measurements within the line of sight of a galaxy can create a double horn or double peak signature \citep[e.g.][]{2014MNRAS.438.1176W}. In such a scenario, we may expect to see at least a correlation between the galaxy inclination and the full width half maximum (FWHM) of the CO emission lines. We, therefore, performed a single Gaussian fit to the CO emission lines and compare the measured FWHM to the galaxy inclination \textit{i}, as estimated in Sect.\,\ref{sssect:aperture}. 
We use the CO(2-1) line since it has a higher S/N ratio in comparison to the CO(1-0) line in the DP sample. For galaxies with no CO(2-1) observations, we use the CO(1-0) line. 
The beam sizes of the CO(1-0) and CO(2-1) observations are different ($23^{\prime\prime}$ and $12^{\prime\prime}$ respectively) for the 35 galaxies observed at IRAM-30m and  for the 17 galaxies obtained from the literature, measured with different telescopes and beam sizes. Given the DP-galaxies redshift distribution, the CO emission lines are not measured at uniform scales.
However, as described in Sect.\,\ref{sssect:aperture}, most measurements include the majority of the molecular gas.
The relation between the CO FWHM and the galaxy inclination for the DP sample is displayed in Fig.\ref{fig:inclination}. 
The mean values and the standard deviation of the CO FWHM for groups of different inclinations are shown. 
Galaxies classified as mergers are presented separately, since the S\'ersic profile  fit to the \textit{r}-band image is not necessarily representing the disc orientation of the galaxies.
Inclination are gathered in three groups: $20^{\circ} < i < 40^{\circ}$,  $40^{\circ} < i < 60^{\circ}$ and $60^{\circ} < i < 80^{\circ}$. 
Galaxies for which we succeed to apply a combined line fit are indicated as black stars (see Sect.\,\ref{ssect:fitting}).

We are not able to find any trend nor any correlation between the measured FWHM and the galaxy inclinations. A large scatter is however expected even in the case of rotating discs, depending on the mass concentration of the galaxies and also on the velocity dispersion of the molecular gas. A rotation curve rises all the more steeply as mass is concentrated, leading to a dependency of the measured velocities on the mass concentration for a given stellar mass (for typical massive galaxies whose mass is dominated by baryonic matter in their central parts). As the CO gas emission tends to be concentrated, the corresponding velocity measurements may probe only a part of the rising of the rotation curve. A more concentrated stellar bulge will thus likely lead to a larger detected FWHM of the CO emission lines. To illustrate this effect, we show on Fig.\,\ref{fig:inclination} a few curves corresponding to different measured rotation velocities, representing measurements for a varying mass concentration at fixed stellar mass, and with two different molecular gas velocity dispersion values. We use the simple estimate:
\begin{equation}
\label{eq:fwhm}
    \mathrm{FWHM_{CO}} = 2.35 \sigma \cos i + 2 v_{rot} \sin i 
\end{equation}
corresponding to the expected width of a double horn velocity profile widened by a gas velocity dispersion $\sigma$, for a disc rotating at $v_{rot}$ and having an inclination $i$. The first term corresponds to the contribution of the velocity of the gas in the orthogonal direction to the disc plane, and the second term to the disc in-plane gas velocity, dominated by rotation. 
We also do not find different effects between galaxies with and without a successful combined fit.
The rotation velocity of disc galaxies depends on galaxy mass \citep{1977A&A....54..661T} but by taking their stellar mass into account, we were still not able to detect any trend.
These findings are in agreement with results on ionised gas velocity dispersions and galaxy inclination of a larger DP sample \citepalias{2020A&A...641A.171M}.

\subsection{Star formation}\label{ssect:sf}
\begin{figure}
\centering 
\includegraphics[width=0.48\textwidth]{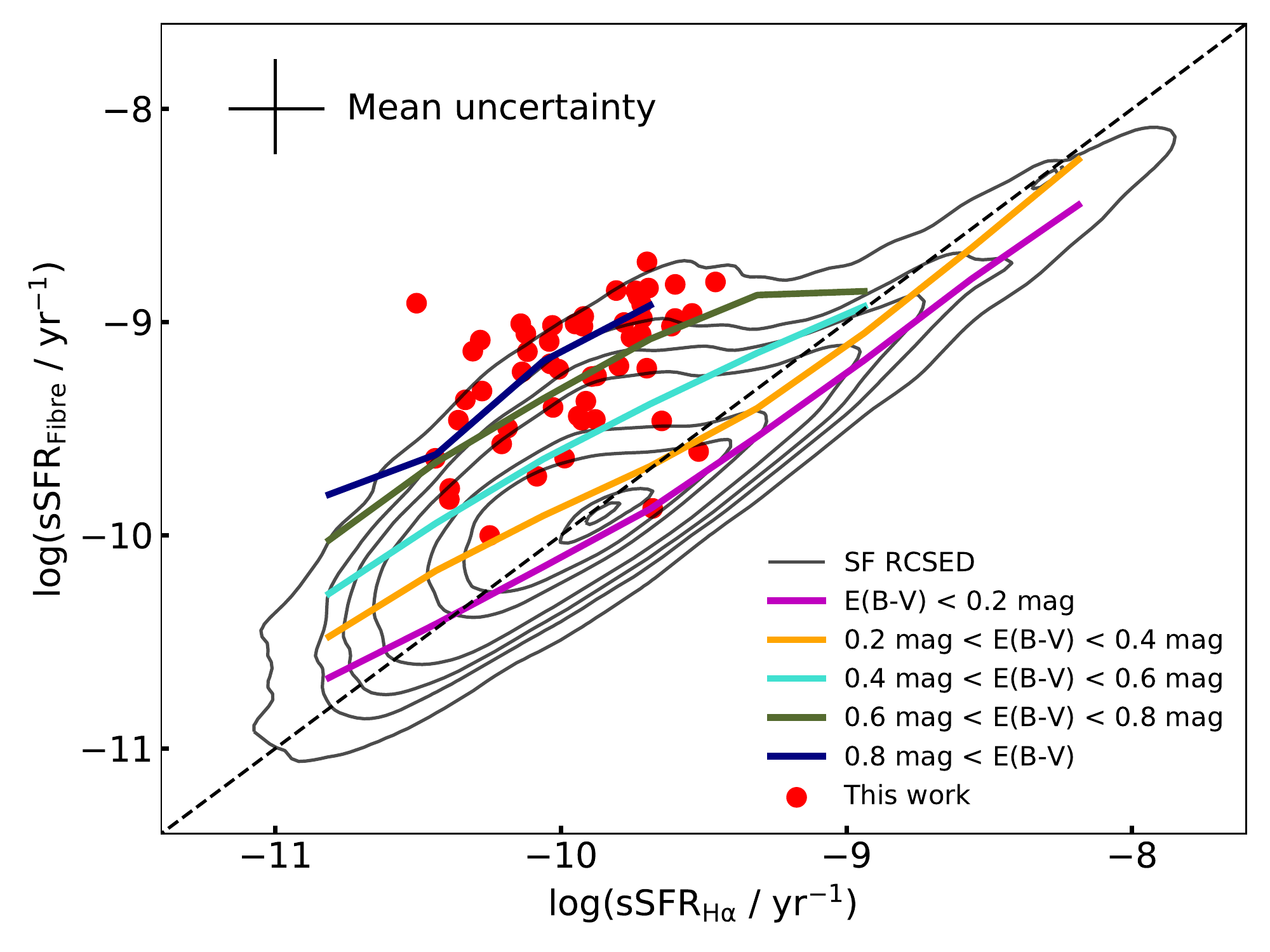}
\caption{Comparison of sSFR estimate inside the SDSS 3$^{\prime \prime}$ fibre. On the $x$-axis is the sSFR estimated from extinction corrected H$\alpha$ luminosities (SFR$_{\rm H\alpha}$), and on the $y$-axis, the sSFR inside the fibre (SFR$_{\rm fibre}$) estimated by \citet{2004MNRAS.351.1151B}. The DP sample is marked with red circles and the galaxies with a S/N$>10$ in the H$\alpha$ line of the RCSED catalogue \citep{2017ApJS..228...14C}, with black contour lines. 
We also show the median of groups of different gas extinction E(B-V), computed following equation\,(\ref{eq:extinction}), with solid thick lines. The black dashed line denotes SFR$_{\rm H\alpha}$ = SFR$_{\rm fibre}$. The black error bar is the mean estimated uncertainty of the star formation rates, including stellar mass uncertainties.} 
\label{fig:fibre_sf}
\end{figure}
To compute the extinction-corrected luminosity of the H$\alpha$ emission line, we follow \citet{2001PASP..113.1449C}: 
\begin{equation}
    {\rm L_{int}(H\alpha) = L_{obs}(H\alpha)\, 10^{0.4 \kappa (H\alpha)E(B-V)}},
\end{equation}
where ${\rm L_{int}(H_\alpha)}$ is the intrinsic and ${\rm L_{obs}(H_\alpha)}$ the observed H$\alpha$ luminosity. $\kappa({\rm H_\alpha})$ is the reddening curve parametrised by \citet{2000ApJ...533..682C} at the H$\alpha$ rest frame wavelength and E(B-V), the colour excess, is computed as:
\begin{equation}\label{eq:extinction}
    {\rm E(B-V) = 1.97\, log_{10} \left[ \frac{(H\alpha / H\beta)_{obs}}{2.86}\right]},
\end{equation}
following \citet{2013AJ....145...47M} and \citet{2013ApJ...763..145D}. The dust extinction estimate is based on the assumption of an intrinsic H$\alpha$/H$\beta$ ratio of 2.86, appropriate for a temperature of ${\rm T=10^4\,K}$ and an electron density of $n_e = 10^2 {\rm cm^{-3}}$ for a case B recombination \citep{2006agna.book.....O}. 
We can thus compute the H$\alpha$-based SFR as ${\rm SFR(H\alpha) =  7.9 \times 10 ^{-42} \times L_{int}(H\alpha)}$ following \citet{2002AJ....124.3135K}.
\begin{figure*}
\centering 
\includegraphics[width=0.98\textwidth]{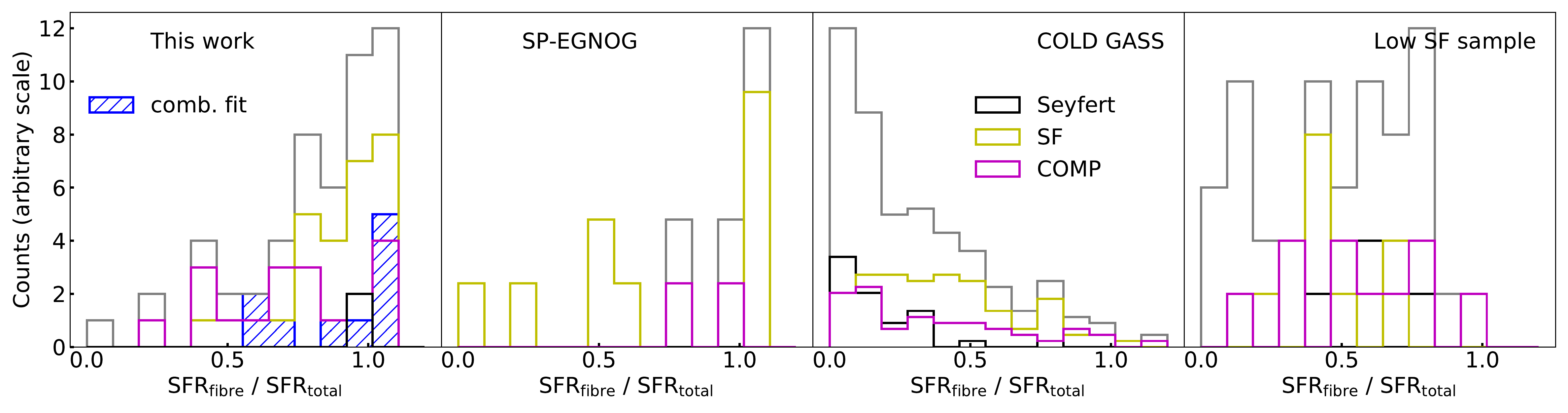}
\caption{Ratio of the SFR inside the $3^{\prime \prime}$ SDSS fibre SFR$_{\rm fibre}$ and the total SFR ${\rm SFR_{total}}$ \citep{2004MNRAS.351.1151B}. We show this relation for galaxies processed by \citet{2004MNRAS.351.1151B}, which are from left to right the DP galaxies, 14 galaxies of the SP-EGNOG survey, 161 galaxies of the COLD GASS sample, and 74 galaxies of the low SF sample. Subsets of BPT classification of SF are in yellow, COMP in magenta and AGN in black, and the total histogram in grey. For the DP sample, the subset of galaxies with successful combined fit is indicated with a hatched blue histogram. The scales of the histograms are in arbitrary units.}
\label{fig:central_sb}%
\end{figure*}

We compute the SFR$_{\rm H\alpha}$ inside the SDSS 3$^{\prime\prime}$ fibre for both emission line components of the DP sample but also for the total emission line luminosity using the non-parametric emission line fit provided by \citet{2017ApJS..228...14C}. To assess the quality of this estimate, we compare the SFR$_{\rm H\alpha}$ estimate to the SFR estimate of the SDSS fibre SFR$_{\rm fibre}$ provided by \citet{2004MNRAS.351.1151B}, which is based on an emission line modelling to avoid creating biases in the SFR estimated from the emission lines as a function of metallicity or stellar mass. This approach also takes the diffuse emission inside a galaxy into account. In Fig.\,\ref{fig:fibre_sf}, we show the specific star formation rate sSFR = SFR / M$_{*}$ inside the 3$^{\prime\prime}$ fibre using the stellar mass estimate \citep{2003MNRAS.341...33K}, considering the SFR$_{\rm H\alpha}$ on the $x$-axis and the SFR$_{\rm fibre}$ on the $y$-axis. We present galaxies from the RCSED catalogue with a ${\rm S/N > 10}$ in the H$\alpha$ emission line with black contours and show the mean values of groups of different extinction E(B-V) computed following Eq. \,\ref{eq:extinction}. We show the DP sample with red dots using the SFR estimate with the non-parametric emission-line fit to account for the entire system. We find the sSFR$_{\rm H\alpha}$ to be underestimated of around 1\,dex in comparison to the sSFR$_{\rm fibre}$ for the DP sample. 
This systematic effect correlates with the measured dust extinction. We observe a mean value of ${\rm E(B-V) = 0.66 \pm 0.19 \, mag}$ for the DP sample, which is in agreement with the observed offset for SF galaxies of the RCSED with comparable gas extinction values. Although we corrected the H$\alpha$ luminosity for extinction, strong dust obscuration can shield parts of the optical light from star formation sites. This phenomenon was discussed in greater detail in \citet{1996ARA&A..34..749S} and references therein. To calculate the SFR correctly, the estimate of the H$\alpha$ luminosity has to be combined with IR estimates as performed e.g. in \citet{2010ApJ...723..530P}.
This means that the calculated SFR$_{\rm H\alpha}$ for both components is systematically underestimated but still provides an estimate enabling us to compare the SF contribution of both components.

Besides the SFR$_{\rm fibre}$, \citet{2004MNRAS.351.1151B} estimated the SFR$_{\rm total}$ for the entire galaxy, enabling to test if SF is concentrated in the central parts of the galaxy or is equally distributed in the disc. 
In Fig.\,\ref{fig:central_sb}, we show the ratio of $\mathcal{R} = {\rm SFR_{fibre} / SFR_{total}}$ for the DP sample, the EGNOG sample, the COLD GASS sample and the low SF sample. We also show subsets of BPT classifications as discussed in Sect.\,\ref{sssect:bpt}. This diagnostic method is only meaningful for galaxies at lower redshift since the $3^{\prime \prime}$ SDSS fibre covers the entire galaxy at higher redshift. We thus exclude the galaxies of $z \sim 0.5$ of the EGNOG sample from this study. We also do not show galaxies of the M sample, neither from the ULIRG sample, since these galaxies have either no SDSS spectral observation or do not show any difference between fibre and total SFR due to their high redshift. Note that the SFR$_{\rm total}$ is calculated using star formation history modelling which relies on a first estimate of SFR$_{\rm fibre}$ from \citet{2004MNRAS.351.1151B}. This can result in a SFR$_{\rm total}$ estimate smaller than the SFR$_{\rm fibre}$, creating ratios slightly greater than one.

For the DP and SP-EGNOG samples, we find a tendency towards a ratio of 1 ($\mathcal{R}_{\rm DP} = 0.81 \pm 0.25$ and $\mathcal{R}_{\rm SP-EGNOG} = 0.77 \pm 0.32$, respectively). 
The galaxies of the DP sample with a successful combined fit (see Sect.\,\ref{ssect:fitting}) show an even higher mean ratio of $\mathcal{R}_{\rm DP} = 0.90 \pm 0.19$, indicating that the majority of their star formation is happening in the centre.
In contrast to that, we find the opposite effect with no central enhancement of SF for galaxies of the COLD GASS sample ($\mathcal{R}_{\rm COLD\,\, GASS} = 0.27 \pm 0.24$). The low SF sample exhibits a very broad distribution ($\mathcal{R}_{\rm low\,\, SF} = 0.48 \pm 0.26$). 
Given the measurement uncertainties, we can observe that the DP  and the SP-EGNOG samples are clearly biased in favour of large values of $\mathcal{R}$, in particular with respect to the COLD GASS sample.

\section{Results}\label{sect:results}
In Sect. \ref{ssect:kinematics},  the ionised and molecular gas kinematics measured in single apertures are compared. Sect. \ref{ssect:co_radio} is focused on the correlation between the molecular gas and the radio continuum. The position of all samples on the Kennicutt-Schmidt relation are discussed in Sect. \ref{ssect:ks}. Last,  the variation of the molecular gas fraction and depletion time with redshift and with the relative distance to the MS are studied in Sect. \ref{ssect:scaling_relations}. 

\subsection{Kinematical arguments: mergers, rotating discs and outflows}
\label{ssect:kinematics}
\begin{figure}
\centering 
\includegraphics[width=0.48\textwidth]{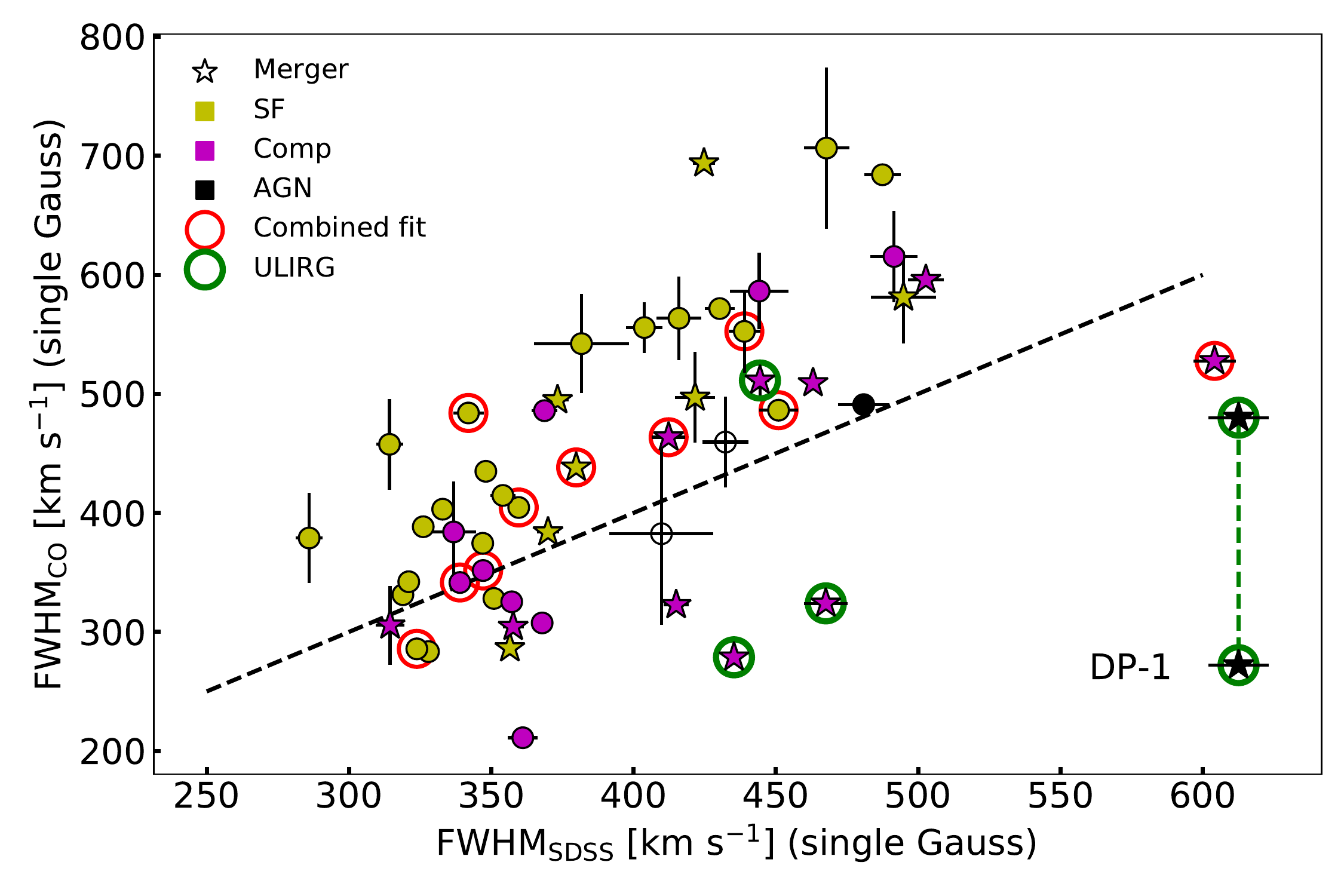}
\caption{Comparison of the FWHM of the ionised and molecular gas emission lines, estimated from a single Gaussian function for the DP sample galaxies. 
The FWHM of the ionised gas emission lines ($x$-axis) measured inside the $3^{\prime\prime}$ SDSS fibre, are estimated by \citet{2017ApJS..228...14C}. The FWHM of the CO line ($x$-axis) are measured as described in Sect.\,\ref{sssect:inclination}. The black dashed line denotes $y=x$. Error bars are estimated from the single Gaussian line fitting.
Stars indicate galaxies classified as mergers. 
The markers are coloured according to the BPT classification (See Sect.\,\ref{sssect:bpt}): SF in yellow, COMP in magenta and AGN
in black. 
We mark galaxies for which we succeed to apply a combined line fit (see Sect.\,\ref{ssect:fitting}) by red circles. 
The four ULIRGs of the DP sample are marked with green circles. For the galaxy DP-1, the CO-FWHM is estimated both inside the $3^{\prime\prime}$ SDSS fibre and for the entire galaxy, and the two points are connected with a green dashed line.
}
\label{fig:fwhm}%
\end{figure}
Since the measurements of ionised and molecular gas used in this work do not originate from the same area, we compare the FWHM values of the ionised gas and the CO lines in Fig.\,\ref{fig:fwhm}. 
We show the uncertainties with error bars, estimated from the fit. Their size can be in some cases smaller than the marker, which is due to high S/N values.

We find that the CO FWHM values are on average larger than those of the ionised gas. This can be explained since the CO measurement probes a larger area. For a typical rotation curve rising with radius before reaching a plateau, the widths of the emission-lines depend on the part of the rotation curve encompassed by the fibre or beam. If the SDSS fibre encompasses a smaller extent of the rising part of the rotation curve than the CO beam, the ionised emission-line is expected to be narrower. The width of the ionised gas emission lines can however be similar to the CO ones, even if the ionised gas extends further away than the SDSS fibre, if the galaxy mass is concentrated enough for the SDSS fibre to encompass the beginning of the plateau of the rotation curve.

In Sect.\,\ref{ssect:fitting}, we performed a combined fit, to select galaxies which show similar kinematics in the ionised and molecular gas. These are highlighted in Fig.\,\ref{fig:fwhm}. The majority of these galaxies are situated near the $y=x$ line, but two galaxies (DP-20 and DP-33) which have CO FWHM values of about 100\,km\,s$^{-1}$ larger than the FWHM estimated for the ionised gas. While they met the selection criteria discussed in Sect.\,\ref{ssect:fitting}, they give an idea of the expected scatter.
In parallel, many galaxies are very close to the $y=x$ line but do not meet the combined fit criteria, due to low S/N values and different double peak ratios.

We also observe galaxies with larger FWHM values for the ionised gas than for the molecular gas. This is expected due to ionised gas outflows driven by a central AGN. 
As displayed in in Fig.\,\ref{fig:fwhm}, most galaxies classified as composite or AGN are near the $y=x$ line or below. 
In particular, for DP-1, the resolved (ULIRG) galaxy of the DP sample, we observe that the FWHM of the molecular gas inside the 3$^{\prime\prime}$ fibre is more than twice smaller than for the optical spectra. When one integrates over the whole galaxy, the molecular gas is still well below the $y=x$ line.
Also, the other galaxies lying below $y=x$ line, including the two of the four ULIRGs marked with green circles, show a broader ionised gas velocity distribution. This is not expected as due to rotation, but is compatible with AGN outflows which could account for velocities undetected in CO.
Interestingly, while mergers are spread all over Fig.\,\ref{fig:fwhm}, the majority of the galaxies below the $y=x$ line are mergers, suggesting a link between the AGN feedback and the merging process.

In contrast to that, we find that the majority of the galaxies classified as SF lie above the $y=x$ line, but fewer of these SF galaxies are classified as mergers.

\subsection{CO and radio luminosity correlation}\label{ssect:co_radio}
\begin{figure*}
\centering 
\includegraphics[width=0.98\textwidth]{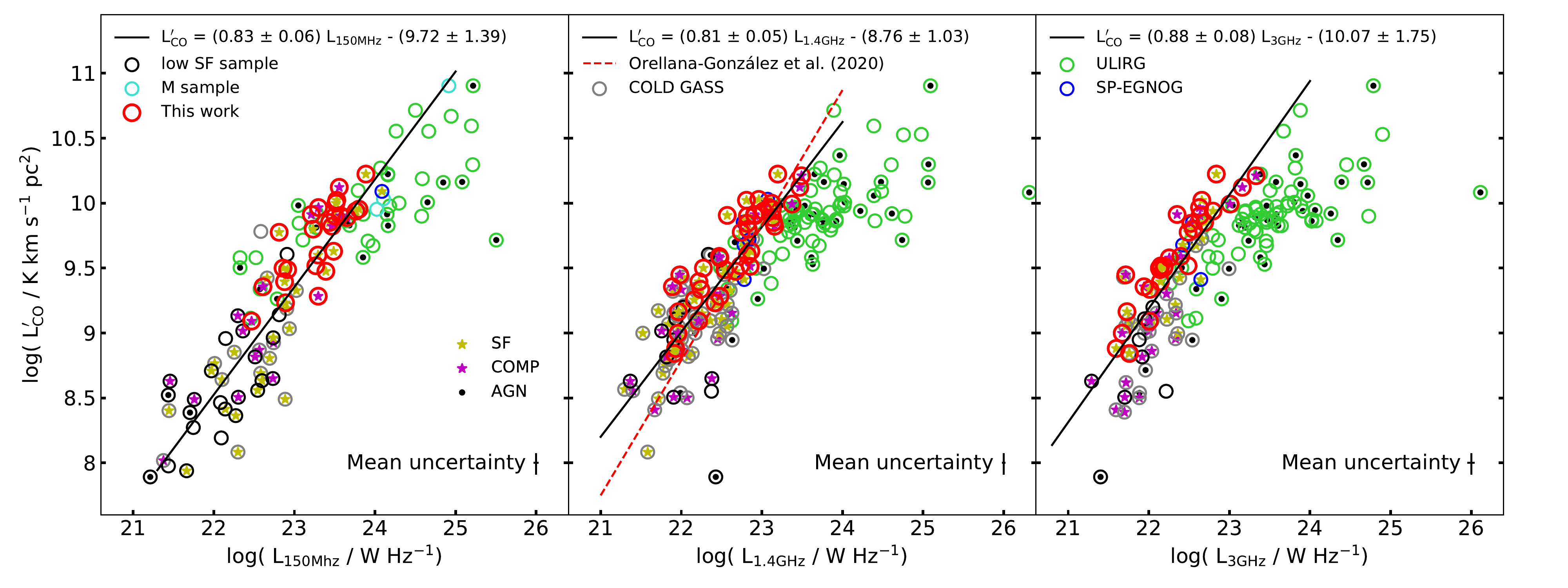}
\caption{Correlation between ${\rm L'_{CO}}$ and radio luminosity at 150\,MHz (left panel), 1.4\,GHz (middle panel) and 3\,GHz (right panel). We show  k-corrected radio luminosity following Eq.\,\ref{eq:radio_lum}. We show with circles the DP (red), the EGNOG (blue), ULIRG (green), the low SF (black), the COLD GASS (grey) and the M samples (turquoise). We mark galaxies which are classified as active/AGN with a black dot in the centre and use a yellow (resp. purple) star to highlight galaxies classified as SF (resp. COMP). In the middle panel we show the best fit for nearby galaxies found by \citet{2020MNRAS.495.1760O} with a red dashed line. We fit a straight line in all three relations to all galaxies classified as SF or COMP and display the best fit with a black solid line. In each panel, we show the average uncertainties in the lower right corner with error bars.}
\label{fig:radio}%
\end{figure*}
Star formation sites accelerate electrons and positrons in supernovae remnants to high energies, emitting synchrotron radiation when interacting with the galaxies magnetic field \citep[e.g.][]{1992ARA&A..30..575C}. It is thus possible to directly trace the star formation with radio continuum observations, which is a well-established technique at 1.4\,GHz \citep{1992ARA&A..30..575C, 2003ApJ...586..794B, 2006ApJ...643..173S, 2011ApJ...737...67M} and has proved to be valid also at ${\rm 150 \,MHz}$, as shown in \citet{2017MNRAS.469.3468C} and \citet{2019A&A...631A.109W}. In contrast, electrons and positrons can also be accelerated in relativistic jets of AGN and shock regions as discussed in e.g. \citet{1989A&A...219...63M}. In extended radio lobes, these high energy particles interact with the magnetic field creating sychrotron emission \citep[see e.g.][]{2012MNRAS.427.3196K}. Both mechanisms result in a spectrum described by a power-law of ${\rm S(\nu) \propto \nu^{\alpha}}$, where ${\rm S(\nu)}$ is the radio flux and $\alpha$ the spectral index.    

SF depends directly on the molecular gas reservoir and thus another way to exploit this connection is the relation between the radio luminosity and L$'_{\rm CO}$. This relation has been known for a long time using the 1.4\,GHz radio continuum, and dates back to the beginning of CO observations \citep{1977ApJ...213..673R, 1984ApJ...283...81I, 2002A&A...385..412M}. Recently, \citet{2020MNRAS.495.1760O} quantified this relation as ${\rm L'_{CO} = (1.04\pm 0.02)\,L_{1.4\,GHz} - 14.09 \pm 0.21}$ for galaxies at $z < 0.27$ for more than 5 orders of magnitude of the luminosities. We here aim to test this relation for the selected CO samples by distinguishing between star-forming galaxies and those with AGN contribution. 

In Fig.\,\ref{fig:radio}, we show the correlation between ${\rm L'_{CO}}$ and radio luminosity at 150\,MHz, 1.4\,GHz and 3\,GHz. The average uncertainties of the observed radio fluxes are $7\,\%$ at 150\,MHz, $5\,\%$ at 1.4\,GHz and $11\,\%$ at 3\,GHz and are indicated by an error bar.
We mark active galaxies with dots, SF galaxies with yellow stars and COMP galaxies with purple stars. We find a good agreement with the empirical correlation found by \citet{2020MNRAS.495.1760O} for ${\rm L_{1.4\,GHz}}$ and observe a similar behaviour for the SF-COMP subsamples at 3\,GHz and 150\,MHz. 
We note that galaxies classified as AGN do not follow such a linear relation. The ULIRG sample especially shows a clear excess in radio luminosity in comparison to other galaxies with comparable ${\rm L'_{CO}}$ measurements.
This might be an indicator that the radio continuum emission is dominated by the AGN and is thus not correlated with the molecular gas anymore. We fit a straight line to all three relations by only using the SF+COMP subsamples and show the fit results in Table\,\ref{table:radio_fit}. For the ${\rm L'_{CO} - L_{1.4\,GHz}}$ relations, we find a less steeper slope ($0.80\pm0.05$) than \citet{2020MNRAS.495.1760O} ($1.04\pm0.02$). 
However, taking the scatter of 0.32 into account, these two estimates are still comparable.
Interestingly, we find similar parameters for the ${\rm L'_{CO} - L_{150\,MHz}}$, the ${\rm L'_{CO} - L_{1.4\,GHz}}$, and the ${\rm L'_{CO} - L_{3\,GHz}}$ relations with nearly the exact same slope.
\begin{table}
\caption{Fit results of ${\rm L'_{CO}}$ - radio correlation }\label{table:radio_fit}
\centering
\begin{tabular}{ l c c c c}        
\hline\hline
\multicolumn{1}{c}{Relation} & \multicolumn{1}{c}{Slope} & \multicolumn{1}{c}{Intercept} & \multicolumn{1}{c}{$\sigma$} & \multicolumn{1}{c}{P} \\ 
\hline
${\rm L'_{CO} - L_{150\,MHz}}$ & 0.82 $\pm$ 0.06 & -9.54 $\pm$ 1.39 & 0.32 & 0.88 \\
${\rm L'_{CO} - L_{1.4\,GHz}}$ & 0.80 $\pm$ 0.05 & -8.62 $\pm$ 1.03 & 0.26 & 0.87 \\
${\rm L'_{CO} - L_{3\,GHz}}$ & 0.87 $\pm$ 0.08 & -10.01 $\pm$ 1.75 & 0.23 & 0.83 \\
\hline
\end{tabular}
\tablefoot{Best fit results for a linear fit of CO luminosties as a function of radio luminosities. We show the slope, the intercept, the scatter $\sigma$, and the Pearson coefficient P.}
\end{table}

\subsection{Kennicutt-Schmidt relation}\label{ssect:ks}
\begin{figure}[h]
\centering 
\includegraphics[width=0.48\textwidth]{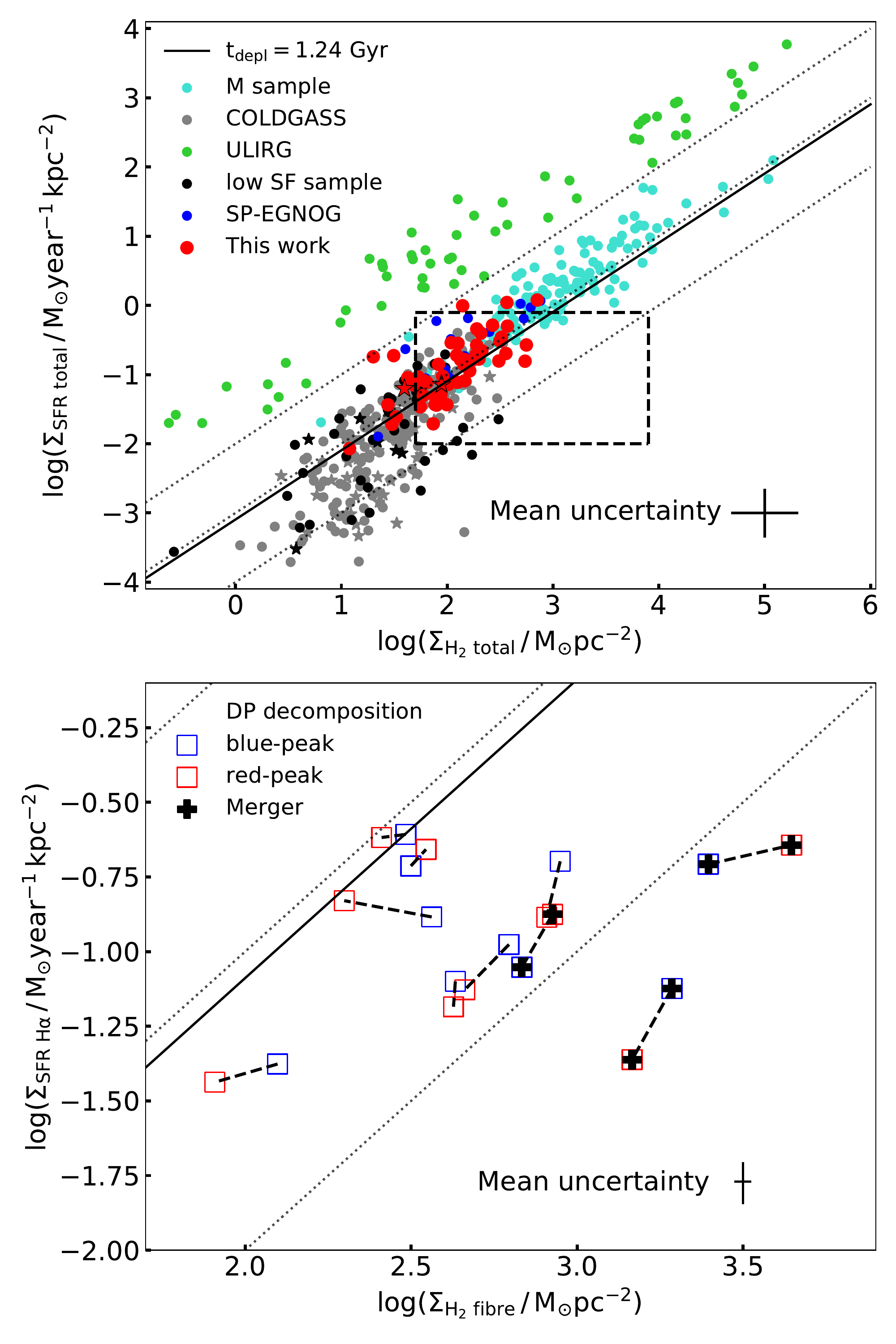}
\caption{Kennicutt-Schmidt (KS) relation for the CO samples. On the \textit{top} panel, we show the KS relation using the total molecular gas mass and the SFR of the entire galaxies, and normalise both quantities using the half-light radii. The DP sample is indicated with red dots, the EGNOG sample with blue dots, the COLD GASS sample with grey dots, the low SF sample with black dots, the M sample with turquoise dots, and the ULIRG sample with green dots. Galaxies of the low SF and COLD GASS samples showing AGN activity are marked with stars (see Sect.\,\ref{sssect:bpt}). 
On the \textit{bottom} panel, we show the decomposition of the 10 DP galaxies, for which we succeeded in performing a combined fit (see Sect.\,\ref{ssect:fitting}).
We show the decomposition on a zoom-in of the \textit{top} panel which we mark by a dashed box.
We display blue (resp. red) squares for the blueshifted (resp. redshifted) component and connect the two components with a black dashed line. We use the SFR estimated by H$\alpha$ emission of each component and the observed H$_2$ mass of each component, and normalise them to the surface of the $3^{\prime \prime}$ SDSS fibre. The three galaxies, classified as mergers, are marked with black plus signs.
In both panels, dotted lines denote constant t$_{\rm depl}$ of 0.1, 1 and 10 Gyr. The solid black line corresponds to a constant t$_{\rm depl}$ of 1.24\,Gyr estimated using \citet{2018ApJ...853..179T} for the mean redshift and stellar mass of the DP sample. In both panels, error bars indicate the mean estimated uncertainties. However, in the \textit{lower} panel, uncertainties estimated from the surface measurement are not included.
}
\label{fig:ks}%
\end{figure}
Figure\,\ref{fig:ks} displays the empirical Kennicutt-Schmidt (KS) relation relating the gas density and SFR through a power law \citep{1959ApJ...129..243S,1998ApJ...498..541K}. We plot on the $x$-axis the SFR surface density ${\rm \Sigma_{SFR} = SFR / \pi R^2}$ and on the $y$-axis the molecular gas surface density ${\rm \Sigma_{H_2} = M_{H_2} / \pi R^2}$, where $R$ is the half-light radius provided in the literature from optical high resolution images, if available, or computed from a 2D S\'ersic profile which is adjusted to the legacy $r$-band image as explained in Sect.\,\ref{sssect:aperture}. In the \textit{top} panel of Fig.\,\ref{fig:ks}, we show the unresolved KS relation. We use the same SFR estimates as used for the MS offset estimate in Sect.\,\ref{ssect:ms}. 
We plot straight lines of constant depletion times ${\rm t_{depl} = M_{H_2} / SFR}$ of 0.1, 1 and 10\,Gyr. 
We also mark the MS depletion time of 1.24\,Gyr,  computed by \citet{2018ApJ...853..179T},  for the mean redshift ($z=0.10$) and stellar mass (${\rm log(M_*/M_{\odot}) = 11.0}$) of the DP sample.
We display the mean uncertainties with error bars which include an average surface estimation uncertainty of 0.2\,dex \citep{2012ApJS..203...24V}.

The DP sample has a mean depletion time of $1.1\pm0.8$\,Gyr and the SP-EGNOG sample, of $0.7\pm0.4$\,Gyr. These are close to the depletion times expected for galaxies situated on the MS.
These two samples are filling a slight under-density of measurements between the region dominated by nearby galaxies of the COLD GASS sample and the region of the M sample at higher redshifts. The majority of the galaxies of the low SF sample and the COLD GASS sample follows the same relation as the M sample, but with some galaxies shifted towards lower star formation efficiencies (with $\rm t_{depl}$ about 10\,Gyr). 

We marked galaxies classified as AGN in Sect.\,\ref{sssect:bpt} with stars, and, in fact, the majority of galaxies with very high depletion times are classified as AGN, which is consistent with a scenario where the AGN is quenching ongoing star formation \citep[e.g.][]{2015MNRAS.452.1841S}.
Nevertheless, some galaxies with large depletion times do not host any detected AGN activity. This might be due to the exhaustion of their gas reservoir or to a hidden AGN.
In contrast, all ULIRGs have significantly smaller depletion times of ${\rm \sim 0.01\,Gyr}$ and show the largest range of ${\rm \Sigma_{H_2}}$ measurements.
Furthermore, as we discussed in Sect.\,\ref{ssect:existing_co}, their SFR might be overestimated due to AGN contribution, since it was computed using ${\rm L_{FIR}}$, shifting the galaxies towards regions of smaller depletion times.

In the \textit{bottom} panel of Fig.\,\ref{fig:ks}, we show the KS relation for the 10 DP galaxies with a successful combined fit (see Sect.\,\ref{ssect:fitting}). 
Since we find similar gas distributions between the ionised and the molecular gas, we may assume that the majority of the detected molecular gas (see Sect.\,\ref{ssect:fitting}) is situated in the central part of the galaxy, fuelling central star formation, as discussed in Sect.\,\ref{ssect:sf}. We show on the $y$-axis ${\rm \Sigma_{SFR\,H\alpha}}$, the SFR surface density estimated using the extinction-corrected H$\alpha$ luminosity of each peak component (see Sect.\,\ref{ssect:sf}). As discussed in Sect.\,\ref{ssect:sf}, this SFR estimate is systematically underestimated of about 1\,dex for the DP sample. On the $x$-axis, we show the individual H$_2$ mass surface densities ${\rm \Sigma_{H_2\,fibre}}$, without applying any aperture correction. Both surface densities are calculated for the SDSS $3^{\prime\prime}$ fibre. We also display the mean uncertainties as in the \textit{top} panel. However, no uncertainties for the surface measurements are included, which leads to significantly smaller error bars. We show the redshifted (resp. blueshifted) peak with a red (resp. blue) square and connect them with a dashed line. 
The three galaxies which are classified as mergers are marked with black plus signs. Interestingly, we find these galaxies to be shifted towards higher molecular gas surface densities than the non-mergers.
As discussed in Sect.\,\ref{ssect:sf}, the SFR$_{\rm H\alpha}$ estimates are most likely underestimated, causing a shift towards larger t$_{\rm depl}$ as observed in the \textit{top panel} of Fig.\,\ref{fig:ks}. Since the SFR$_{\rm H\alpha}$ is systematically underestimated, we are only able to effectively compare the relative difference in t$_{\rm depl}$ of the two components.

If an evenly distributed rotating gaseous disc were at the origin of a DP structure, we would expect to observe similar SF efficiencies in each component, comparable t$_{\rm depl}$ values and both components aligned with the lines of constant t$_{\rm depl}$ in the KS-relation. 
Alternatively, if the DP structure derives from two different gas populations, two different SF efficiencies would be expected, with two different values of t$_{\rm depl}$ and thus the two components would not be aligned with the lines of constant t$_{\rm depl}$. 
We note that due to the logarithmic scale in Fig.\,\ref{fig:ks},
t$_{\rm depl}$ scales also in a logarithmic manner perpendicular to the lines of constant t$_{\rm depl}$. Hence, the difference in t$_{\rm depl}$ between the two components is expected the largest for the galaxies near the line of t$_{\rm depl}=10\,{\rm Gyr}$, and classified as mergers. Indeed, these galaxies show a difference in t$_{\rm depl}$ between 0.2 and 0.5\,Gyr. In parallel, a difference in t$_{\rm depl}$ of up to 0.3\,Gyr is estimated for the non-merger galaxies.

These trends suggest that the DP signature can probe two different star formation sites in the merger galaxies, but also a central disc of homogeneous star formation in the LTG and S0 galaxies. However, taking the uncertainties into account, it is difficult to be conclusive on these two scenarios.
To correctly classify multiple components in a galaxy, observations with optical integral field spectroscopy and spatially resolved measurements of the molecular gas are needed. Such high resolution observations might also reveal perturbed merger remnants or contracted gaseous discs in the remaining DP galaxies, for which we fail to perform a combined fit. 

\subsection{Molecular gas mass fraction and depletion time}\label{ssect:scaling_relations}
\begin{figure*}
\centering 
\includegraphics[width=0.98\textwidth]{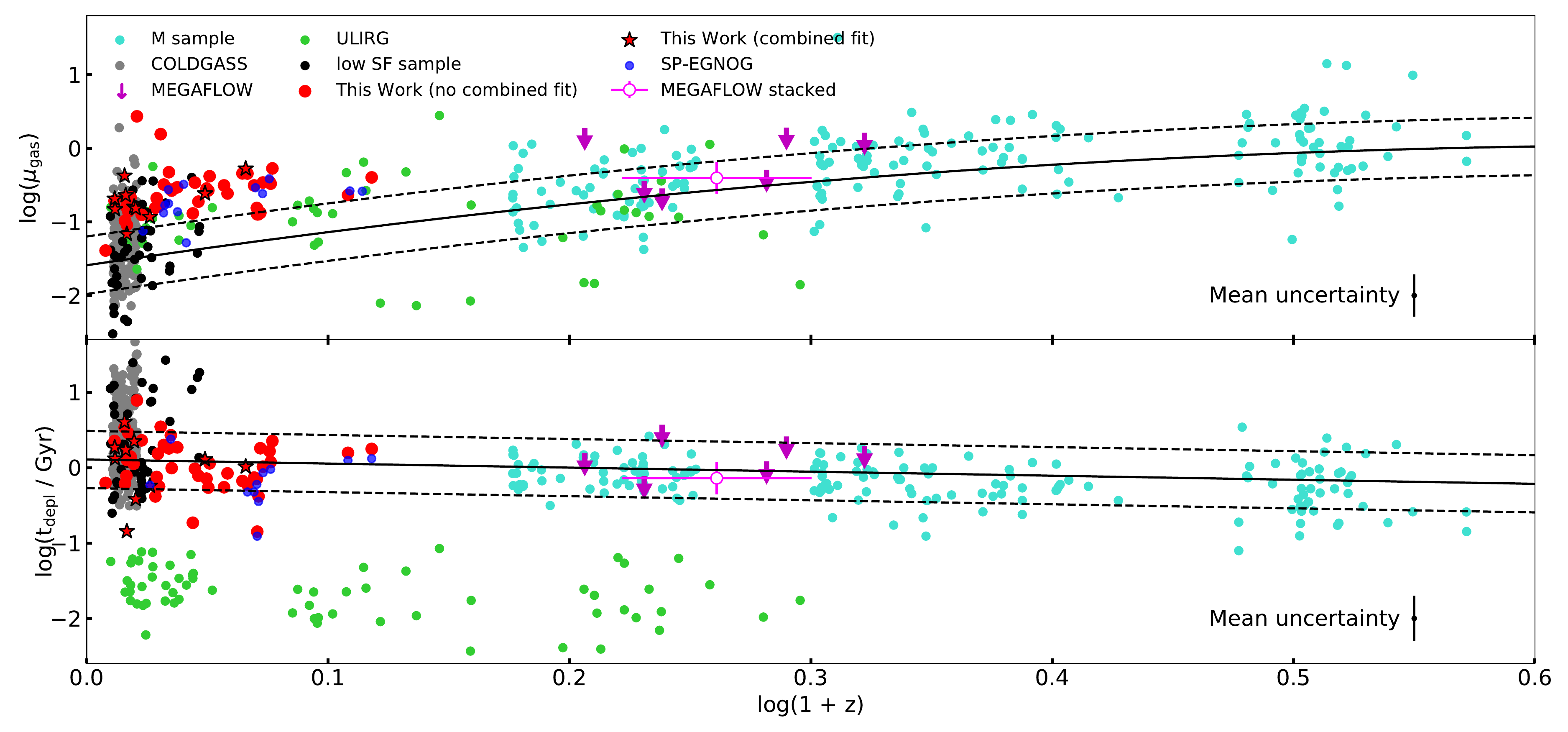}
\caption{Gas fraction ${\rm \mu_{gas} = M_{H2} / M_{*}}$ (top panel) and depletion time ${\rm t_{depl} = M_{H2} / SFR}$ (bottom panel) as a function of ${\rm log( 1 + z)}$. We show the observed DP galaxies (red), the EGNOG sample (blue), the COLD GASS sample (grey), the low SF sample (black), the ULIRG sample (green), the M sample (turquoise) and the detection limits of the MEGAFLOW galaxies (magenta). With an empty circle, we show the estimate based on stacking from \citet{2021MNRAS.501.1900F}. The solid black lines represent the scaling relations expected for MS galaxies found by \citet{2018ApJ...853..179T}, scaled to the mean stellar mass and size of the DP sample (${\rm log(M_*/M_{\odot}) = 11.0}$ and ${\rm R_e = 8.3 kpc}$). The dashed line mark the scatter of 0.4\,dex found for the M sample.}
\label{fig:scaling}%
\end{figure*}
\begin{figure*}
\centering 
\includegraphics[width=0.98\textwidth]{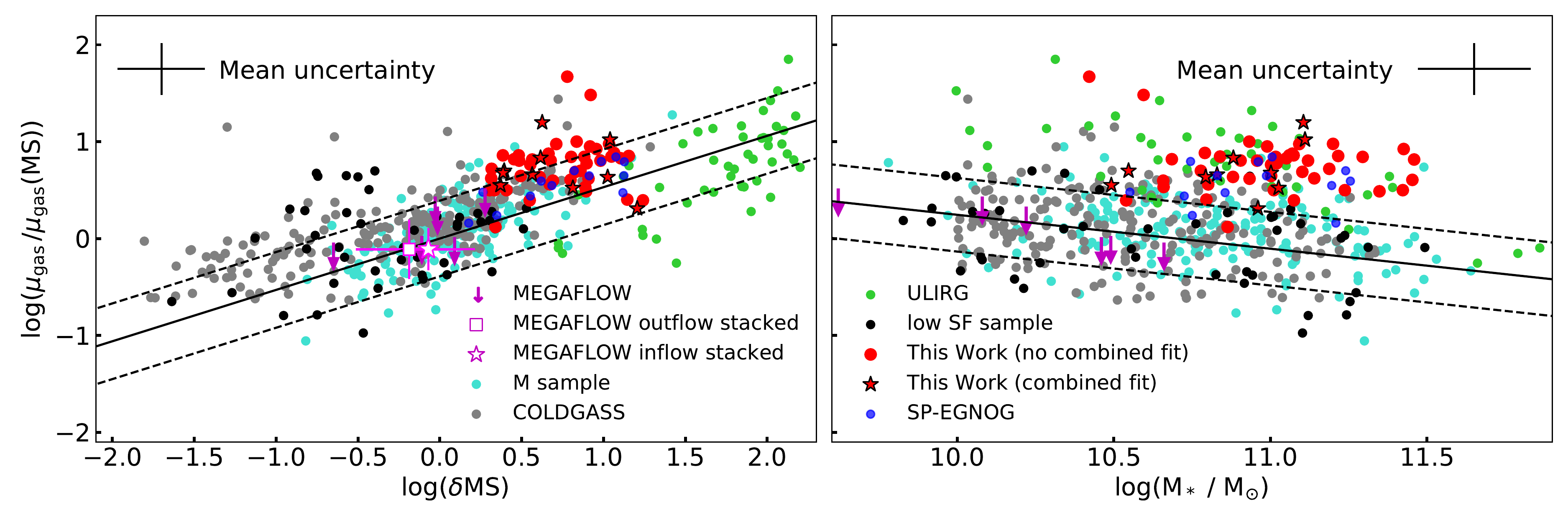}
\caption{
Gas fraction divided by their average values on the MS ${\rm \mu_{gas} / \mu_{gas}(MS)}$ as a function of as a function of their relative distance to the main sequence ${\rm \delta MS = SFR / SFR_{MS}}$ (left panel) and stellar mass M$_*$ (right panel). 
We use the predicted ${\rm \mu_{gas}(MS)}$ according to \citet{2018ApJ...853..179T} and use the parametrisation of ${\rm SFR_{MS}}$ following \citet{2014ApJS..214...15S}.
We show the observed DP galaxies (red), the SP-EGNOG sample (blue), the COLD GASS sample (grey), the low SF sample (black), the ULIRG sample (green), the M sample (turquoise). For the MEGAFLOW sample, we show the detection limits of each galaxy (magenta) and, with an empty square (resp. star), the estimate reached by stacking for galaxies identified with an outflow (resp. inflow) \citep{2021MNRAS.501.1900F}. 
The solid black lines represent the scaling relations
found by \citet{2018ApJ...853..179T} and the dashed line mark the scatter of 0.4\,dex found for the M sample.}
\label{fig:scaling:specific}%
\end{figure*}
The evolutionary state of a galaxy is strongly dependent on its star formation rate relying on the molecular gas reservoir. We thus characterise the CO samples using the molecular gas-to-stellar mass ratio (${\rm \mu_{gas} = M_{H_2} / M_{\odot}}$) and the depletion time  (${\rm t_{depl} = M_{H_2} / SFR}$). In Fig.\,\ref{fig:scaling}, we show $\mu_{\rm gas}$ and t$_{\rm depl}$ as a function of log(1 + z) for a redshift range of $0 < z < 3$. We present the DP sample, the M sample, the COLD GASS sample, the low SF sample, the SP-EGNOG sample, the ULIRG sample and the estimated limits for the MEGAFLOW sample and their estimate obtained through stacking.
To discuss the different measurements, we make use of the best fitting unified scaling relations for $\mu_{\rm gas}$ and t$_{\rm depl}$ found by \citet{2018ApJ...853..179T}:
\begin{equation}\label{eq:mu}
\begin{split}
    {\rm log(\mu_{gas})} = 
    0.12 
    - {\rm 3.62 \, (log(1 + z) - 0.66)^{2}}
    + {\rm 0.53 \, log(\delta MS)}\\
    - {\rm 0.35 \, log(\delta M_{*})}
    + {\rm 0.11 \, log(\delta R)}
\end{split}
\end{equation}
and
\begin{equation}\label{eq:tdepl}
\begin{split}
    {\rm log(t_{depl})} = 
    0.09 
    - {\rm 0.62 \, log(1 + z)}
    - {\rm 0.44 \, log(\delta MS)}\\
    - {\rm 0.09 \, log(\delta M_{*})}
    + {\rm 0.11 \, log(\delta R)}
\end{split}
\end{equation}
where ${\rm \delta MS = SFR / SFR_{MS}}$ is the distance to the MS, ${\rm \delta M_*} = M_* / 5\times10^{10} M_{\odot}$ the correction to the fiducial stellar mass of $5\times 10^{10} {\rm M_{\odot}}$ and ${\rm \delta R = R_e / R_{0,e}}$ the correction of the effective radius ${\rm R_e}$ to the average radius of star forming galaxies of ${\rm R_{0,e} = 8.9 kpc (1 + z)^{-0.75}(M_*/5\times 10^{10} M_{\odot})^{0.23}}$.
In Fig.\,\ref{fig:scaling}, we show the expected evolution of $\mu_{\rm gas}$ and t$_{\rm depl}$ for MS galaxies as a function of redshift and of the mean stellar mass of the DP sample (${\rm M_* = 10^{11} M_{\odot}}$) and the mean size of ${\rm R_e = 8.3 kpc}$. These relations allow one to visualise how the molecular gas fraction decreases steeply since $z \sim 3$, whereas in parallel the depletion time slightly increases, showing that ongoing star formation depends on the available molecular gas.
We find that the M sample is still well described by the empirical line and thus present the scatter of 0.4\,dex for the expected scaling relation.
We choose this scaling relation since the majority of the literature samples discussed here (COLD GASS, SP-EGNOG, parts of the ULIRG and the M sample) were used to obtain this relation valid over large parts of cosmic time.

Galaxies of the low SF sample have a large scatter in $\mu_{\rm gas}$ and in depletion time. Some of these galaxies are quenched due to exhausted gas reservoirs or stopped their SF due to AGN activity. This allows to probe a wide range of gas fractions and depletion times. A similar behaviour is also observed for the COLD GASS sample. We find that galaxies with long depletion times of these two samples are dominated by AGN.
In contrast, the ULIRG sample show a large scatter of gas fractions of 1 to 2\,dex and very short depletion times of around 0.01\,Gyr, which might be also an effect of overestimation of their SFR. As discussed in Sect.\,\ref{ssect:ms}, their stellar mass can be overestimated due to the presence of strong AGN. 
While their possibly underestimated mass fraction of gas $\mu_{\rm gas}$ follows the scaling relation, their depletion time, 2 dex below the scaling relation, is probably underestimated. 

The DP and SP-EGNOG samples are shifted above the expected gas fraction, indicating an unusual large gas reservoir but their depletion time is in the expected range at low redshift. 
We find two of the DP galaxies (DP-17 and DP-24) to be shifted 2\,dex over the expected gas fraction. DP-24 is a ULIRG merger selected from \citet{2009AJ....138..858C} and DP-17 is a major merger galaxy selected from the COLD GASS sample. We thus assume that the large $\mu_{\rm gas}$ values are due to an overestimated CO-to-molecular gas mass conversion factor (see Sect.\,\ref{ssect:co_h2}). 
We also identify 3 DP galaxies (DP-10, 30 and 44) and one galaxy of the SP-EGNOG (A06) with unexpected small t$_{\rm depl}$ values of $\sim 0.1 {\rm Gyr}$. We find all of these galaxies to show a large SFR, between 35 and $106\,{\rm M_{\odot} yr^{-1}}$.

In Fig.\,\ref{fig:scaling:specific}, we show $\mu_{\rm gas}$ divided by the expected value provided by equation\,\ref{eq:mu} as a function of their relative distance to the MS (${\rm \delta MS}$) (left panel) and their stellar mass (right panel). We computed the expected SFR value of the MS (${\rm SFR_{MS}}$) following \citet{2014ApJS..214...15S}.
In order to compare these values to the expected scaling relations established by \citet{2018ApJ...853..179T}, we show black lines indicating the relative deviation of $\mu_{\rm gas}$ from the expected value of the main sequence $\mu_{\rm gas}({\rm MS})$. 
Therefore, we compute $\mu_{\rm gas}$, according to equation\,\ref{eq:mu}, with  the fiducial values despite different ${\rm \delta MS}$ (resp. ${\rm M_*}$) values for the left (resp. right) panel, while the redshift term cancels out. We then divide these values by $\mu_{\rm gas}({\rm MS})$ calculated for the same arbitrary redshift, ${\rm \delta MS} = 1$ and the fiducial values. The resulting function for the left panel is ${\rm\mu_{gas} / \mu_{gas}({MS}) = 0.53\, log(\delta{MS})}$ and ${\rm \mu_{gas} / \mu_{gas}({MS}) = -0.35\, log(\delta{M_{*}})}$ for the right panel. Dashed lines show the scatter found for the M sample.

We present the same samples as for Fig.\,\ref{fig:scaling} but show for the MEGAFLOW sample the stacked estimates for galaxies with identified outflows and inflows separately. As discussed in \citet{2021MNRAS.501.1900F}, gas fraction and depletion times are well compatible with the established values of the MS with a mass-selected sample \citep{2018ApJ...853..179T}. This indicates that galaxies with identified accretion may not have a specifically large gas content, at odds with the theoretical expectations of the quasi equilibrium and compaction models where accretion of gas is seen as a gas replenishment.
Galaxies from the M sample are scattered around the expected values on the MS, as well as the majority of the COLD GASS galaxies. Some galaxies of the COLD GASS and the low SF samples are shifted below the MS but with significant large amounts of molecular gas and large ${\rm t_{depl}}$ values. As discussed in Sect.\,\ref{ssect:ks}, these outliers are mostly dominated by an AGN. 
The ULIRG sample exhibits a fraction of molecular gas $\mu_{\rm gas}$, expected for their distance to the MS according to \citet{2018ApJ...853..179T}. 

Galaxies from the SP-EGNOG sample are slightly shifted  above the expected MS $\mu_{\rm gas}$ values (except A06) but is still compatible with the scatter of the MS.
For the DP sample galaxies, due to their selection criteria, they belong to the upper MS. Most of them exhibit a larger molecular gas reservoir than the MS scaling relation found by \citet{2018ApJ...853..179T}.
As explained above, two galaxies (DP-17 and DP-24) are shifted more than 1\,dex above the expected value.
In contrast to that, we find 3 galaxies with smaller gas mass fraction than expected. Interestingly, these galaxies (DP-10, 30 and 44) are the exact same galaxies which are shifted to significantly shorter t$_{\rm depl}$ values (see Fig.\,\ref{fig:scaling}) with strong star formation rates. This might be an indicator of a late-stage starburst and an almost depleted gas reservoir.
We see for the majority of the DP sample that the expected $\mu_{\rm gas}$ values are even higher relative to their shift above the main sequence. This means in parallel that we observe larger t$_{\rm depl}$ values than we would expect for their distance to the MS. Thus, the gas is not consumed as efficiently as expected which could be due to recent gas accretion, corresponding to an early-stage starburst. Alternatively, the large bulges characterising the DP galaxies might stabilise the gas and reduce the star formation efficiency.

On the right panel of Fig.\,\ref{fig:scaling:specific}, we observe the COLD GASS, low SF and the M samples to be scattered around the expected values for the MS. However, we observe a strong deviation from the scaling relations for the SP-EGNOG, ULIRG and DP samples. These samples show larger gas mass fractions than expected for galaxies of their mass on the MS. 
The shift above the MS observed for these galaxies might be related to galaxy interactions and mergers which increased their molecular gas reservoir significantly and fuelled their star formation.

\section{Discussion}\label{sect:discussion}
The results are discussed here to account for the properties of the DP sample. Sect. \ref{ssect:SF-gal} is devoted to galaxies which are mainly star-formation dominated with a remarkable radio-continuum-CO correlation. Sect. \ref{ssect:discussion:rotating:disc} discusses the effect of bars.  Sect. \ref{ssect:discussion:central:sf} suggests that DP galaxies are akin to the compaction phase observed after mergers in high-z galaxies. 

\subsection{Star-formation dominated galaxies}
\label{ssect:SF-gal}
The evolutionary state of galaxies is mostly determined by their growth rate and star formation efficiency. The relation between star formation rate and molecular gas densities is important, since it quantifies the efficiency of the process \citep[see][and references therein]{2008AJ....136.2846B}. 
In a recent study, \citet{2021MNRAS.500.1261C} showed that L$_{\rm CO}$ and IR emission at ${\rm 12\,\mu m}$ describe an even more robust correlation than L$_{\rm CO}$ and SFR. This provides a better estimator to predict the molecular gas mass for different kinds of galaxies. This is neither significantly affected by the presence of an obscured AGN nor relying on a correct choice of the CO-to-H$_2$ conversion factor. 

Ongoing star formation is also measurable in the IR regime where dust grains are heated from the ultraviolet light emitted by young stars. As discussed in Sect.\,\ref{ssect:co_radio}, electrons are accelerated in supernova remnants of massive young stars, emitting synchrotron radiation. The underlying process of these two emissions is star formation and the radio continuum–infrared (RC–IR) correlation has been studied extensively for star-forming galaxies \citep[e.g.][]{2003ApJ...586..794B, 2008MNRAS.386..953I, 2010A&A...518L..31I, 2014MNRAS.445.2232S, 2015ApJ...805...31L}. 
Although the RC–IR correlation has been known for a long time to be one of the tightest in galaxy physics, \citet{2021MNRAS.504..118M} emphasised the poor match of IR and radio samples, that could bias the calibration. They find a slightly non-linear correlation of slope 1.11 $\pm$ 0.01.

In order to extend the RC–IR correlation, \citet{2020MNRAS.495.1760O} found a three dimensional connection L$_{\rm CO}$, L$_{\rm 1.4\,GHz}$ and the infrared luminosity L$_{\rm IR}$ for galaxies with a redshift smaller than $z < 0.27$. They excluded quasar-like objects to focus on star formation activity. To further explore this relation, we tested the correlation between L$_{\rm CO}$ and the radio continuum luminosity at 150\,MHz, 1.4\,GHz and 3\,GHz. 
We find as well a linear relation for galaxies classified as SF or COMP with the BPT diagram (see Sect.\,\ref{sssect:bpt}) for all three radio continuum measurements. We performed a linear fit and find a slightly flatter relation between L$_{\rm CO}$ and L$_{\rm 1.4\,GHz}$ than \citet{2020MNRAS.495.1760O}. Furthermore, we find nearly the same slope for L$_{\rm CO}$-L$_{\rm 150\,MHz}$ relation ($0.80\pm0.06$), the L$_{\rm CO}$-L$_{\rm 1.4\,GHz}$ relation ($0.79\pm0.04$) and the L$_{\rm CO}$-L$_{\rm 3\,GHz}$ relation ($0.87\pm0.07$). 

We also find that these linear relations are not valid for the majority of ULIRGs, which are mostly active galaxies such as quasars or AGN. Galaxies with high IR luminosities were observed to be nearly all advanced mergers with circum-nuclear starburst and AGN activity \citep{1996ARA&A..34..749S}. Such galaxies might characterise an important stage of quasar formation and powerful radio galaxies, which is compatible with the large offset that the ULIRG sample shows between CO and radio continuum luminosities. It is still under debate to which relative fractions the ongoing starburst and the AGN are contributing to the IR and radio continuum luminosities \citep[e.g.][]{2018MNRAS.480.3562D}.

Hence, the slope for SF and COMP galaxies measured constant over a wide range of radio wavelengths might be an indicator that the underlying process is dominated by star formation.
This is also in agreement with DP galaxies having comparable depletion times of around 1\,Gyr (see Sect.\,\ref{ssect:ks}). In particular, the good agreement between results at 1.4\,GHz, 3\,GHz and 150\,MHz suggests that the latter observable is a robust tracer for star formation \citep{2017MNRAS.469.3468C, 2019A&A...631A.109W}. 
Hence, the presented DP sample is dominated by star formation with no significant AGN contribution, as confirmed by the BPT classification discussed in Sect. \ref{sssect:bpt}. Indeed, only two galaxies (DP-1 and DP-22), respectively a merger and a S0 galaxy, exhibit AGN excitation on the BPT diagram, but they have a high star formation rate and large molecular gas content. 

\subsection{Gas infall due to bars}
\label{ssect:discussion:rotating:disc}
Bars are well known to effectively transport gas inwards and create a central starburst. The torques they exert may lead to the creation of star-forming rings in the central parts of galaxies or to accumulation of gas in the very centre (see \citet{1996FCPh...17...95B} for a review). \citet{1999ApJ...525..691S} showed, using a sample of nearby galaxies, that the central molecular gas concentration is higher in barred systems than in unbarred galaxies. This leads to a central star formation enhancement, which they estimated to be larger than $0.1-1\,{\rm M_{\odot} yr^{-1}}$. By comparing barred, unbarred, and interacting systems, \citet{2019MNRAS.484.5192C} found that cold gas is transported inwards by a bar or tidal interaction, which leads to the growth and rejuvenation of star formation in the central region. \citet{2011MNRAS.416.2182E} found that bars are responsible for 3.5 times more triggered central star formation than galaxy-galaxy interactions.

However, in our sample, we only find DP-13 and DP-14 to show a dominant bar. As discussed in Sect.\,\ref{sssect:environment_morph}, the present sample is characterised by bulge-dominated systems and mergers. This suggests that the central star formation enhancement and the higher gas concentration which we observe are most likely related to galaxy interactions and mergers, as we discuss in the next section.

\subsection{Central star formation and compaction phase?}
\label{ssect:discussion:central:sf}
As mentioned in Sect.\,\ref{sect:introduction}, galaxy interactions and galaxy mergers can trigger star formation \citep{1986ApJ...301...57B, 2002MNRAS.331..333P}. Relying on larger galaxy samples with 10$^5$ SDSS DR4 galaxy pairs, \citet{2008MNRAS.385.1903L} found a clear star formation enhancement triggered by galaxy interactions. Based on a systematic search for galaxy pairs in the SDSS DR7, \citet{2011MNRAS.412..591P} found evidence for a central starburst induced by galaxy interactions. 
As discussed in \citet{2009Natur.457..451D}, the merger mechanism forms steady streams enhancing the growth of the central spheroid, leading to earlier Hubble Types. This is a different evolution from one of violent mergers which strongly modify the morphology, and is in agreement with the hierarchical bulge growth described in \citet{2007A&A...476.1179B}. 

We find that 19\,\% of the DP sample galaxies show the same kinematic signature in the molecular as in the ionised gas, indicating that in these galaxies, most of the molecular gas is located in the very central region of radius $3^{\prime\prime}$. We, furthermore, observe a central star formation enhancement for the majority of the DP sample and detect a significant gas reservoir. 
These findings confirm a compact central star formation site supporting an effective molecular gas transportation into the galaxy centre. 
This scenario is reminiscent of the gas compaction phase suggested for $z = 2-4$ galaxies by observations \citep[e.g.][]{2013ApJ...765..104B, 2017ApJ...840...47B} and simulations \citep{2015MNRAS.450.2327Z, 2016MNRAS.458..242T,2016MNRAS.457.2790T} according to which galaxies experience a central enhancement of SF due to gas contraction at their centres, before inside-out depletion and quenching.
This model was further described over large scales of cosmic time by \citet{2016MNRAS.457.2790T}, with repetitive compaction and depletion phases shifting galaxies up and down the main sequence before finally quenching. 
Since our galaxies are 0.3\,dex above the MS, the central SF enhancement and the similar kinematic distribution in the ionised and molecular gas might be a sign of an ongoing compaction phase. In such a scenario, a recent minor merger, a galaxy interaction or a disc instability funnelled gas into the central parts, igniting star formation. 

In Sect.\,\ref{ssect:scaling_relations}, we find that the DP and SP-EGNOG samples have significantly larger molecular gas fractions, by on average 0.8\,dex, than main-sequence galaxies of the same mass and redshift ranges studied by \citet{2018ApJ...853..179T}. This discrepancy can be accounted for by a conversion factor of $\alpha_{\rm CO} = 0.80 {\rm \,M_{\odot} / (K\,km\,s^{-1}\, pc^2)}$ adopted for ULIRGs by \citet{1997ApJ...478..144S}.
We also observe a central star formation enhancement for the SP-EGNOG sample, which is indeed very similar to the DP sample in terms of stellar mass and redshift.
In Fig.\,\ref{fig:scaling:specific}, we observe both samples to be situated between the population of the M sample and the extreme case of the ULIRG sample. Both samples are defined to be situated at the upper MS and above (see Sect.\,\ref{ssect:ms}) but are also showing an increase of molecular gas mass fraction with only a slight decrease in depletion time.
The galaxies of the MEGAFLOW sample, which are galaxies showing in- and outflows in the circumgalactic medium, are maintaining their star formation as described in the quasi-equilibrium model \citep{2021MNRAS.501.1900F}. These galaxies have SFE compatible with star formation efficiencies measured for the MS.
The offset of the DP sample, observed in Fig.\,\ref{fig:scaling:specific}, suggests that large amounts of gas were recently accreted, possibly through a merger event, and were effectively funnelled into the central regions, where we observe the majority of the ongoing star formation. 

\section{Conclusions}\label{sect:conclusion}
We present new observations of the molecular gas content for 35 DP emission line galaxies with ongoing star formation, situated more than 0.3\,dex above the MS. We consider in addition 17 DP galaxies from existing CO samples matching the same criterion, leading to a sample of 52 galaxies. We try to fit the same double Gaussian parameters to the central optical emission lines and to the CO lines integrated over the entire galaxy. We succeed in finding similar kinematic signatures for these two measurements in 10 ($19\,\%$) DP galaxies. By comparing the SFR inside the SDSS fibre and in the total galaxy, we find a significant central enhancement of star formation for this DP sample. 
We discuss the possibility of a rotating gaseous disc creating a DP signature. By comparing the emission line width of the CO gas and the galaxy inclination, we do not find any correlation but, considering the scatter expected due to galaxy mass concentration or molecular gas velocity dispersion, the lack of correlation does not allow us to exclude this origin for the observed DP. A deep gravitational potential can in fact be the origin of the DP. This might be the result of a recent minor merger event or galaxy-galaxy interaction which funnelled gas into the central regions.
The DP signature might also be the result of cold gas accretion from cosmic filaments, which recently fell into the galaxy centre.
These scenarios account for the observed increase of molecular gas and its funnelling into the central region where the majority of the stars are formed. 
This is also in agreement with the observed excess of dust extinction in the centre.

Such a recently ignited star formation is traced by radio continuum emission at 150\,MHz, 1.4\,GHz and 3\,GHz which are all three linearly correlated in log with ${\rm L'_{CO}}$ with the same slope. This is a signature of synchrotron emission, mostly dominated by star formation. Within this interpretation, the possible merger-induced central star formation is happening without a simultaneous increase in AGN activity.

Arguments are discussed whether the observed central star formation and the large molecular gas reservoir are the results of a recent merger. Bar structures in galaxies can also effectively migrate gas inwards, which cannot be the case for the presented galaxy samples as it lacks these morphological types. However, we cannot exclude the possibility that we observe gas-rich spiral galaxies with a central molecular disc formed due to large scale instabilities. In such a scenario, we would observe a central rotating disc which might not be aligned with the host galaxy orientation. To further probe our findings and to distinguish between rotating disc and merger-induced central star formation, high resolution observations of the molecular and ionised gas would be necessary. A kinematic decomposition of spatially extended gas would enable us to further characterise the dynamics of these systems and draw conclusions on the origin of double-peak emission line galaxies. In addition, we could also explore spatially resolved star formation and compare its efficiency with findings for regular spiral galaxies \citep{2008AJ....136.2846B, 2008AJ....136.2782L} in order to conclude on the underlying process of star formation and galaxy growth.

\begin{acknowledgements}
We thank the anonymous referee who helped us to improve our emission line fitting procedure which finally strengthened the arguments in our analysis and discussion. We thank Susanne Maschmann for helpful advise on the English language.

This work is based on observations carried out under the two project numbers 198-19 and 166-20 with the IRAM 30m telescope at Pico Veleta in Spain. IRAM is supported by INSU/CNRS (France), MPG (Germany) and IGN (Spain).

This paper makes use of the following ALMA data: ADS/JAO.ALMA\#2015:1:00113:S.ALMA is a partnership of ESO (representing its member states), NSF (USA) and NINS (Japan), together with NRC (Canada), MOST and ASIAA (Taiwan), and KASI (Republic of Korea), in cooperation with the Republic of Chile. The Joint ALMA Observatory is operated by ESO, AUI/NRAO and NAOJ.

Funding for the Sloan Digital Sky Survey IV has been provided by the Alfred P. Sloan Foundation, the U.S. Department of Energy Office of Science, and the Participating Institutions. SDSS-IV acknowledges
support and resources from the Center for High-Performance Computing at
the University of Utah. The SDSS web site is www.sdss.org.

SDSS-IV is managed by the Astrophysical Research Consortium for the 
Participating Institutions of the SDSS Collaboration including the 
Brazilian Participation Group, the Carnegie Institution for Science, 
Carnegie Mellon University, the Chilean Participation Group, the French Participation Group, Harvard-Smithsonian Center for Astrophysics, 
Instituto de Astrof\'isica de Canarias, The Johns Hopkins University, Kavli Institute for the Physics and Mathematics of the Universe (IPMU) / 
University of Tokyo, the Korean Participation Group, Lawrence Berkeley National Laboratory, 
Leibniz Institut f\"ur Astrophysik Potsdam (AIP),  
Max-Planck-Institut f\"ur Astronomie (MPIA Heidelberg), 
Max-Planck-Institut f\"ur Astrophysik (MPA Garching), 
Max-Planck-Institut f\"ur Extraterrestrische Physik (MPE), 
National Astronomical Observatories of China, New Mexico State University, 
New York University, University of Notre Dame, 
Observat\'ario Nacional / MCTI, The Ohio State University, 
Pennsylvania State University, Shanghai Astronomical Observatory, 
United Kingdom Participation Group,
Universidad Nacional Aut\'onoma de M\'exico, University of Arizona, 
University of Colorado Boulder, University of Oxford, University of Portsmouth, 
University of Utah, University of Virginia, University of Washington, University of Wisconsin, 
Vanderbilt University, and Yale University.

The Legacy Surveys (http://legacysurvey.org/) consist of three individual and complementary projects: the Dark Energy Camera Legacy Survey (DECaLS; NOAO Proposal ID \# 2014B-0404; PIs: David Schlegel and Arjun Dey), the Beijing-Arizona Sky Survey (BASS; NOAO Proposal ID \# 2015A-0801; PIs: Zhou Xu and Xiaohui Fan), and the Mayall $z$-band Legacy Survey (MzLS; NOAO Proposal ID \# 2016A-0453; PI: Arjun Dey). DECaLS, BASS and MzLS together include data obtained, respectively, at the Blanco telescope, Cerro Tololo Inter-American Observatory, National Optical Astronomy Observatory (NOAO); the Bok telescope, Steward Observatory, University of Arizona; and the Mayall telescope, Kitt Peak National Observatory, NOAO. The Legacy Surveys project is honoured to be permitted to conduct astronomical research on Iolkam Du`ag (Kitt Peak), a mountain with particular significance to the Tohono O'odham Nation.
\\

LOFAR is the Low Frequency Array designed and constructed by ASTRON. 
It has observing, data processing, and data storage facilities in several countries, which are owned by various parties (each with their own funding sources), and which are collectively operated by the International LOFAR Telescope (ILT) foundation under a joint scientific policy. 
The ILT resources have benefited from the following recent major funding sources: CNRS-INSU, Observatoire de Paris and Universit\'e d'Orl\'eans, France; BMBF, MIWF-NRW, MPG, Germany; Science Foundation Ireland (SFI), Department of Business, Enterprise and Innovation (DBEI), Ireland; NWO, The Netherlands; The Science and Technology Facilities Council, UK; Ministry of Science and Higher Education, Poland; The Istituto Nazionale di Astrofisica (INAF), Italy.
This research made use of the Dutch national e-infrastructure with support of the SURF Cooperative (e-infra 180169) and the LOFAR e-infra group. The J\"ulich LOFAR Long Term Archive and the German LOFAR network are both coordinated and operated by the J\"ulich Supercomputing Centre (JSC), and computing resources on the supercomputer JUWELS at JSC were provided by the Gauss Centre for Supercomputing e.V. (grant CHTB00) through the John von Neumann Institute for Computing (NIC).

This research made use of the University of Hertfordshire high-performance computing facility and the LOFAR-UK computing facility located at the University of Hertfordshire and supported by STFC [ST/P000096/1], and of the Italian LOFAR IT computing infrastructure supported and operated by INAF, and by the Physics Department of Turin university (under an agreement with Consorzio Interuniversitario per la Fisica Spaziale) at the C3S Supercomputing Centre, Italy.

\end{acknowledgements}

\bibliographystyle{aa}
\bibliography{Mybiblio}

\appendix

\section{Spectra}\label{sect:spectra}
In Fig.\,\ref{fig:spec} to \ref{fig:spec_5} we show all galaxies of the DP sample. We present their $70^{\prime\prime} \times 70^{\prime\prime}$ legacy survey snapshot \citep{2019AJ....157..168D}, the ionised gas emission lines H$\alpha$ and the [NII]$\lambda 6550, 6585$ doublet, and the CO spectra. We further show the fit results as discussed in Sect.\,\ref{ssect:fitting}.
\begin{figure*}[h]
    \makebox[\linewidth][c]{
    \includegraphics[width=0.98\textwidth]{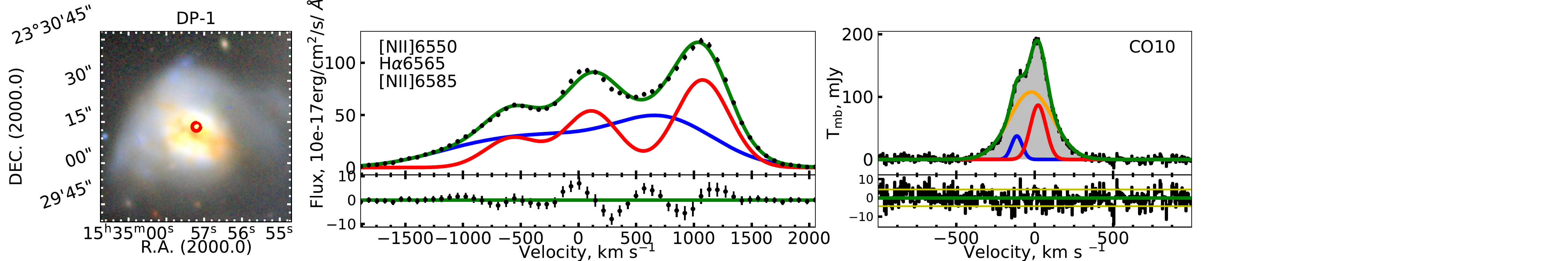}
    }
    \makebox[\linewidth][c]{
    \includegraphics[width=0.98\textwidth]{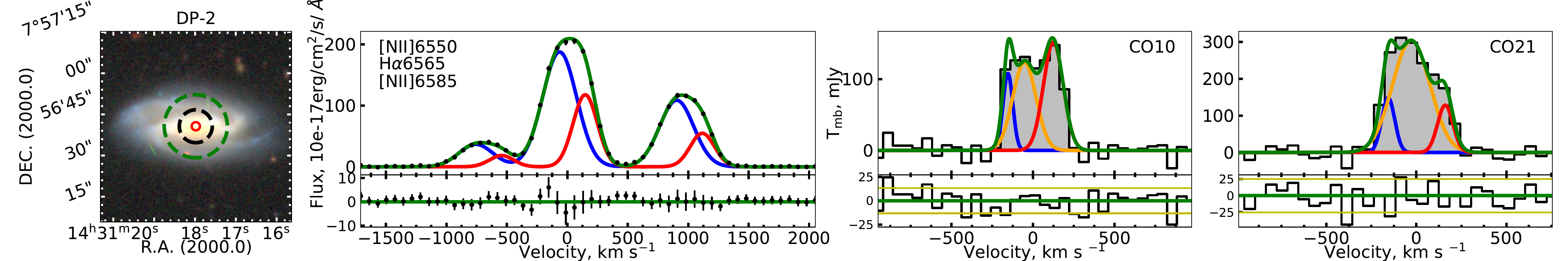}
    }
    \makebox[\linewidth][c]{
    \includegraphics[width=0.98\textwidth]{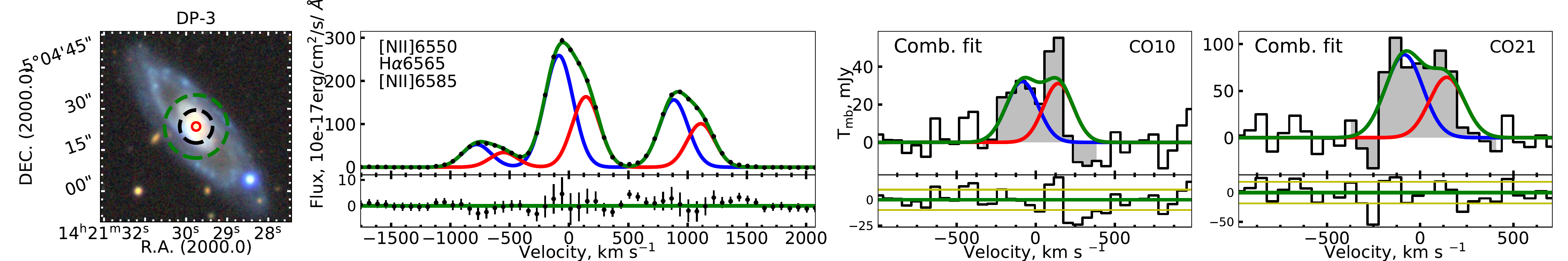}
    }
    \makebox[\linewidth][c]{
    \includegraphics[width=0.98\textwidth]{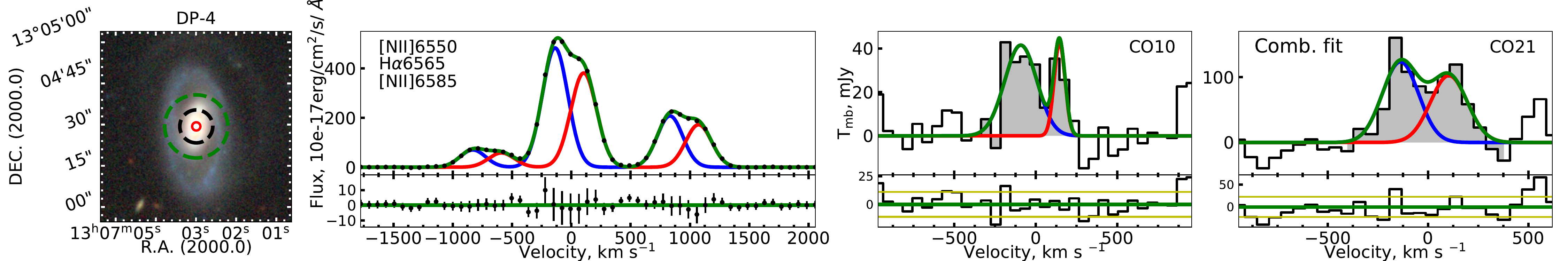}
    }
    \makebox[\linewidth][c]{
    \includegraphics[width=0.98\textwidth]{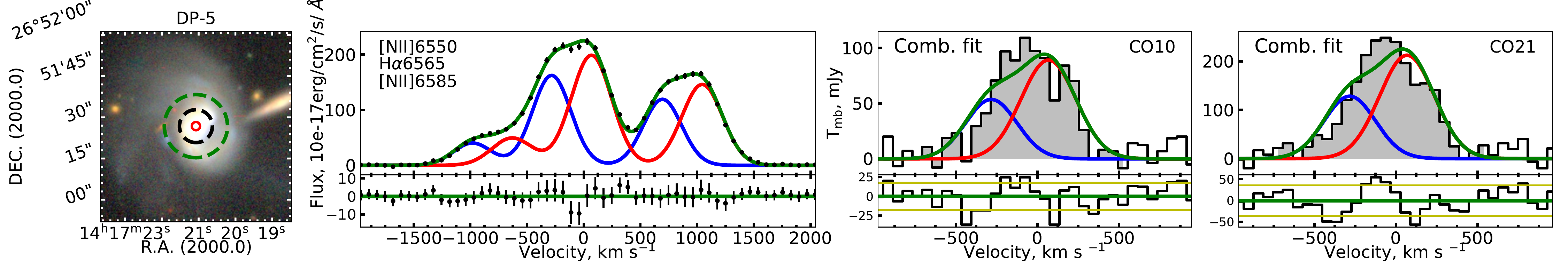}
    }
    \makebox[\linewidth][c]{
    \includegraphics[width=0.98\textwidth]{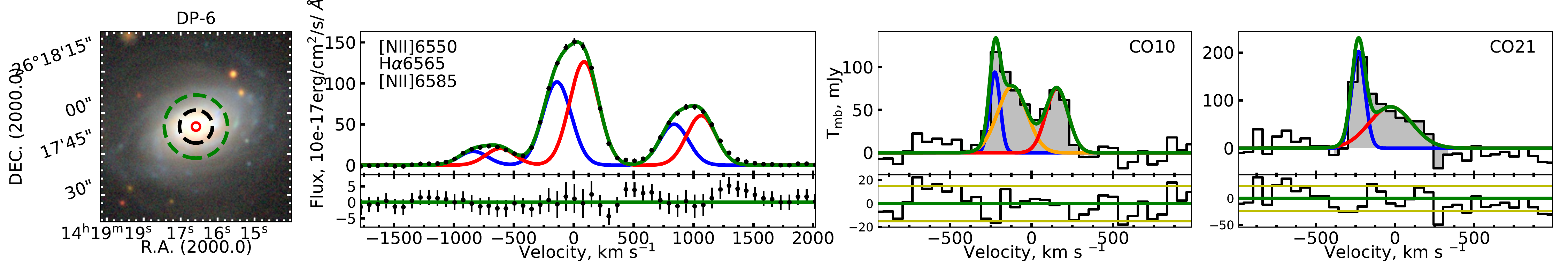}
    }
    \makebox[\linewidth][c]{
    \includegraphics[width=0.98\textwidth]{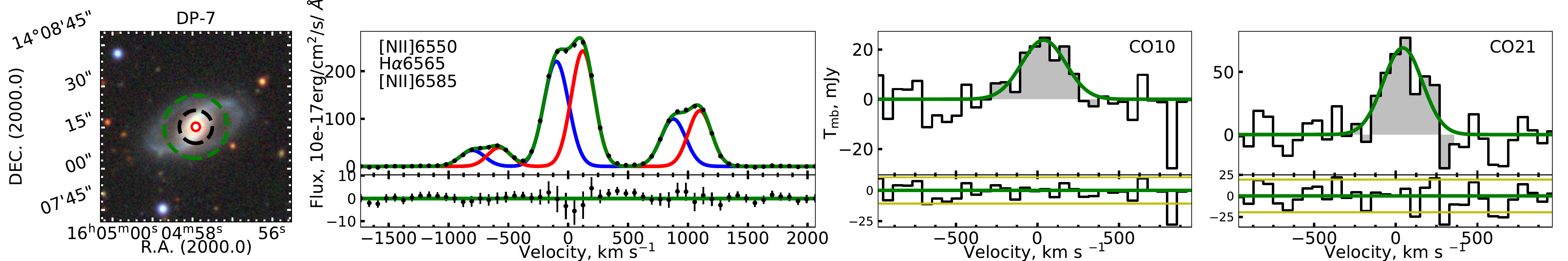}
    }

   \caption{Fit results of ionised gas emission lines and CO(1-0)/CO(2-1) lines. We show on the left the $70^{\prime\prime} \times 70^{\prime\prime}$ legacy survey snapshots \citep{2019AJ....157..168D} and mark the position of the SDSS 3$^{\prime\prime}$ fibre in red, the IRAM CO(1-0) (resp. CO(2-1)) beam of 23$^{\prime\prime}$ (resp. 12$^{\prime\prime}$) with a green (resp. black) dashed line. For interferometry observations conducted by \citet{2013ApJ...768..132B}, we show the beam with a blue ellipse. For DP-24, DP-25 and DP-31, we show the FCRAO CO(1-0) beam of 50$^{\prime\prime}$. For DP-1, we extracted a CO(1-0) signal from ALMA interferometry observation for the exact same area as the SDSS 3$^{\prime\prime}$ fibre. Next to the snapshots, we show the H$\alpha$ emission line and the [NII]$\lambda6550/6585$ doublet fitted with a double Gaussian function. On the right-hand side, we show the CO(1-0) and the CO(2-1) line, if observed, fitted by a single, a double or a triple Gaussian functions. In case of a successful combined fit as described in Sect.\,\ref{ssect:fitting}, we indicate this in the top left of the CO panels. For a detailed description of the fitting procedure, see Sec.\,\ref{ssect:fitting}.}
    \label{fig:spec}
\end{figure*}
\setcounter{figure}{0}
\begin{figure*}[h]
    \makebox[\linewidth][c]{
    \includegraphics[width=0.98\textwidth]{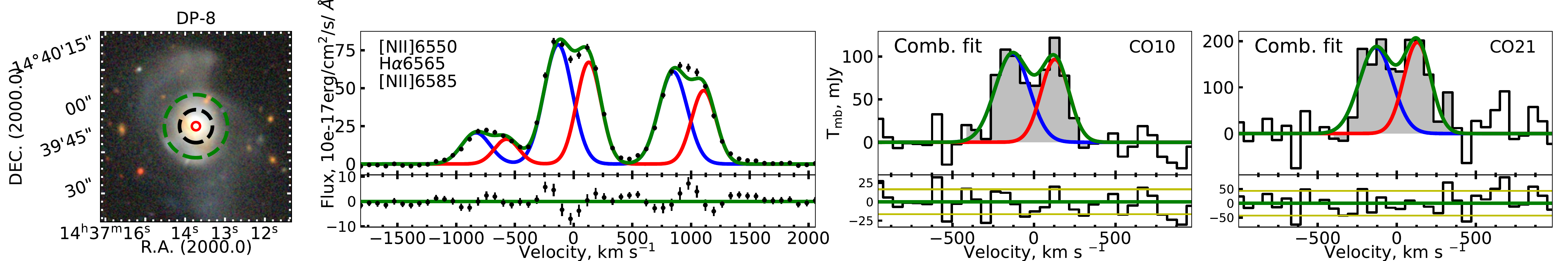}
    }
    \makebox[\linewidth][c]{
    \includegraphics[width=0.98\textwidth]{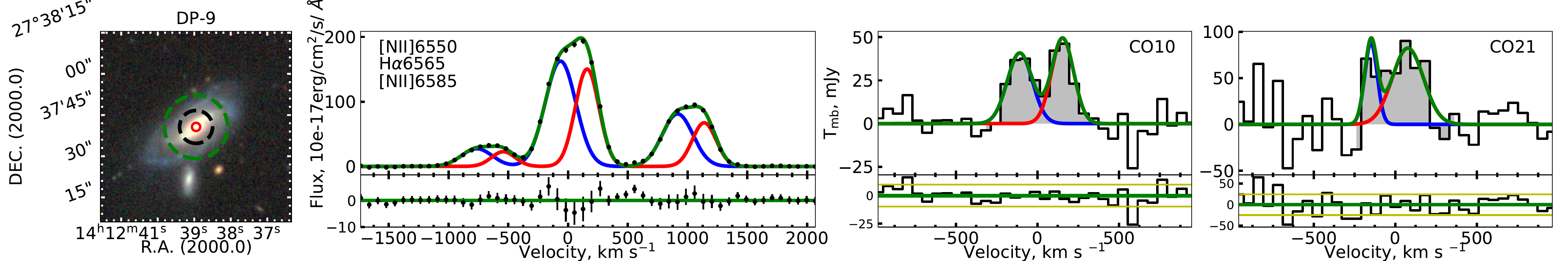}
    }
    \makebox[\linewidth][c]{
    \includegraphics[width=0.98\textwidth]{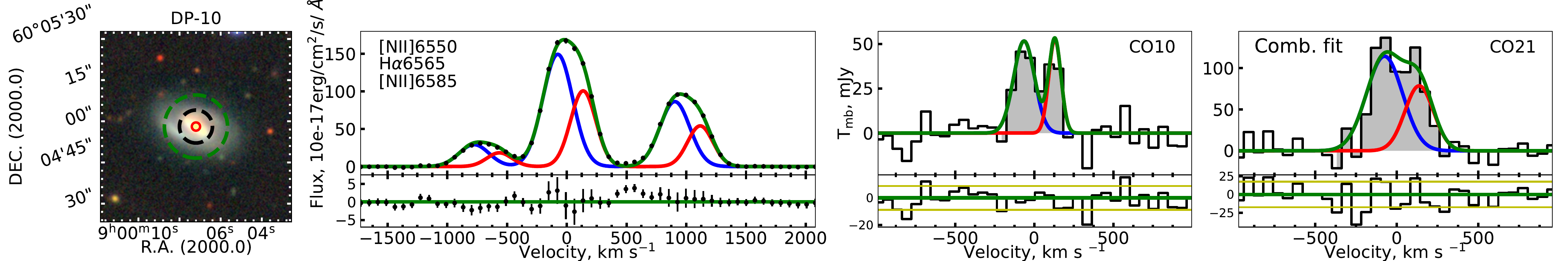}
    }
    \makebox[\linewidth][c]{
    \includegraphics[width=0.98\textwidth]{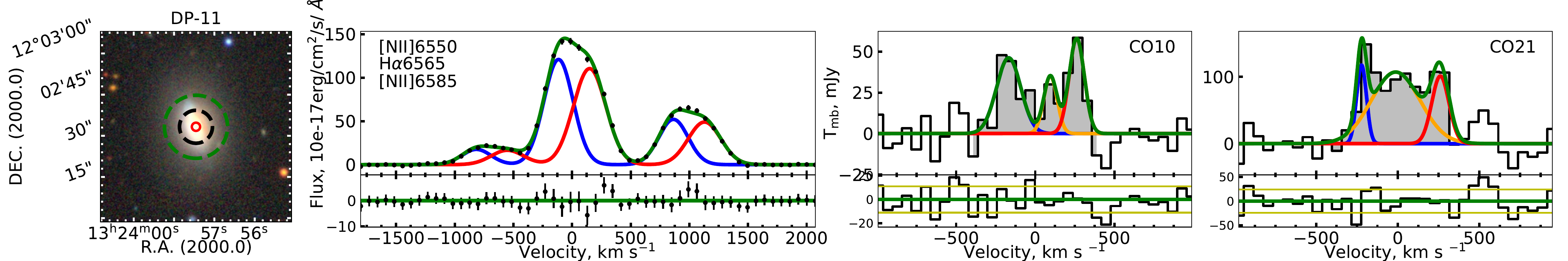}
    }
    \makebox[\linewidth][c]{
    \includegraphics[width=0.98\textwidth]{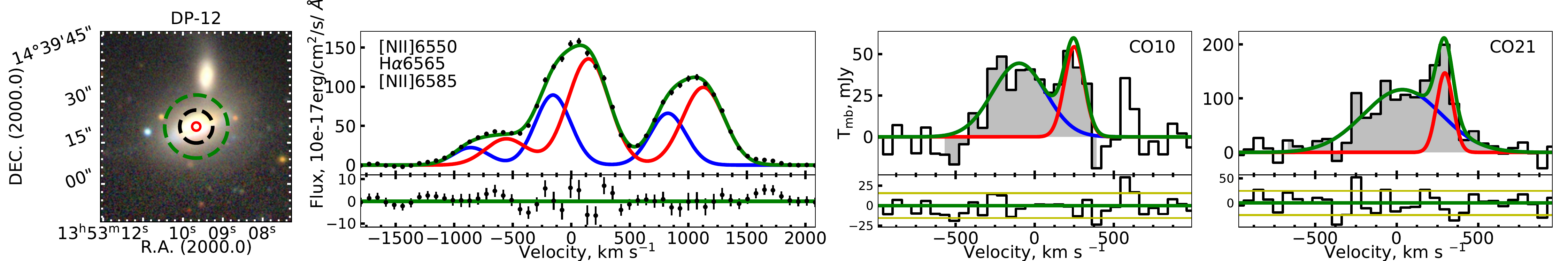}
    }
    \makebox[\linewidth][c]{
    \includegraphics[width=0.98\textwidth]{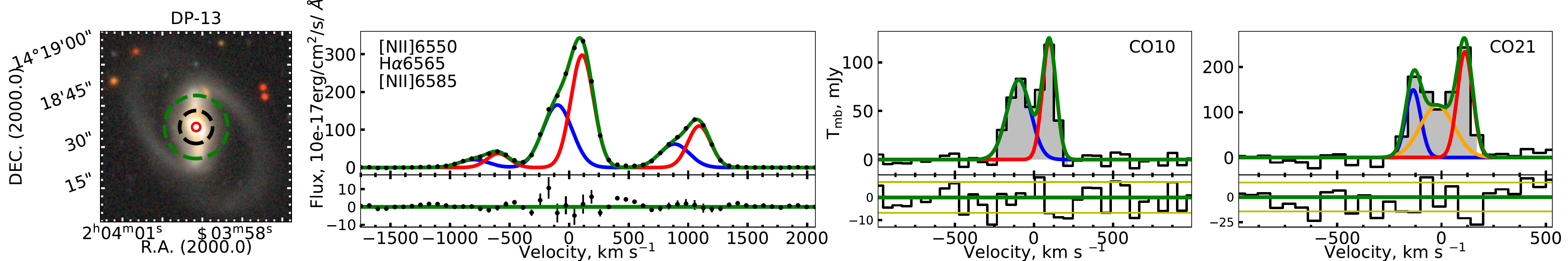}
    }
    \makebox[\linewidth][c]{
    \includegraphics[width=0.98\textwidth]{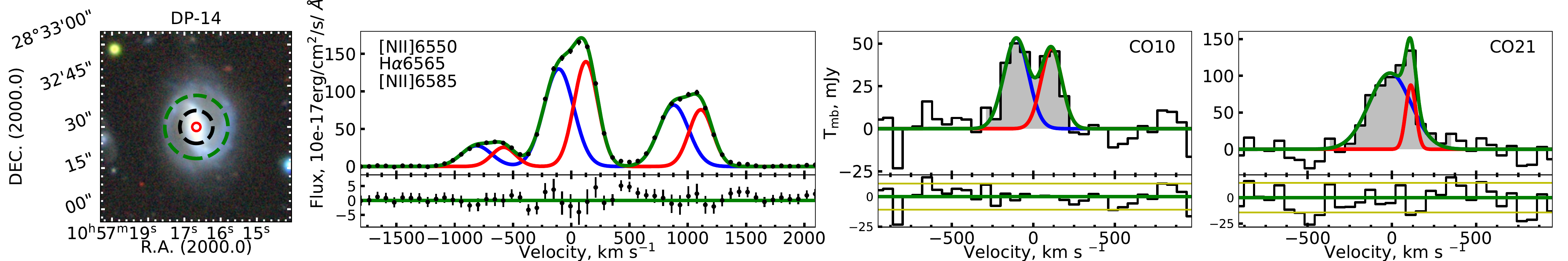}
    }
    \makebox[\linewidth][c]{
    \includegraphics[width=0.98\textwidth]{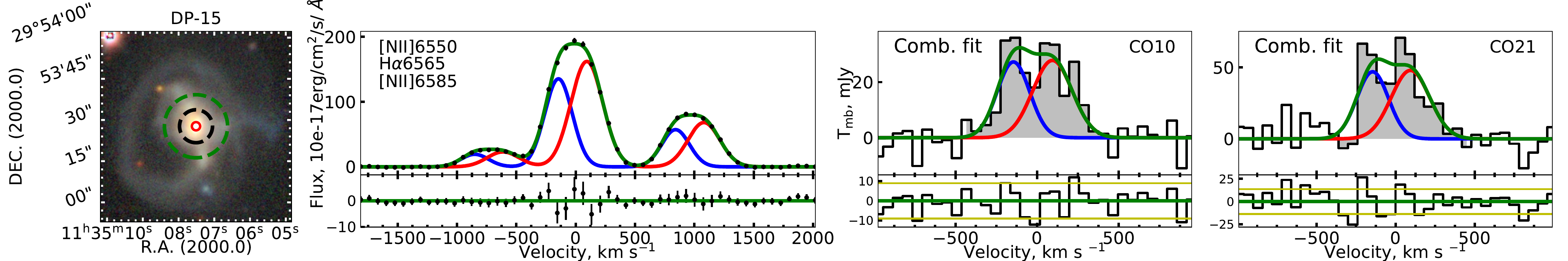}
    }
   \caption{Continued}
    \label{fig:spec_1}
\end{figure*}
\setcounter{figure}{0}
\begin{figure*}[h]
    \makebox[\linewidth][c]{
    \includegraphics[width=0.98\textwidth]{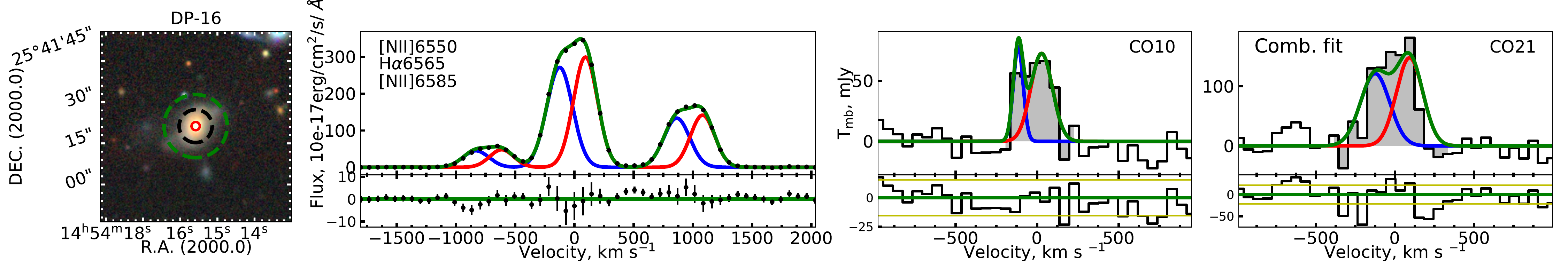}
    }
    \makebox[\linewidth][c]{
    \includegraphics[width=0.98\textwidth]{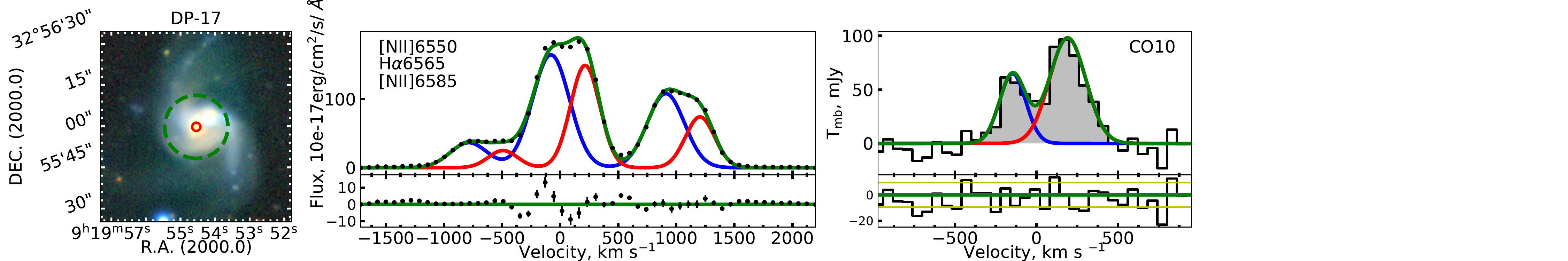}
    }
    \makebox[\linewidth][c]{
    \includegraphics[width=0.98\textwidth]{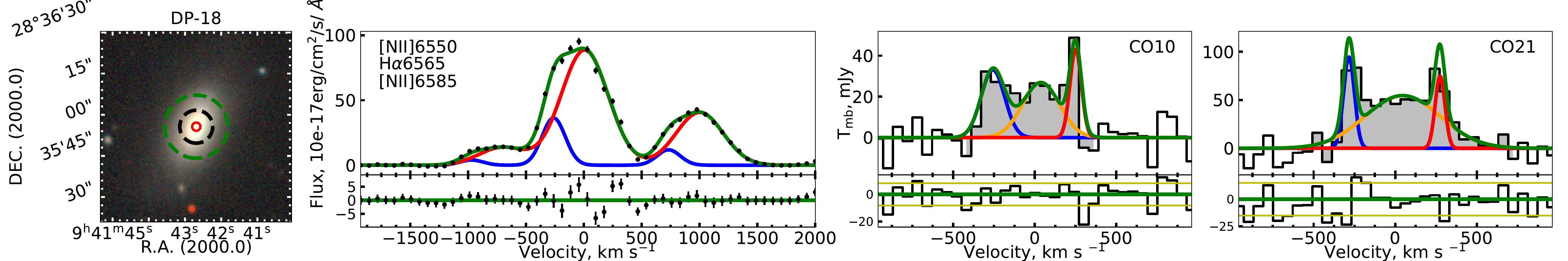}
    }
    \makebox[\linewidth][c]{
    \includegraphics[width=0.98\textwidth]{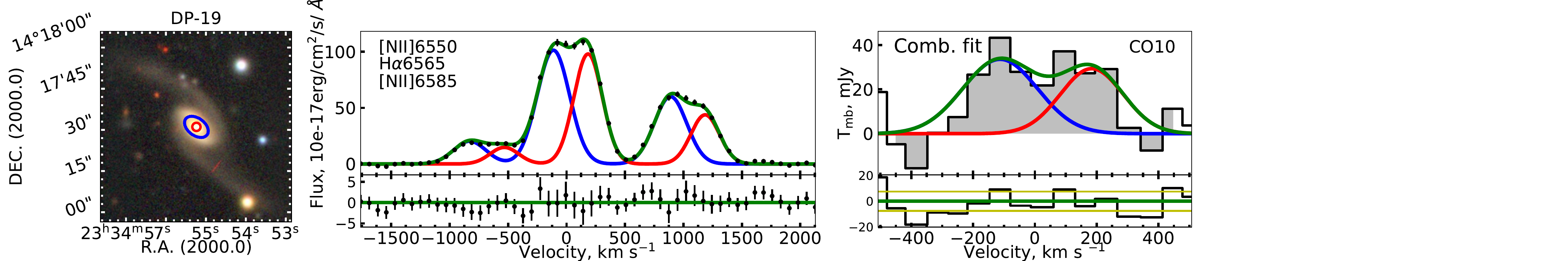}
    }
    \makebox[\linewidth][c]{
    \includegraphics[width=0.98\textwidth]{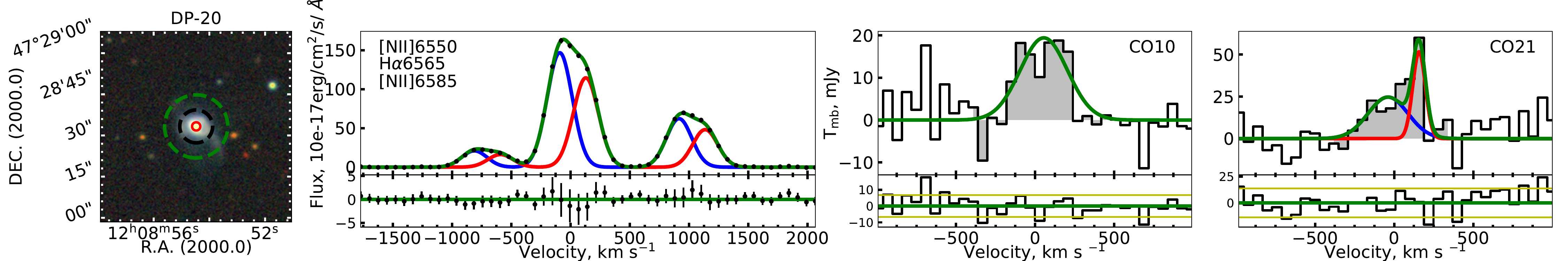}
    }
    \makebox[\linewidth][c]{
    \includegraphics[width=0.98\textwidth]{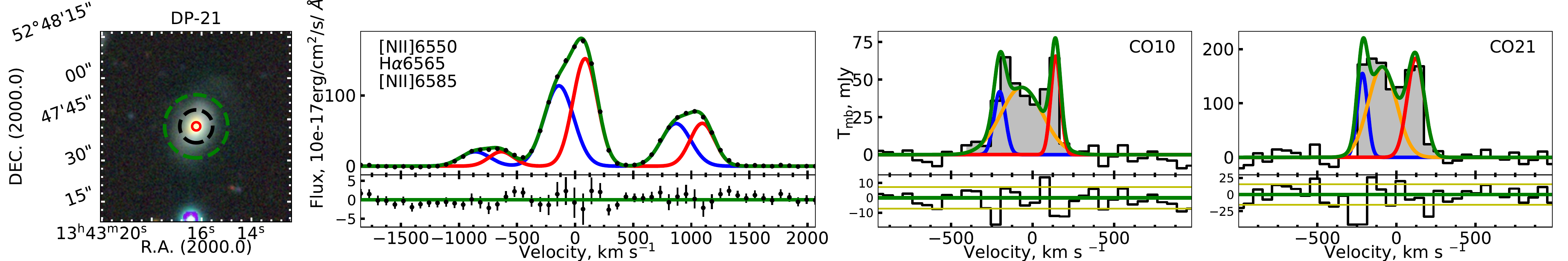}
    }
    \makebox[\linewidth][c]{
    \includegraphics[width=0.98\textwidth]{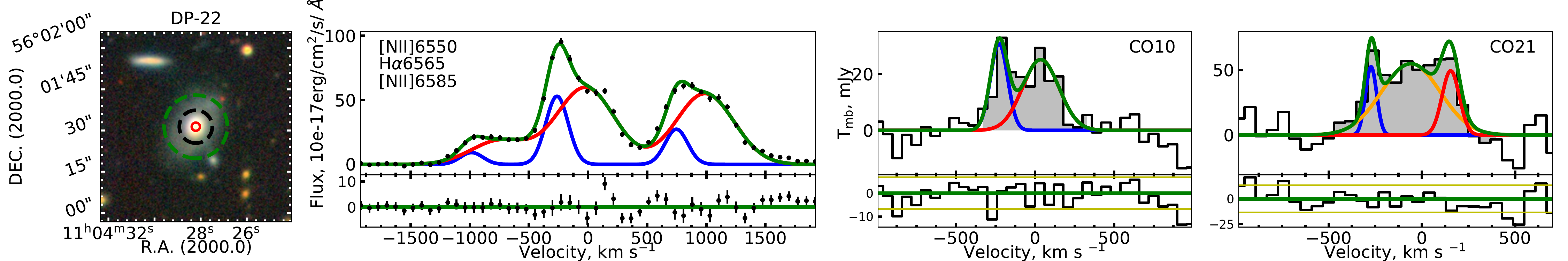}
    }
    \makebox[\linewidth][c]{
    \includegraphics[width=0.98\textwidth]{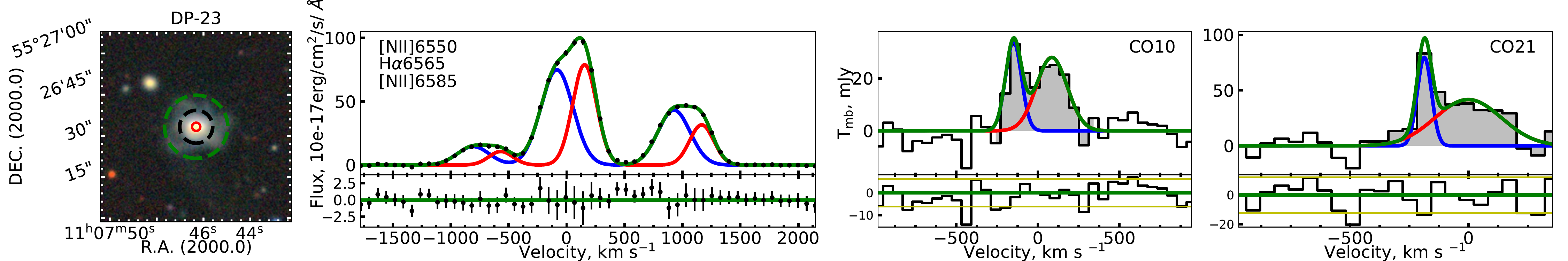}
    }
   \caption{Continued}
    \label{fig:spec_2}
\end{figure*}
\setcounter{figure}{0}
\begin{figure*}[h]
    \makebox[\linewidth][c]{
    \includegraphics[width=0.98\textwidth]{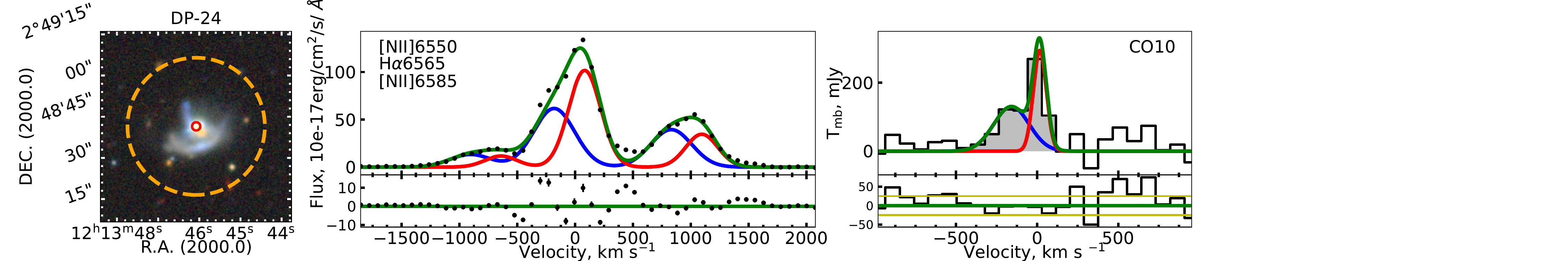}
    }
    \makebox[\linewidth][c]{
    \includegraphics[width=0.98\textwidth]{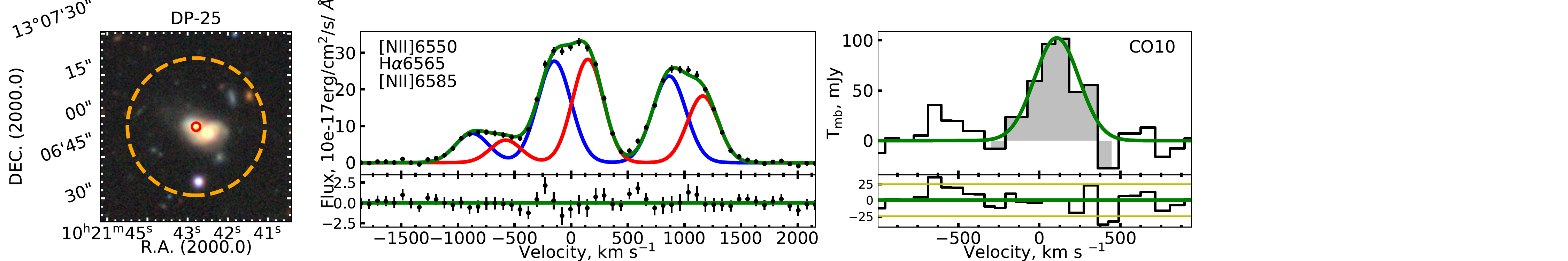}
    }
    \makebox[\linewidth][c]{
    \includegraphics[width=0.98\textwidth]{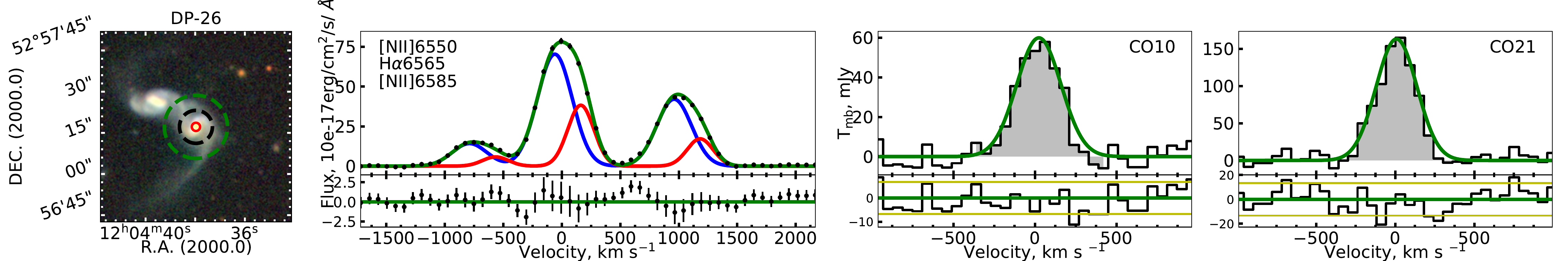}
    }
    \makebox[\linewidth][c]{
    \includegraphics[width=0.98\textwidth]{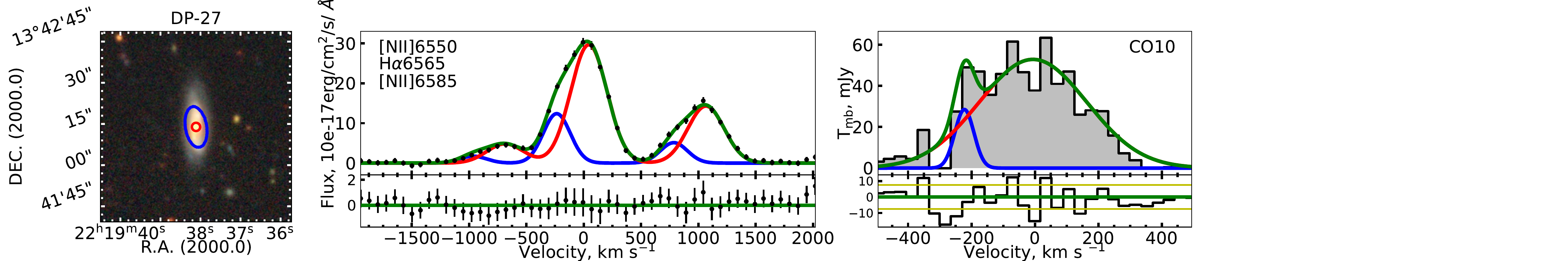}
    }
    \makebox[\linewidth][c]{
    \includegraphics[width=0.98\textwidth]{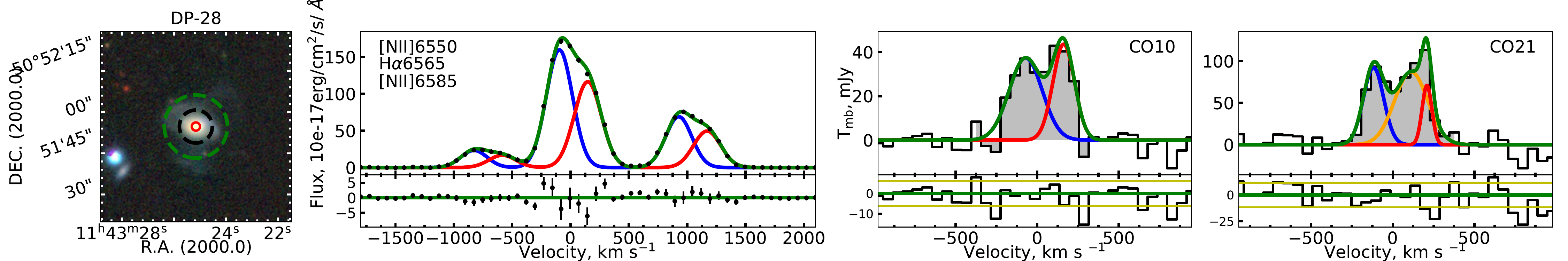}
    }
    \makebox[\linewidth][c]{
    \includegraphics[width=0.98\textwidth]{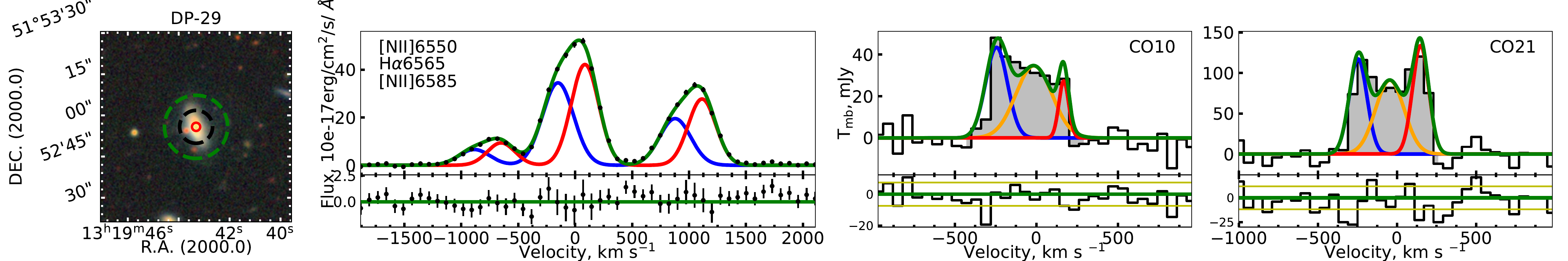}
    }
    \makebox[\linewidth][c]{
    \includegraphics[width=0.98\textwidth]{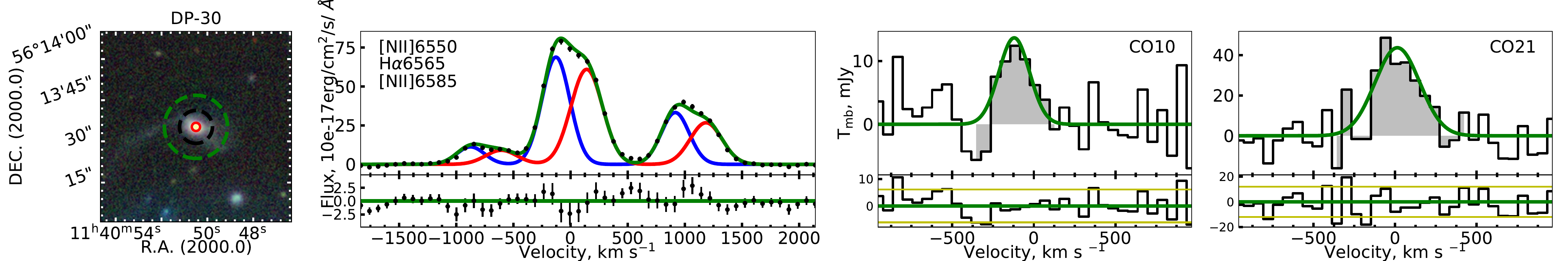}
    }
    \makebox[\linewidth][c]{
    \includegraphics[width=0.98\textwidth]{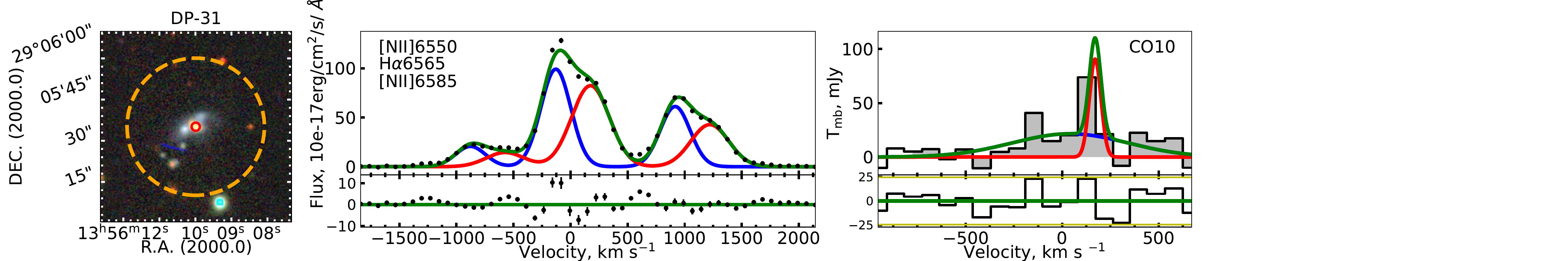}
    }
   \caption{Continued}
    \label{fig:spec_3}
\end{figure*}
\setcounter{figure}{0}
\begin{figure*}[h]
    \makebox[\linewidth][c]{
    \includegraphics[width=0.98\textwidth]{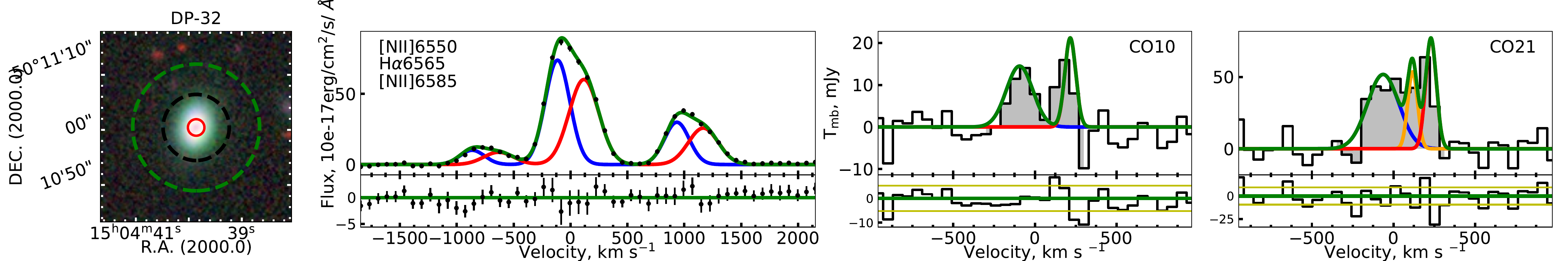}
    }
    \makebox[\linewidth][c]{
    \includegraphics[width=0.98\textwidth]{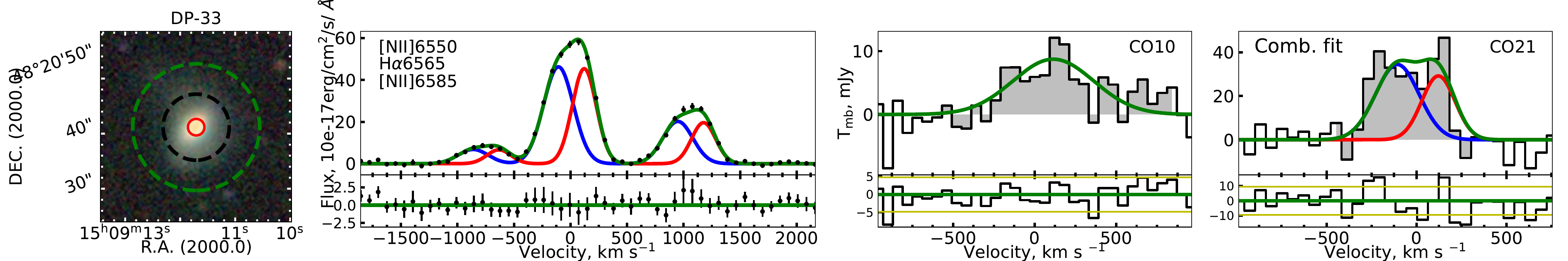}
    }
    \makebox[\linewidth][c]{
    \includegraphics[width=0.98\textwidth]{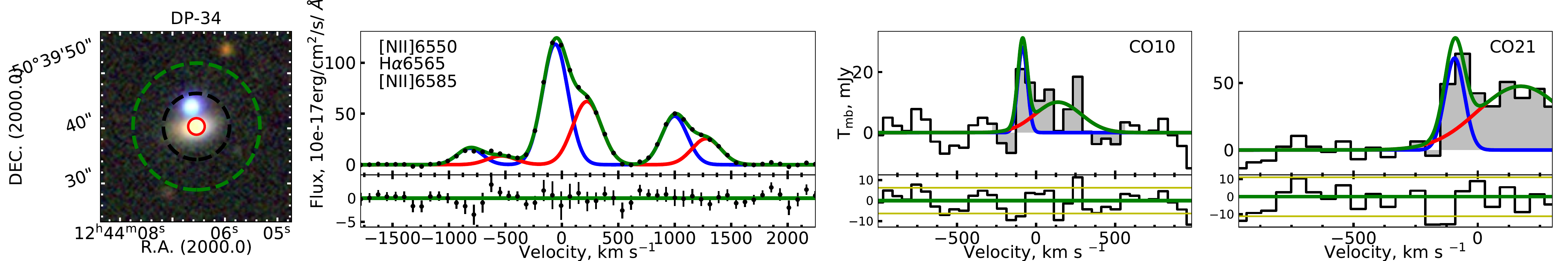}
    }
    \makebox[\linewidth][c]{
    \includegraphics[width=0.98\textwidth]{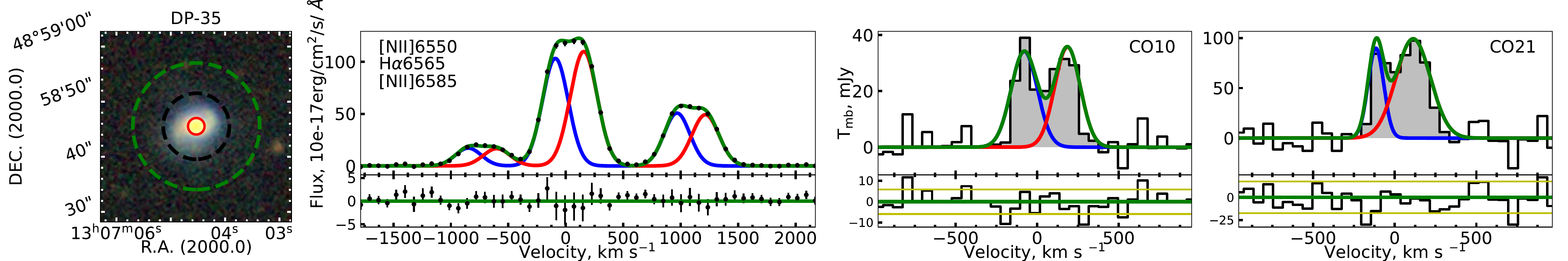}
    }
    \makebox[\linewidth][c]{
    \includegraphics[width=0.98\textwidth]{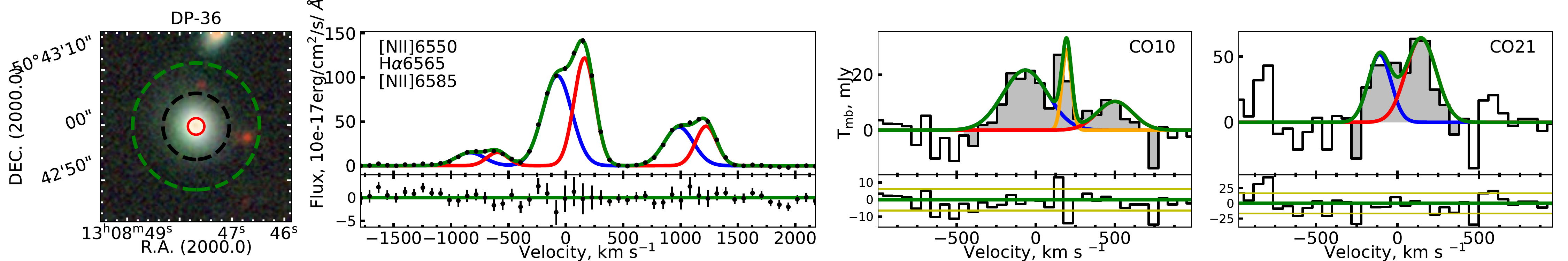}
    }
    \makebox[\linewidth][c]{
    \includegraphics[width=0.98\textwidth]{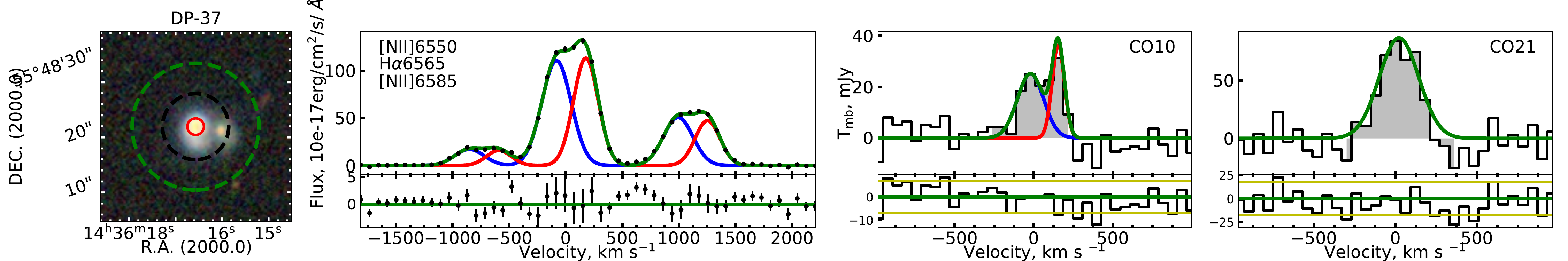}
    }
    \makebox[\linewidth][c]{
    \includegraphics[width=0.98\textwidth]{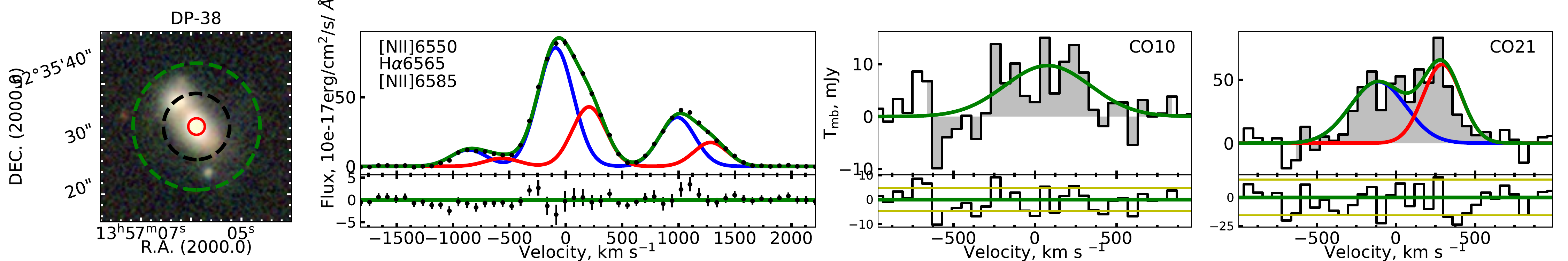}
    }
    \makebox[\linewidth][c]{
    \includegraphics[width=0.98\textwidth]{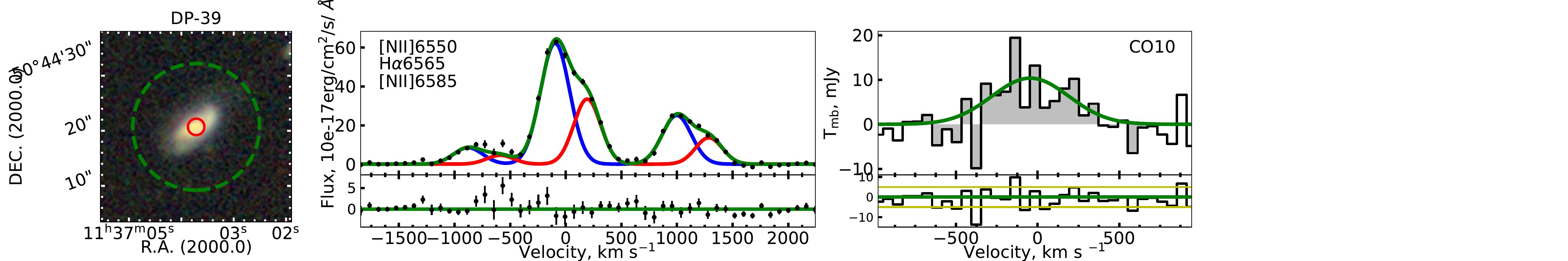}
    }
   \caption{Continued}
    \label{fig:spec_4}
\end{figure*}
\setcounter{figure}{0}
\begin{figure*}[h]
    \makebox[\linewidth][c]{
    \includegraphics[width=0.98\textwidth]{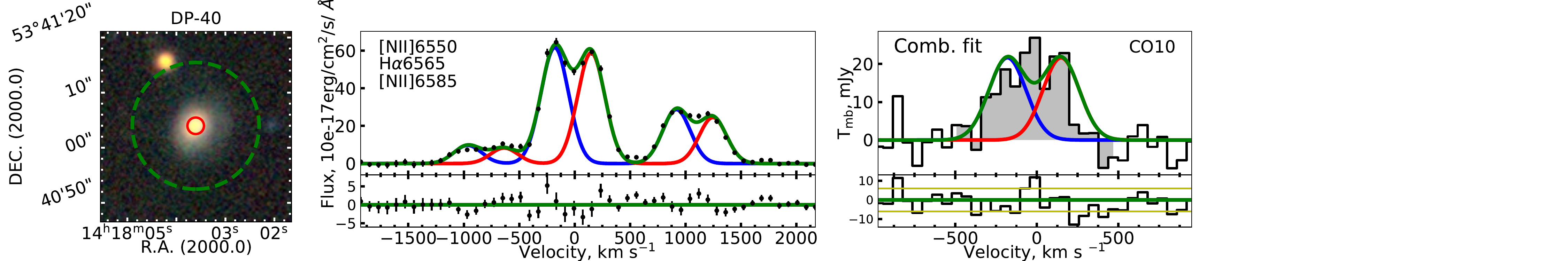}
    }
    \makebox[\linewidth][c]{
    \includegraphics[width=0.98\textwidth]{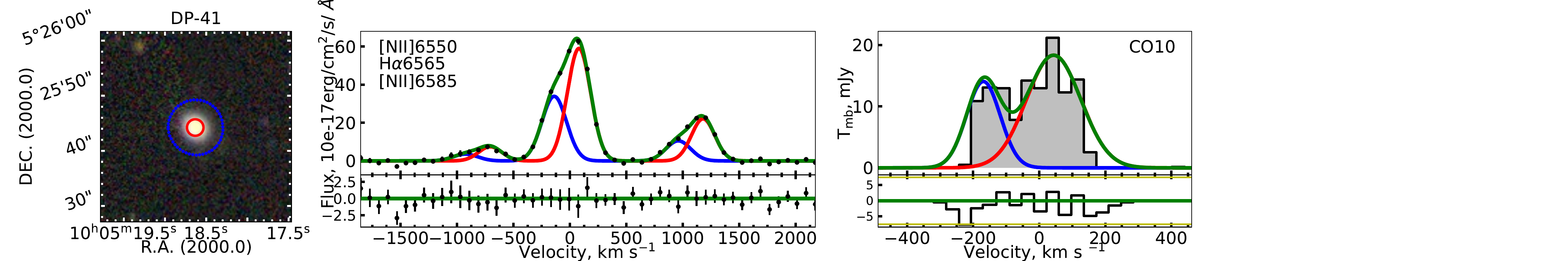}
    }    
    \makebox[\linewidth][c]{
    \includegraphics[width=0.98\textwidth]{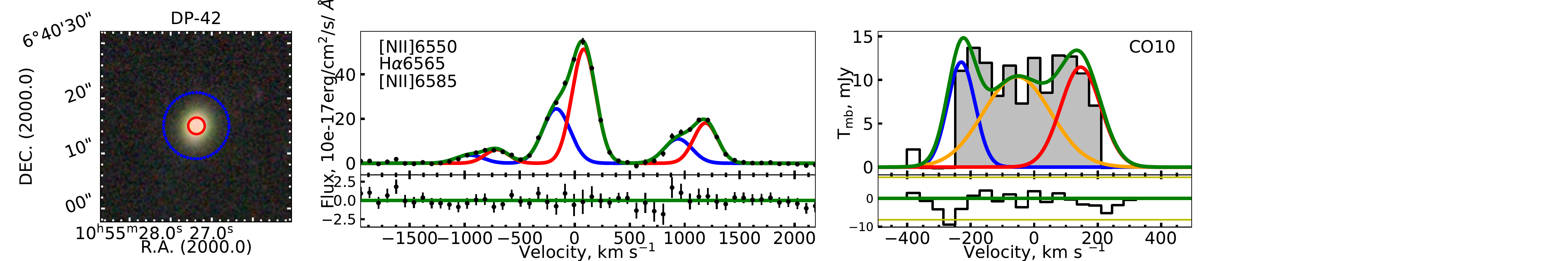}
    }
    \makebox[\linewidth][c]{
    \includegraphics[width=0.98\textwidth]{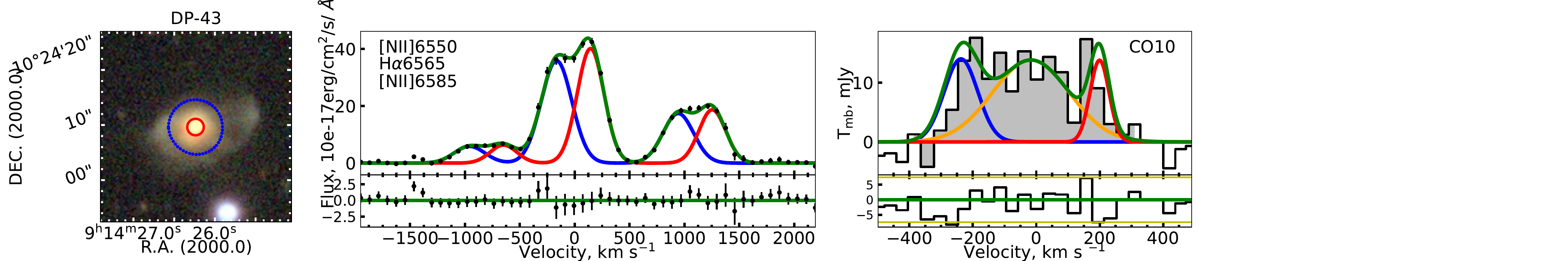}
    }
    \makebox[\linewidth][c]{
    \includegraphics[width=0.98\textwidth]{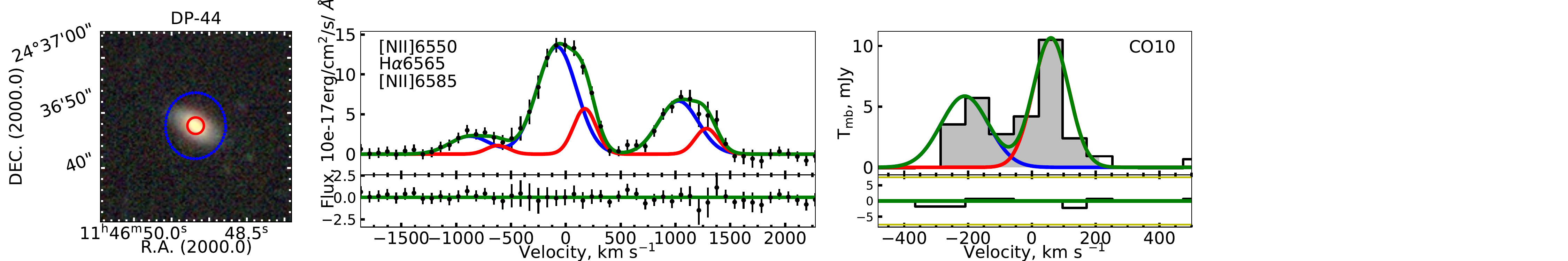}
    }
    \makebox[\linewidth][c]{
    \includegraphics[width=0.98\textwidth]{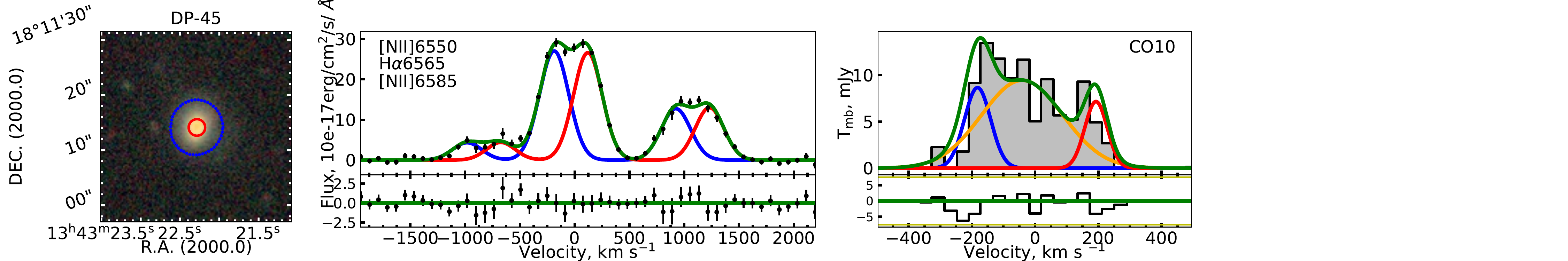}
    }
    \makebox[\linewidth][c]{
    \includegraphics[width=0.98\textwidth]{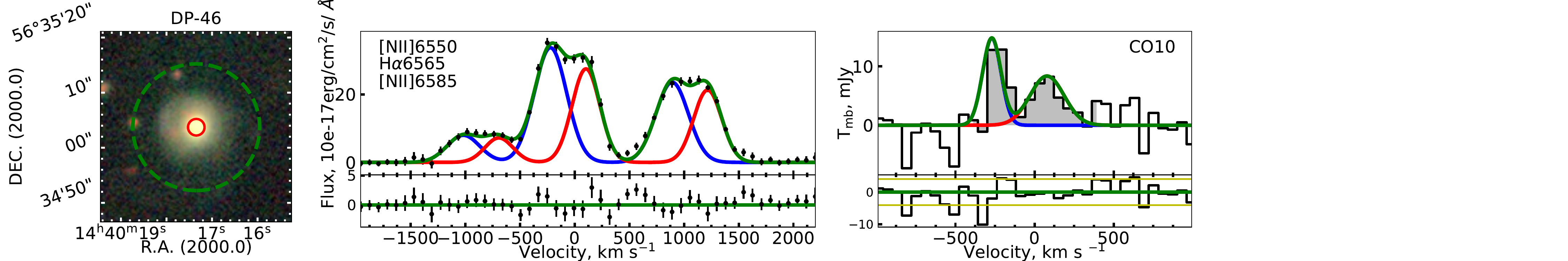}
    }
    \makebox[\linewidth][c]{
    \includegraphics[width=0.98\textwidth]{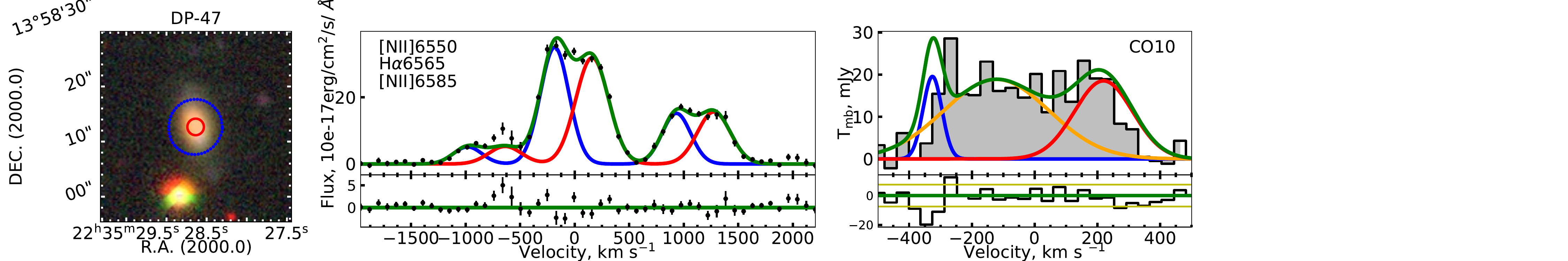}
    }
   \caption{Continued}
    \label{fig:spec_5}
\end{figure*}
\setcounter{figure}{0}
\begin{figure*}[h]
    \makebox[\linewidth][c]{
    \includegraphics[width=0.98\textwidth]{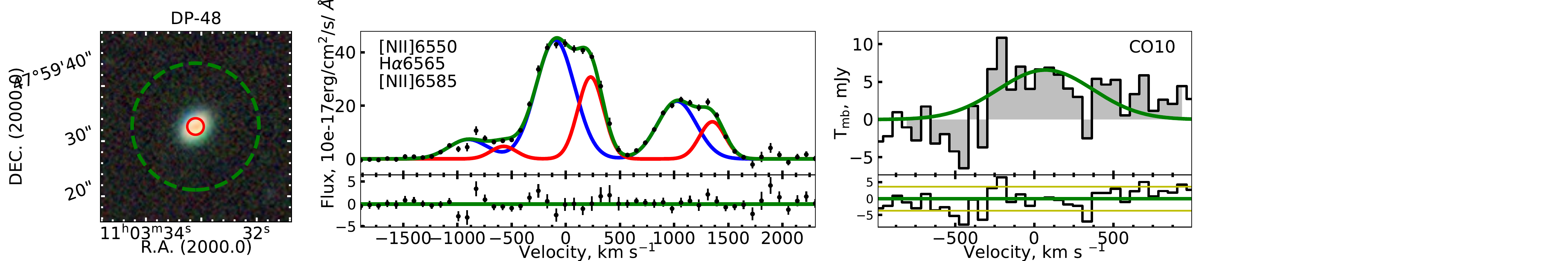}
    }
    \makebox[\linewidth][c]{
    \includegraphics[width=0.98\textwidth]{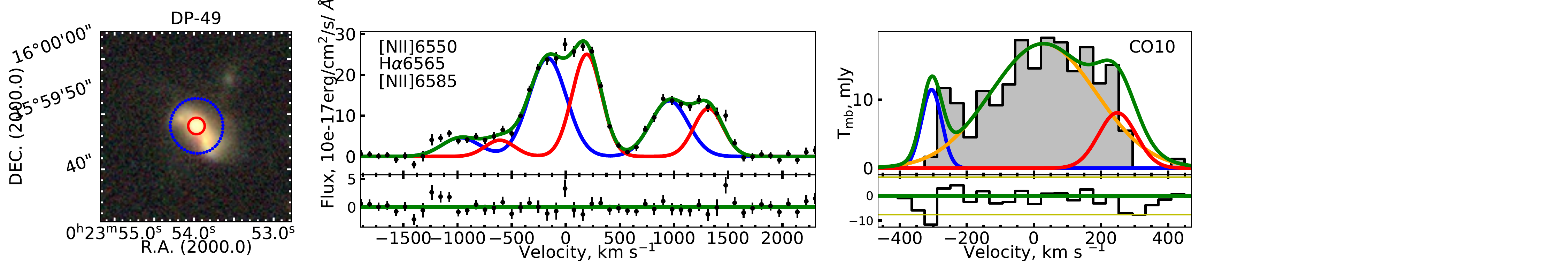}
    }    
    \makebox[\linewidth][c]{
    \includegraphics[width=0.98\textwidth]{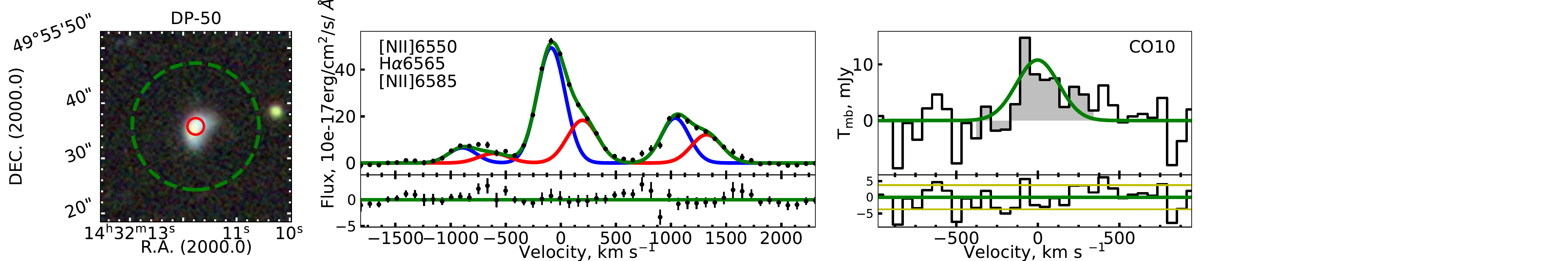}
    }
    \makebox[\linewidth][c]{
    \includegraphics[width=0.98\textwidth]{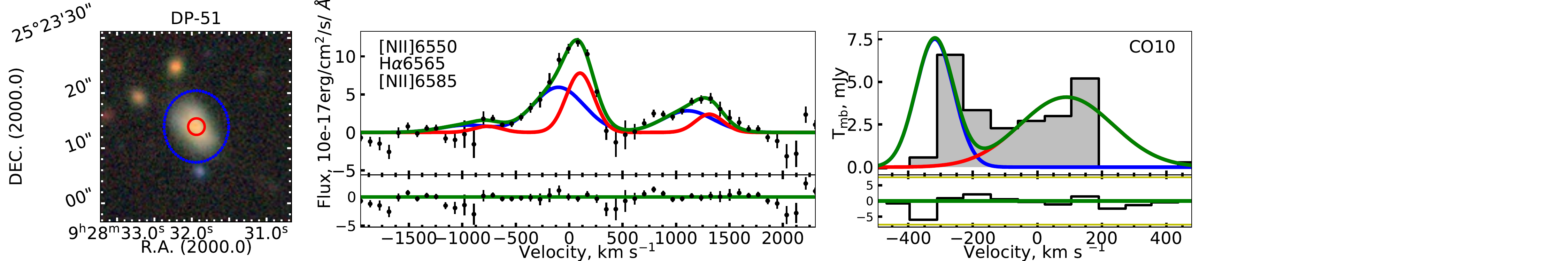}
    }
    \makebox[\linewidth][c]{
    \includegraphics[width=0.98\textwidth]{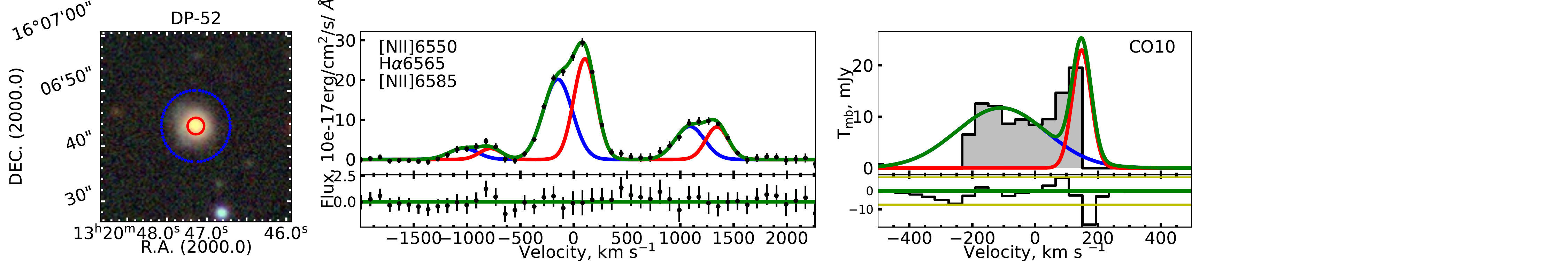}
    }
   \caption{Continued}
    \label{fig:spec_6}
\end{figure*}

\section{Observation Tables}\label{ssect:observation_tables}
In Table\,\ref{table:CO(1-0)-parameters} (resp. \ref{table:CO(2-1)-parameters}), we present the parameters for the CO(1-0) (resp. CO(2-1)) line obtained by our fitting procedure described in Sect.\,\ref{ssect:fitting}. In Table\,\ref{table:molecular_mass} we present the CO-to-M$_{\rm H_2}$ conversion factor, the molecular gas mass, the aperture correction factor, the estimated molecular gas mass fraction, and the depletion time.  
\begin{table*}
\caption{Observed CO(1-0) line parameters}\label{table:CO(1-0)-parameters}
\centering
\begin{tabular}{ l c c c c c c c c c c}        
\hline\hline
\multicolumn{1}{c}{ID} & \multicolumn{1}{c}{Comb. fit} & \multicolumn{1}{c}{I$_{1 \, \, \rm CO(1-0)}$} & \multicolumn{1}{c}{$\mu_1$} & \multicolumn{1}{c}{$\sigma_1$} & \multicolumn{1}{c}{I$_{2 \, \, \rm CO(1-0)}$} & \multicolumn{1}{c}{$\mu_2$} & \multicolumn{1}{c}{$\sigma_2$} & \multicolumn{1}{c}{I$_{3 \, \, \rm CO(1-0)}$} & \multicolumn{1}{c}{$\mu_3$} & \multicolumn{1}{c}{$\sigma_3$} \\ 
\hline
\multicolumn{1}{c}{} & \multicolumn{1}{c}{} & \multicolumn{1}{c}{Jy km s$^-1$} & \multicolumn{1}{c}{km s$^-1$} & \multicolumn{1}{c}{km s$^-1$} & \multicolumn{1}{c}{Jy km s$^-1$} & \multicolumn{1}{c}{km s$^-1$} & \multicolumn{1}{c}{km s$^-1$} & \multicolumn{1}{c}{Jy km s$^-1$} & \multicolumn{1}{c}{km s$^-1$} & \multicolumn{1}{c}{km s$^-1$} \\ 
\hline
DP-1$^{a}$ & 0 & 3.3 $\pm$ 0.1 & -113 $\pm$ 1 & 34 $\pm$ 1 & 37.5 $\pm$ 0.2 & -22 $\pm$ 1 & 138 $\pm$ 1 & 10.9 $\pm$ 0.7 & 21 $\pm$ 1 & 50 $\pm$ 1 \\
DP-2 & 0 & 8.1 $\pm$ 0.1 & -154 $\pm$ 1 & 30 $\pm$ 1 & 22.1 $\pm$ 0.3 & -52 $\pm$ 1 & 71 $\pm$ 1 & 23.1 $\pm$ 0.5 & 124 $\pm$ 1 & 60 $\pm$ 1 \\
DP-3 & 1 & 8.1 $\pm$ 0.4 & -85 $\pm$ 8 & 100 $\pm$ 4 & 7.4 $\pm$ 0.5 & 143 $\pm$ 11 & 94 $\pm$ 6 & -  & -  & -  \\
DP-4 & 0 & 10.3 $\pm$ 0.1 & -92 $\pm$ 1 & 99 $\pm$ 1 & 3.6 $\pm$ 0.2 & 145 $\pm$ 1 & 33 $\pm$ 1 & -  & -  & -  \\
DP-5 & 1 & 21.1 $\pm$ 1.3 & -283 $\pm$ 17 & 157 $\pm$ 9 & 37.6 $\pm$ 1.9 & 67 $\pm$ 15 & 168 $\pm$ 8 & -  & -  & -  \\
DP-6 & 0 & 7.6 $\pm$ 0.2 & -225 $\pm$ 1 & 32 $\pm$ 1 & 17.5 $\pm$ 0.4 & -118 $\pm$ 1 & 90 $\pm$ 1 & 12.9 $\pm$ 0.2 & 155 $\pm$ 1 & 68 $\pm$ 1 \\
DP-7 & 0 & 8.1 $\pm$ 0.2 & 38 $\pm$ 3 & 134 $\pm$ 3 & -  & -  & -  & -  & -  & -  \\
DP-8 & 1 & 27.4 $\pm$ 1.7 & -132 $\pm$ 8 & 105 $\pm$ 6 & 20.8 $\pm$ 1.5 & 129 $\pm$ 8 & 85 $\pm$ 6 & -  & -  & -  \\
DP-9 & 0 & 7.8 $\pm$ 0.2 & -108 $\pm$ 1 & 76 $\pm$ 1 & 8.4 $\pm$ 0.2 & 154 $\pm$ 1 & 67 $\pm$ 1 & -  & -  & -  \\
DP-10 & 0 & 8.6 $\pm$ 0.1 & -64 $\pm$ 1 & 66 $\pm$ 1 & 5.1 $\pm$ 0.2 & 130 $\pm$ 1 & 38 $\pm$ 1 & -  & -  & -  \\
DP-11 & 0 & 8.9 $\pm$ 0.2 & -167 $\pm$ 1 & 76 $\pm$ 1 & 3.7 $\pm$ 0.2 & 95 $\pm$ 1 & 42 $\pm$ 1 & 7.0 $\pm$ 0.2 & 261 $\pm$ 1 & 47 $\pm$ 1 \\
DP-12 & 0 & 18.7 $\pm$ 0.3 & -98 $\pm$ 2 & 168 $\pm$ 2 & 8.5 $\pm$ 0.2 & 250 $\pm$ 1 & 62 $\pm$ 2 & -  & -  & -  \\
DP-13$^{b}$ & 0 & 15.4 $\pm$ 0.2 & -96 $\pm$ 1 & 74 $\pm$ 1 & 13.3 $\pm$ 0.2 & 96 $\pm$ 1 & 43 $\pm$ 1 & -  & -  & -  \\
DP-14 & 0 & 10.0 $\pm$ 0.3 & -104 $\pm$ 1 & 75 $\pm$ 2 & 7.5 $\pm$ 0.2 & 110 $\pm$ 1 & 63 $\pm$ 2 & -  & -  & -  \\
DP-15 & 1 & 6.8 $\pm$ 0.4 & -145 $\pm$ 11 & 99 $\pm$ 5 & 8.5 $\pm$ 0.5 & 93 $\pm$ 12 & 121 $\pm$ 6 & -  & -  & -  \\
DP-16 & 0 & 5.9 $\pm$ 0.1 & -115 $\pm$ 1 & 30 $\pm$ 1 & 12.1 $\pm$ 0.2 & 27 $\pm$ 1 & 66 $\pm$ 1 & -  & -  & -  \\
DP-17$^{b}$ & 0 & 13.1 $\pm$ 0.2 & -146 $\pm$ 1 & 80 $\pm$ 1 & 27.5 $\pm$ 0.3 & 190 $\pm$ 1 & 111 $\pm$ 1 & -  & -  & -  \\
DP-18 & 0 & 6.0 $\pm$ 0.2 & -256 $\pm$ 2 & 71 $\pm$ 2 & 7.6 $\pm$ 0.3 & 38 $\pm$ 4 & 112 $\pm$ 2 & 3.8 $\pm$ 0.2 & 250 $\pm$ 1 & 35 $\pm$ 2 \\
DP-19$^{c}$ & 1 & 10.2 $\pm$ 0.6 & -112 $\pm$ 9 & 120 $\pm$ 6 & 7.5 $\pm$ 0.5 & 184 $\pm$ 8 & 101 $\pm$ 5 & -  & -  & -  \\
DP-20 & 0 & 7.3 $\pm$ 0.2 & 54 $\pm$ 4 & 149 $\pm$ 3 & -  & -  & -  & -  & -  & -  \\
DP-21 & 0 & 3.7 $\pm$ 0.2 & -201 $\pm$ 1 & 35 $\pm$ 1 & 14.4 $\pm$ 0.1 & -67 $\pm$ 1 & 128 $\pm$ 1 & 4.9 $\pm$ 0.4 & 141 $\pm$ 3 & 30 $\pm$ 1 \\
DP-22 & 0 & 4.2 $\pm$ 0.3 & -227 $\pm$ 2 & 54 $\pm$ 2 & 7.3 $\pm$ 0.3 & 35 $\pm$ 4 & 115 $\pm$ 2 & -  & -  & -  \\
DP-23 & 0 & 4.2 $\pm$ 0.3 & -149 $\pm$ 3 & 48 $\pm$ 2 & 6.6 $\pm$ 0.4 & 84 $\pm$ 4 & 94 $\pm$ 2 & -  & -  & -  \\
DP-24$^{d}$ & 0 & 35.6 $\pm$ 0.7 & -158 $\pm$ 1 & 108 $\pm$ 1 & 29.6 $\pm$ 0.6 & 16 $\pm$ 1 & 40 $\pm$ 1 & -  & -  & -  \\
DP-25$^{d}$ & 0 & 35.3 $\pm$ 0.4 & 107 $\pm$ 1 & 137 $\pm$ 1 & -  & -  & -  & -  & -  & -  \\
DP-26 & 0 & 20.4 $\pm$ 0.3 & 24 $\pm$ 1 & 136 $\pm$ 1 & -  & -  & -  & -  & -  & -  \\
DP-27$^{c}$ & 0 & 2.1 $\pm$ 0.2 & -222 $\pm$ 1 & 30 $\pm$ 2 & 22.5 $\pm$ 0.3 & -6 $\pm$ 1 & 169 $\pm$ 2 & -  & -  & -  \\
DP-28 & 0 & 9.5 $\pm$ 0.3 & -70 $\pm$ 4 & 101 $\pm$ 3 & 7.3 $\pm$ 0.3 & 162 $\pm$ 2 & 66 $\pm$ 3 & -  & -  & -  \\
DP-29 & 0 & 7.3 $\pm$ 0.3 & -244 $\pm$ 2 & 67 $\pm$ 2 & 9.6 $\pm$ 0.1 & -15 $\pm$ 1 & 111 $\pm$ 2 & 2.1 $\pm$ 0.4 & 167 $\pm$ 5 & 30 $\pm$ 2 \\
DP-30 & 0 & 3.2 $\pm$ 0.2 & -120 $\pm$ 5 & 93 $\pm$ 3 & -  & -  & -  & -  & -  & -  \\
DP-31$^{d}$ & 0 & 16.2 $\pm$ 0.5 & 40 $\pm$ 9 & 299 $\pm$ 3 & 6.8 $\pm$ 2.5 & 170 $\pm$ 2 & 30 $\pm$ 3 & -  & -  & -  \\
DP-32 & 0 & 2.9 $\pm$ 0.2 & -94 $\pm$ 4 & 80 $\pm$ 4 & 1.8 $\pm$ 0.2 & 216 $\pm$ 2 & 33 $\pm$ 4 & -  & -  & -  \\
DP-33 & 0 & 5.4 $\pm$ 0.5 & 114 $\pm$ 13 & 247 $\pm$ 16 & -  & -  & -  & -  & -  & -  \\
DP-34 & 0 & 2.1 $\pm$ 0.1 & -85 $\pm$ 1 & 30 $\pm$ 1 & 3.7 $\pm$ 0.2 & 138 $\pm$ 9 & 144 $\pm$ 1 & -  & -  & -  \\
DP-35 & 0 & 6.9 $\pm$ 0.3 & -79 $\pm$ 2 & 80 $\pm$ 2 & 7.0 $\pm$ 0.2 & 185 $\pm$ 2 & 78 $\pm$ 2 & -  & -  & -  \\
DP-36 & 0 & 7.7 $\pm$ 0.3 & -66 $\pm$ 4 & 141 $\pm$ 3 & 2.2 $\pm$ 0.1 & 194 $\pm$ 1 & 30 $\pm$ 3 & 2.9 $\pm$ 0.2 & 499 $\pm$ 3 & 112 $\pm$ 3 \\
DP-37 & 0 & 5.2 $\pm$ 0.4 & -20 $\pm$ 4 & 82 $\pm$ 5 & 3.6 $\pm$ 0.2 & 154 $\pm$ 1 & 39 $\pm$ 5 & -  & -  & -  \\
DP-38 & 0 & 6.4 $\pm$ 0.3 & 77 $\pm$ 11 & 262 $\pm$ 9 & -  & -  & -  & -  & -  & -  \\
DP-39 & 0 & 6.1 $\pm$ 0.3 & -45 $\pm$ 11 & 235 $\pm$ 9 & -  & -  & -  & -  & -  & -  \\
DP-40 & 1 & 6.2 $\pm$ 0.2 & -179 $\pm$ 2 & 113 $\pm$ 2 & 6.1 $\pm$ 0.2 & 148 $\pm$ 2 & 112 $\pm$ 2 & -  & -  & -  \\
DP-41$^{c}$ & 0 & 1.8 $\pm$ 0.2 & -169 $\pm$ 4 & 51 $\pm$ 4 & 3.8 $\pm$ 0.3 & 43 $\pm$ 4 & 82 $\pm$ 4 & -  & -  & -  \\
DP-42$^{c}$ & 0 & 1.3 $\pm$ 0.2 & -229 $\pm$ 5 & 43 $\pm$ 4 & 2.8 $\pm$ 0.6 & -54 $\pm$ 23 & 106 $\pm$ 4 & 1.8 $\pm$ 0.4 & 146 $\pm$ 11 & 63 $\pm$ 4 \\
DP-43$^{c}$ & 0 & 1.9 $\pm$ 0.3 & -237 $\pm$ 6 & 53 $\pm$ 7 & 4.2 $\pm$ 0.6 & -16 $\pm$ 13 & 120 $\pm$ 7 & 1.0 $\pm$ 0.1 & 199 $\pm$ 2 & 30 $\pm$ 7 \\
DP-44$^{c}$ & 0 & 1.1 $\pm$ 0.3 & -210 $\pm$ 15 & 74 $\pm$ 14 & 1.5 $\pm$ 0.2 & 59 $\pm$ 8 & 55 $\pm$ 14 & -  & -  & -  \\
DP-45$^{c}$ & 0 & 0.9 $\pm$ 0.2 & -182 $\pm$ 7 & 41 $\pm$ 6 & 3.0 $\pm$ 0.6 & -40 $\pm$ 21 & 128 $\pm$ 6 & 0.6 $\pm$ 0.3 & 192 $\pm$ 6 & 35 $\pm$ 6 \\
DP-46 & 0 & 2.2 $\pm$ 0.2 & -270 $\pm$ 4 & 58 $\pm$ 3 & 2.3 $\pm$ 0.3 & 78 $\pm$ 9 & 107 $\pm$ 3 & -  & -  & -  \\
DP-47$^{c}$ & 0 & 1.5 $\pm$ 0.1 & -325 $\pm$ 2 & 30 $\pm$ 1 & 7.9 $\pm$ 0.6 & -121 $\pm$ 12 & 167 $\pm$ 1 & 4.3 $\pm$ 0.4 & 219 $\pm$ 7 & 92 $\pm$ 1 \\
DP-48 & 0 & 4.9 $\pm$ 0.5 & 70 $\pm$ 19 & 299 $\pm$ 28 & -  & -  & -  & -  & -  & -  \\
DP-49$^{c}$ & 0 & 0.9 $\pm$ 0.1 & -305 $\pm$ 3 & 30 $\pm$ 2 & 7.2 $\pm$ 0.4 & 27 $\pm$ 10 & 157 $\pm$ 2 & 1.1 $\pm$ 0.3 & 248 $\pm$ 9 & 56 $\pm$ 2 \\
DP-50 & 0 & 3.5 $\pm$ 0.4 & -1 $\pm$ 12 & 129 $\pm$ 13 & -  & -  & -  & -  & -  & -  \\
DP-51$^{c}$ & 0 & 1.1 $\pm$ 0.4 & -318 $\pm$ 14 & 58 $\pm$ 16 & 1.5 $\pm$ 0.5 & 90 $\pm$ 38 & 144 $\pm$ 16 & -  & -  & -  \\
DP-52$^{c}$ & 0 & 4.2 $\pm$ 0.3 & -108 $\pm$ 9 & 143 $\pm$ 8 & 1.7 $\pm$ 0.3 & 148 $\pm$ 1 & 30 $\pm$ 8 & -  & -  & -  \\
\hline
\end{tabular}
\tablefoot{The denotations $a$, $b$, $c$, and $d$ are the same as in table \ref{table:sample}.We present CO(1-0) fitting parameters from the fitting procedure described in Sect.\,\ref{ssect:fitting}. We note if we performed a combined fit using the kinematic parameters from the optical ionised gas emission lines from the SDSS spectrum with the flag combined fit. We further present the intensity I$_{CO(1-0)}$, the peak position $\mu$ and the Gaussian $\sigma$ for each line component.}
\end{table*}
\begin{table*}
\caption{Observed CO(2-1) line parameters}\label{table:CO(2-1)-parameters}
\begin{tabular}{ l c c c c c c c c c c}        
\hline\hline
\multicolumn{1}{c}{ID} & \multicolumn{1}{c}{Comb. fit} & \multicolumn{1}{c}{I$_{1 \, \, \rm CO(2-1)}$} & \multicolumn{1}{c}{$\mu_1$} & \multicolumn{1}{c}{$\sigma_1$} & \multicolumn{1}{c}{I$_{2 \, \, \rm CO(2-1)}$} & \multicolumn{1}{c}{$\mu_2$} & \multicolumn{1}{c}{$\sigma_2$} & \multicolumn{1}{c}{I$_{3 \, \, \rm CO(2-1)}$} & \multicolumn{1}{c}{$\mu_3$} & \multicolumn{1}{c}{$\sigma_3$} \\ 
\hline
\multicolumn{1}{c}{} & \multicolumn{1}{c}{} & \multicolumn{1}{c}{Jy km s$^-1$} & \multicolumn{1}{c}{km s$^-1$} & \multicolumn{1}{c}{km s$^-1$} & \multicolumn{1}{c}{Jy km s$^-1$} & \multicolumn{1}{c}{km s$^-1$} & \multicolumn{1}{c}{km s$^-1$} & \multicolumn{1}{c}{Jy km s$^-1$} & \multicolumn{1}{c}{km s$^-1$} & \multicolumn{1}{c}{km s$^-1$} \\ 
\hline
DP-2 & 0 & 13.4 $\pm$ 0.2 & -156 $\pm$ 1 & 36 $\pm$ 1 & 79.0 $\pm$ 0.6 & -28 $\pm$ 1 & 103 $\pm$ 1 & 13.7 $\pm$ 0.3 & 160 $\pm$ 1 & 42 $\pm$ 1 \\
DP-3 & 1 & 22.3 $\pm$ 1.0 & -85 $\pm$ 8 & 100 $\pm$ 4 & 15.3 $\pm$ 1.0 & 143 $\pm$ 11 & 94 $\pm$ 6 & -  & -  & -  \\
DP-4 & 1 & 27.9 $\pm$ 0.9 & -137 $\pm$ 4 & 90 $\pm$ 2 & 23.6 $\pm$ 0.9 & 103 $\pm$ 5 & 91 $\pm$ 3 & -  & -  & -  \\
DP-5 & 1 & 50.5 $\pm$ 3.0 & -283 $\pm$ 17 & 157 $\pm$ 9 & 89.7 $\pm$ 4.4 & 67 $\pm$ 15 & 168 $\pm$ 8 & -  & -  & -  \\
DP-6 & 0 & 21.6 $\pm$ 0.2 & -232 $\pm$ 1 & 42 $\pm$ 1 & 29.4 $\pm$ 0.3 & -28 $\pm$ 1 & 135 $\pm$ 1 & -  & -  & -  \\
DP-7 & 0 & 20.9 $\pm$ 0.2 & 42 $\pm$ 1 & 120 $\pm$ 1 & -  & -  & -  & -  & -  & -  \\
DP-8 & 1 & 49.3 $\pm$ 3.0 & -132 $\pm$ 8 & 105 $\pm$ 6 & 42.5 $\pm$ 3.1 & 129 $\pm$ 8 & 85 $\pm$ 6 & -  & -  & -  \\
DP-9 & 0 & 8.5 $\pm$ 0.1 & -149 $\pm$ 1 & 38 $\pm$ 1 & 19.8 $\pm$ 0.2 & 76 $\pm$ 1 & 96 $\pm$ 1 & -  & -  & -  \\
DP-10 & 1 & 31.8 $\pm$ 1.3 & -75 $\pm$ 8 & 110 $\pm$ 4 & 17.3 $\pm$ 1.0 & 137 $\pm$ 9 & 87 $\pm$ 5 & -  & -  & -  \\
DP-11 & 0 & 8.8 $\pm$ 0.1 & -222 $\pm$ 1 & 30 $\pm$ 1 & 40.0 $\pm$ 0.4 & -13 $\pm$ 1 & 149 $\pm$ 1 & 13.3 $\pm$ 0.2 & 260 $\pm$ 1 & 52 $\pm$ 1 \\
DP-12 & 0 & 69.5 $\pm$ 0.4 & 35 $\pm$ 1 & 238 $\pm$ 1 & 18.7 $\pm$ 0.2 & 295 $\pm$ 1 & 50 $\pm$ 1 & -  & -  & -  \\
DP-13$^{b}$ & 0 & 13.6 $\pm$ 0.3 & -135 $\pm$ 1 & 36 $\pm$ 1 & 23.2 $\pm$ 0.3 & -20 $\pm$ 1 & 80 $\pm$ 1 & 21.6 $\pm$ 0.5 & 112 $\pm$ 3 & 36 $\pm$ 1 \\
DP-14 & 0 & 32.1 $\pm$ 0.1 & -11 $\pm$ 1 & 123 $\pm$ 1 & 6.6 $\pm$ 0.3 & 112 $\pm$ 1 & 30 $\pm$ 1 & -  & -  & -  \\
DP-15 & 1 & 11.8 $\pm$ 0.7 & -145 $\pm$ 11 & 99 $\pm$ 5 & 14.6 $\pm$ 0.7 & 93 $\pm$ 12 & 121 $\pm$ 6 & -  & -  & -  \\
DP-16 & 1 & 27.9 $\pm$ 1.2 & -125 $\pm$ 6 & 91 $\pm$ 3 & 31.0 $\pm$ 1.2 & 92 $\pm$ 5 & 83 $\pm$ 3 & -  & -  & -  \\
DP-18 & 0 & 7.9 $\pm$ 0.2 & -283 $\pm$ 1 & 33 $\pm$ 1 & 31.6 $\pm$ 0.4 & 45 $\pm$ 2 & 230 $\pm$ 1 & 5.6 $\pm$ 0.1 & 274 $\pm$ 1 & 30 $\pm$ 1 \\
DP-20 & 0 & 7.8 $\pm$ 0.3 & -38 $\pm$ 4 & 126 $\pm$ 3 & 5.2 $\pm$ 0.2 & 157 $\pm$ 1 & 40 $\pm$ 3 & -  & -  & -  \\
DP-21 & 0 & 12.6 $\pm$ 0.2 & -215 $\pm$ 1 & 32 $\pm$ 1 & 37.5 $\pm$ 0.4 & -89 $\pm$ 1 & 89 $\pm$ 1 & 24.7 $\pm$ 0.3 & 121 $\pm$ 1 & 53 $\pm$ 1 \\
DP-22 & 0 & 3.9 $\pm$ 0.1 & -272 $\pm$ 1 & 30 $\pm$ 1 & 21.7 $\pm$ 0.4 & -59 $\pm$ 2 & 157 $\pm$ 1 & 5.7 $\pm$ 0.2 & 154 $\pm$ 1 & 46 $\pm$ 1 \\
DP-23 & 0 & 6.0 $\pm$ 0.1 & -186 $\pm$ 1 & 30 $\pm$ 1 & 14.6 $\pm$ 0.4 & -1 $\pm$ 3 & 139 $\pm$ 1 & -  & -  & -  \\
DP-26 & 0 & 52.9 $\pm$ 0.3 & 8 $\pm$ 1 & 129 $\pm$ 1 & -  & -  & -  & -  & -  & -  \\
DP-28 & 0 & 14.4 $\pm$ 0.3 & -117 $\pm$ 1 & 62 $\pm$ 1 & 22.2 $\pm$ 0.4 & 112 $\pm$ 1 & 101 $\pm$ 1 & 5.3 $\pm$ 0.1 & 209 $\pm$ 1 & 30 $\pm$ 1 \\
DP-29 & 0 & 16.5 $\pm$ 0.3 & -246 $\pm$ 1 & 56 $\pm$ 1 & 20.7 $\pm$ 0.7 & -45 $\pm$ 1 & 90 $\pm$ 1 & 17.3 $\pm$ 0.3 & 148 $\pm$ 1 & 51 $\pm$ 1 \\
DP-30 & 0 & 15.0 $\pm$ 0.3 & 13 $\pm$ 1 & 137 $\pm$ 1 & -  & -  & -  & -  & -  & -  \\
DP-32 & 0 & 12.6 $\pm$ 0.2 & -63 $\pm$ 1 & 96 $\pm$ 1 & 4.0 $\pm$ 0.1 & 117 $\pm$ 1 & 30 $\pm$ 1 & 6.3 $\pm$ 0.2 & 229 $\pm$ 1 & 32 $\pm$ 1 \\
DP-33 & 1 & 10.4 $\pm$ 1.0 & -106 $\pm$ 23 & 120 $\pm$ 11 & 6.8 $\pm$ 0.6 & 121 $\pm$ 16 & 92 $\pm$ 8 & -  & -  & -  \\
DP-34 & 0 & 6.8 $\pm$ 0.2 & -93 $\pm$ 1 & 39 $\pm$ 1 & 20.7 $\pm$ 0.5 & 172 $\pm$ 2 & 174 $\pm$ 1 & -  & -  & -  \\
DP-35 & 0 & 10.8 $\pm$ 0.2 & -116 $\pm$ 1 & 47 $\pm$ 1 & 26.4 $\pm$ 0.3 & 112 $\pm$ 1 & 106 $\pm$ 1 & -  & -  & -  \\
DP-36 & 0 & 9.0 $\pm$ 0.2 & -109 $\pm$ 2 & 69 $\pm$ 1 & 15.1 $\pm$ 0.3 & 144 $\pm$ 1 & 93 $\pm$ 1 & -  & -  & -  \\
DP-37 & 0 & 26.5 $\pm$ 0.2 & 21 $\pm$ 1 & 121 $\pm$ 1 & -  & -  & -  & -  & -  & -  \\
DP-38 & 0 & 21.3 $\pm$ 0.6 & -113 $\pm$ 5 & 175 $\pm$ 4 & 18.8 $\pm$ 0.6 & 291 $\pm$ 3 & 120 $\pm$ 4 & -  & -  & -  \\
\hline
\end{tabular}
\tablefoot{The same as Table~\ref{table:CO(1-0)-parameters} but for CO(2-1) measurements}
\end{table*}
\begin{table*}
\caption{Results of molecular gas mass estimation }\label{table:molecular_mass}
\centering
\begin{tabular}{ l c c c c c c c c}        
\hline\hline
\multicolumn{1}{c}{ID} & \multicolumn{1}{c}{L$^{\prime}_{\rm CO(1-0)}$} & \multicolumn{1}{c}{L$^{\prime}_{\rm CO(2-1)}$} & \multicolumn{1}{c}{$\alpha_{CO}$} & \multicolumn{1}{c}{M$_{\rm H_2}$} & \multicolumn{1}{c}{f$_{a \,{\rm CO(1-0)}}$} & \multicolumn{1}{c}{M$_{\rm corr \, H_2}$} & \multicolumn{1}{c}{log($\mu_{\rm gas}$)} & \multicolumn{1}{c}{t$_{\rm depl}$} \\ 
\hline
\multicolumn{1}{c}{} & \multicolumn{1}{c}{10$^8$ L$_{l}^{d}$} & \multicolumn{1}{c}{10$^8$ L$_{l}^{d}$} & \multicolumn{1}{c}{M$_{\odot}$ / (K km s$^-1$ pc$^2$)} & \multicolumn{1}{c}{10$^9$ M$_{\odot}$} & \multicolumn{1}{c}{} & \multicolumn{1}{c}{10$^9$ M$_{\odot}$} & \multicolumn{1}{c}{} & \multicolumn{1}{c}{Gyr} \\ 
\hline
DP-1$^{a}$ & 7.7 $\pm$ 0.1  &  -   & 3.9 & 3.0 & 1.0 & 3.0 & -1.4 & 0.6 \\
DP-2 & 17.5 $\pm$ 0.2  & 8.7 $\pm$ 0.1  & 3.9 & 6.9 & 1.6 & 11.0 & -0.9 & 1.4 \\
DP-3 & 5.2 $\pm$ 0.2  & 3.1 $\pm$ 0.1  & 3.8 & 1.9 & 1.9 & 3.7 & -1.0 & 1.0 \\
DP-4 & 4.7 $\pm$ 0.1  & 4.4 $\pm$ 0.1  & 3.8 & 1.8 & 1.6 & 2.9 & -1.0 & 0.8 \\
DP-5 & 36.2 $\pm$ 1.4  & 21.6 $\pm$ 0.8  & 4.0 & 14.5 & 1.9 & 28.0 & -0.7 & 2.1 \\
DP-6 & 23.6 $\pm$ 0.3  & 7.9 $\pm$ 0.0  & 4.0 & 9.4 & 1.6 & 15.5 & -0.9 & 2.1 \\
DP-7 & 5.1 $\pm$ 0.2  & 3.3 $\pm$ 0.0  & 3.8 & 1.9 & 1.4 & 2.6 & -1.1 & 0.4 \\
DP-8 & 31.7 $\pm$ 1.5  & 15.1 $\pm$ 0.7  & 3.9 & 12.2 & 1.2 & 14.6 & -0.7 & 1.4 \\
DP-9 & 11.1 $\pm$ 0.2  & 4.8 $\pm$ 0.0  & 3.8 & 4.2 & 1.3 & 5.5 & -0.9 & 0.6 \\
DP-10 & 9.5 $\pm$ 0.2  & 8.6 $\pm$ 0.3  & 3.9 & 3.7 & 1.3 & 4.8 & -1.3 & 0.1 \\
DP-11 & 13.7 $\pm$ 0.2  & 10.9 $\pm$ 0.1  & 3.9 & 5.4 & 1.3 & 7.1 & -1.2 & 1.0 \\
DP-12 & 20.4 $\pm$ 0.3  & 16.5 $\pm$ 0.1  & 4.1 & 8.3 & 1.6 & 13.3 & -1.1 & 1.8 \\
DP-13$^{b}$ & 24.0 $\pm$ 0.2  & 12.2 $\pm$ 0.2  & 3.9 & 9.3 & 1.4 & 12.7 & -0.8 & 1.0 \\
DP-14 & 16.9 $\pm$ 0.4  & 9.3 $\pm$ 0.1  & 3.8 & 6.5 & 1.3 & 8.7 & -0.9 & 0.8 \\
DP-15 & 15.0 $\pm$ 0.6  & 6.5 $\pm$ 0.2  & 3.9 & 5.9 & 1.7 & 9.7 & -1.0 & 1.4 \\
DP-16 & 18.9 $\pm$ 0.2  & 15.5 $\pm$ 0.4  & 3.8 & 7.2 & 1.1 & 8.3 & -0.9 & 0.3 \\
DP-17$^{b}$ & 44.9 $\pm$ 0.4  &  -   & 3.8 & 16.9 & 2.1 & 34.7 & 0.1 & 3.8 \\
DP-18 & 23.3 $\pm$ 0.6  & 15.1 $\pm$ 0.1  & 4.0 & 9.3 & 1.4 & 12.7 & -1.0 & 1.7 \\
DP-19$^{c}$ & 31.6 $\pm$ 1.3  &  -   & 3.9 & 12.4 & 1.0 & 12.4 & -0.9 & 0.6 \\
DP-20 & 15.5 $\pm$ 0.5  & 6.9 $\pm$ 0.2  & 3.8 & 5.8 & 1.1 & 6.4 & -0.9 & 0.4 \\
DP-21 & 51.1 $\pm$ 1.0  & 41.4 $\pm$ 0.3  & 4.0 & 20.3 & 1.2 & 23.7 & -0.7 & 0.6 \\
DP-22 & 26.3 $\pm$ 1.0  & 18.0 $\pm$ 0.3  & 3.9 & 10.2 & 1.3 & 12.8 & -0.9 & 1.2 \\
DP-23 & 25.7 $\pm$ 1.1  & 12.2 $\pm$ 0.3  & 3.9 & 9.9 & 1.2 & 11.8 & -0.9 & 0.5 \\
DP-24$^{d}$ & 162.2 $\pm$ 2.2  &  -   & 3.8 & 60.9 & 1.0 & 61.0 & 0.2 & 3.5 \\
DP-25$^{d}$ & 95.9 $\pm$ 1.0  &  -   & 3.9 & 37.8 & 1.0 & 38.7 & -0.5 & 2.0 \\
DP-26 & 63.5 $\pm$ 0.8  & 41.1 $\pm$ 0.2  & 3.8 & 24.4 & 1.3 & 31.4 & -0.4 & 1.4 \\
DP-27$^{c}$ & 80.4 $\pm$ 1.2  &  -   & 3.9 & 31.5 & 1.0 & 31.5 & -0.6 & 2.7 \\
DP-28 & 55.8 $\pm$ 1.4  & 34.9 $\pm$ 0.4  & 3.9 & 21.7 & 1.1 & 24.5 & -0.6 & 0.9 \\
DP-29 & 72.5 $\pm$ 2.0  & 52.0 $\pm$ 0.7  & 3.9 & 28.4 & 1.1 & 31.3 & -0.6 & 1.7 \\
DP-30 & 17.1 $\pm$ 1.0  & 20.1 $\pm$ 0.3  & 3.8 & 6.5 & 1.1 & 7.3 & -0.9 & 0.2 \\
DP-31$^{d}$ & 128.9 $\pm$ 14.1  &  -   & 4.0 & 51.3 & 1.0 & 52.6 & -0.5 & 0.9 \\
DP-32 & 28.3 $\pm$ 1.5  & 34.3 $\pm$ 0.4  & 3.8 & 10.7 & 1.1 & 11.3 & -0.7 & 0.7 \\
DP-33 & 36.7 $\pm$ 3.1  & 29.1 $\pm$ 2.1  & 3.8 & 13.9 & 1.1 & 15.1 & -0.7 & 1.2 \\
DP-34 & 40.3 $\pm$ 1.7  & 47.9 $\pm$ 0.8  & 3.8 & 15.2 & 1.1 & 16.1 & -0.6 & 0.7 \\
DP-35 & 100.1 $\pm$ 2.5  & 66.9 $\pm$ 0.7  & 4.0 & 39.8 & 1.1 & 41.8 & -0.6 & 0.5 \\
DP-36 & 94.0 $\pm$ 2.6  & 44.3 $\pm$ 0.7  & 3.8 & 36.0 & 1.1 & 38.3 & -0.4 & 1.1 \\
DP-37 & 82.7 $\pm$ 4.4  & 62.1 $\pm$ 0.6  & 3.9 & 31.8 & 1.0 & 33.4 & -0.5 & 0.5 \\
DP-38 & 63.5 $\pm$ 3.2  & 99.3 $\pm$ 2.0  & 3.9 & 24.6 & 1.1 & 27.5 & -0.7 & 0.8 \\
DP-39 & 75.9 $\pm$ 4.1  &  -   & 3.8 & 28.6 & 1.0 & 30.0 & -0.4 & 0.6 \\
DP-40 & 158.9 $\pm$ 3.8  &  -   & 3.9 & 61.4 & 1.1 & 64.6 & -0.3 & 1.0 \\
DP-41$^{c}$ & 74.4 $\pm$ 4.2  &  -   & 3.8 & 28.0 & 1.0 & 28.0 & -0.3 & 0.6 \\
DP-42$^{c}$ & 85.6 $\pm$ 10.2  &  -   & 3.8 & 32.7 & 1.0 & 32.7 & -0.5 & 0.7 \\
DP-43$^{c}$ & 106.8 $\pm$ 10.4  &  -   & 4.1 & 43.9 & 1.0 & 43.9 & -0.8 & 0.7 \\
DP-44$^{c}$ & 39.2 $\pm$ 5.4  &  -   & 3.8 & 15.0 & 1.0 & 15.0 & -0.9 & 0.1 \\
DP-45$^{c}$ & 70.4 $\pm$ 10.1  &  -   & 4.0 & 28.2 & 1.0 & 28.2 & -0.9 & 0.4 \\
DP-46 & 69.7 $\pm$ 5.9  &  -   & 4.1 & 28.4 & 1.1 & 31.5 & -0.9 & 1.6 \\
DP-47$^{c}$ & 223.9 $\pm$ 12.0  &  -   & 4.1 & 91.1 & 1.0 & 91.1 & -0.5 & 1.0 \\
DP-48 & 87.5 $\pm$ 9.3  &  -   & 3.8 & 33.3 & 1.0 & 34.1 & -0.5 & 1.6 \\
DP-49$^{c}$ & 164.6 $\pm$ 9.1  &  -   & 4.0 & 65.0 & 1.0 & 65.0 & -0.5 & 1.2 \\
DP-50 & 64.3 $\pm$ 8.2  &  -   & 3.8 & 24.1 & 1.0 & 24.9 & -0.3 & 2.2 \\
DP-51$^{c}$ & 104.1 $\pm$ 24.8  &  -   & 3.8 & 40.0 & 1.0 & 40.0 & -0.6 & 1.6 \\
DP-52$^{c}$ & 292.6 $\pm$ 20.8  &  -   & 3.9 & 115.4 & 1.0 & 115.4 & -0.4 & 1.8 \\
\hline
\end{tabular}
\tablefoot{The denotations $a$, $b$, $c$, and $d$ are the same as in Table \ref{table:sample}. We present the total intrinsic CO(1-0) (resp. CO(2-1)) luminosity L$^{\prime}_{\rm CO(1-0)}$ (resp.  L$^{\prime}_{\rm CO(2-1)}$) with  L$_{l} = {\rm K \, km \, s^{-1} \, pc^2}$, the luminosity-to-molecular gas mass conversion factor $\alpha_{CO}$, the measured molecular gas mass M$_{\rm H_2}$, the aperture correction factor f$_{a \,{\rm CO10}}$, the aperture corrected molecular gas mass M$_{\rm corr \, H_2}$, the mass fraction $\mu_{\rm gas}$ and the depletion time t$_{\rm depl}$.}
\end{table*}

\end{document}